\documentclass[12pt]{article}

\usepackage[english]{babel}
\usepackage[utf8]{inputenc}
\usepackage{mathtools}
\usepackage{graphicx}
\usepackage[colorinlistoftodos]{todonotes}
\usepackage[margin=0.9in]{geometry}
\usepackage{multicol}
\usepackage{amssymb}
\usepackage{amsmath}
\usepackage{mathabx}
\usepackage{physics}
\usepackage{slashed}
\usepackage{hyperref}
\usepackage{csquotes}
\usepackage{tikz-cd}

\newcommand{\be}{\begin{equation}}   
	\newcommand{\ee}{\end{equation}}

\newcommand{\refb}[1]{(\ref{#1})}

\newcommand{\df}{{\rm d}}
\newcommand{\ds}{s^{-1}} 

\numberwithin{equation}{section}

\begin{document}
	\begin{titlepage}
		\rightline{}
		\rightline{August 11, 2026} 
		\rightline{MIT-CTP/6087} 
        \rightline{HU-EP-26/25} 
		\begin{center}
	\vskip 1.1cm
	{\Large \bf {Universal quadratic field equations}}\\[2.0ex] 
    {\Large \bf {via homotopy algebras }}
	\vskip 1.4cm
			
	{\large\bf {Christoph Chiaffrino,\textsuperscript{\textnormal 1} Raji Ashenafi Mamade,\textsuperscript{\textnormal 2}  \\[1.3ex] and Barton Zwiebach\textsuperscript{\textnormal 2}}} 
			\vskip 1cm
			\vskip .3cm
			 \textsuperscript{\textnormal 1}{\it   Institut f\"ur Physik, \\
				Humboldt-Universität zu Berlin,\\
				Zum großen Windkanal 2, 12489 Berlin, Germany}\\
			
			\vskip .5cm
           \textsuperscript{\textnormal 2}{\it   Center for Theoretical Physics -- a Leinweber Institute, 
			\\
				Massachusetts Institute of Technology, \\
				Cambridge MA 02139, USA}\\
            
			\vskip .4cm
			christoph.mario.chiaffrino@hu-berlin.de, raji@mit.edu, zwiebach@mit.edu

			\vskip 1.6cm
		\end{center}

		\begin{quote} 	
			
			\centerline{\textbf{ Abstract}} 
			\bigskip
			We explain how the `bar–cobar'
            construction for homotopy algebras reformulates the equations of motion of arbitrary gauge theories
            as gauge-covariant quadratic equations for an extended set of fields.
            The linear term of the new equations 
            encodes
            the interactions of the original theory, while the quadratic term is universal. The extended fields include type-I multilocal fields,
            which depend on a  
            set of coordinates and
            type-II multilocal fields, which depend on several sets of coordinates. 
            The new equations of motion are 
            the Maurer-Cartan equations of a 
            differential graded associative algebra or Lie algebra. We show that every solution of the new equations is gauge equivalent to a solution with only type-I fields, that
            represents 
            a solution of the original equations of motion. 
            In string theory   
            type-I fields are entangled states of the associated CFT, to be inserted across multiple punctures of a Riemann surface. General type-II states can also represent disconnected Riemann surfaces.

			\medskip

		\end{quote} 
		\vfill
		\setcounter{footnote}{0}
		
		\setcounter{tocdepth}{2}  
		
	\end{titlepage}
		
	\baselineskip 17pt

	\tableofcontents

\section{Introduction and perspectives}

It is well known that classical bosonic open string field theory can be constructed in the framework of differential graded \emph{associative}  algebras~\cite{Witten:1985cc}.  
The action is cubic in the string field and the equations of motion quadratic. The role of the differential is played by the BRST operator and the product is based on an associative `star' multiplication rule for open strings.  As it turned out, this is a special case. 
For classical bosonic closed strings one could have expected a string field theory based on a differential graded {\it Lie algebra}, also with a cubic action and quadratic field equations. Such an action, however, does not seem to exist.   
Bosonic closed string field theory~\cite{Zwiebach:1992ie} is based on an $L_\infty$ algebra, a homotopy 
Lie algebra. The action is nonpolynomial and so are the equations of motion. 
More generally, none of the superstring theories -- Type I, heterotic, or Type II -- has a cubic formulation or quadratic field equations~\cite{Sen:2024nfd}.  More generally, homotopy algebras have proven useful in the studies of perturbative quantum field theories~\cite{Jurco:2020yyu,Chiaffrino:2020akd, Bernardes:2025uzg,Bernardes:2026ofg, Arvanitakis:2020rrk, Arvanitakis:2021ecw, Erbin:2020eyc, Cabus:2025inm, Maccaferri:2023gcg, Doubek:2020rbg, Munster:2011ij,Konosu:2024zrq}.

Cubic open string field theory has enabled the construction of many classical solutions, motivating considerable interest in finding
a cubic closed string field theory. This cannot be achieved if one uses a single closed string field~\cite{Sonoda:1989sj}.
With additional 
auxiliary closed string fields a cubic formulation could be possible.  The same may be true of open string field theories that, based on homotopy associative $A_\infty$ algebras, are also nonpolynomial~\cite{Erler:2013xta,Erler:2015lya,Kunitomo:2015usa}. 
An example is provided by~\cite{Erbin:2023hcs}, where it was shown that the stubbed version of
bosonic open string field theory, which is based on an $A_\infty$ algebra, can be reformulated in terms of a differential graded {\it associative} algebra by introducing an auxiliary open string field. 
There are also other avenues being pursued; instead of looking for a theory based on a strict Lie algebra, one seeks simpler data determining higher interactions. These are 
aims in the bootstrap approach~\cite{Firat:2023glo} and in setting topological recursions for the geometry behind bosonic closed string field theory~\cite{Ishibashi:2024kdv, Firat:2024ajp}. 

Early on, the bar-cobar construction of homotopy algebras was brought to the attention of physicists, suggesting that it could result in cubic formulations of the string field theories, as well as cubic formulations for all types of field theories\footnote{One of us, BZ, first heard this suggestion from James Stasheff in the early 1990s.}. It was also known that it would involve a fantastically large number of extra fields. Perhaps surprisingly, this suggestion has not been investigated in any detail. This is the goal of this paper. 

There are several questions to which we give brief answers.
\begin{enumerate}
    \item {\em What is the nature of the extra fields that are added?}
The extra fields are, in general, multilocal versions of the original fields. Thus, for a scalar field $\phi(x)$ one ends up including new fields
$\phi (x_1, \ldots, x_n)$ with $n \geq 2$, among others. These will be called
{\em type-I} multilocal fields.  Letting 
$\{ x\}_k   \equiv  \{ x^{(k)}_1, \ldots ,  x^{(k)}_k \} $ denote a collection of coordinates, we also have {\em type-II} multilocal fields $\phi ( \{ x\}_{k_1} | \cdots 
| \{ x\}_{k_f})$, using vertical bars to separate the dependence on the 
various collections of coordinates.

\item   {\it In field theories the vector space is graded by ghost number, which defines a degree.
The classical field is constrained to a fixed 
degree, whereas the full space is
relevant to the quantum field.  What happens on the enlarged state space?}
The enlarged state space is also graded, and the new classical field also emerges from a fixed degree subspace.
That fixed degree subspace receives contributions from subspaces of different degrees in the original space
of the field theory. 
We have not explored how the quantum theory is formulated in the enlarged space.

\item
 {\it  How do we see that the equations of motion for the enlarged set of fields are equivalent to the original equations of motion? }
 This is in fact the bulk of the work we do here.  We show that
 the general solution of 
 the equations of motion is a gauge transformation of a solution 
 having only type-I multilocality.  One can then prove that such solutions are
 in one-to-one correspondence with the solutions of the original field equations.

\item  {\it Is an action available for the dynamics of the enlarged set of fields?}
Not from the bar–cobar construction alone: it does not itself provide the cyclic bilinear form needed to write an action.  We believe
it will be challenging to find an action principle for the new equations of motion. In this paper we leave open the possible existence
of a cubic action from which these field equations, or an equivalent version of them, would follow by variation. 

\end{enumerate}

The basic point of the bar-cobar construction can be explained once three definitions are in place.
While more details are given in Section~\ref{dgvsaac}, here are the main facts. 
\begin{itemize}
\item A differential graded 
associative algebra $(A, m_1 , m_2)$.  Here, $(A, m_1) $ is a differential graded vector space, 
with $A$ a graded vector space and $m_1$ (often labeled as `$\df$') a degree minus
one differential:  $m_1 m_1= 0$.  
Moreover, $(A, m_2)$ is an algebra with a degree zero associative product
$m_2: A \otimes A \to A$.  Finally, $m_1$ is a derivation with respect to $m_2$. 

In this structure one can define the homology $H(A)$ based on $m_1$ as well as 
the Maurer-Cartan equation
$m_1 a + m_2 (a,a) = 0$ for elements $a \in A_{-1}$ of degree minus one.  In a physical
theory, the homology is usually identified with the BRST physical states, and the Maurer-Cartan equation is the
classical field equation. We call the set of gauge-equivalence classes of solutions the Maurer–Cartan set. It is the set of genuine deformations of 
the differential graded algebra $A$ and physically represents the theory expanded
around inequivalent consistent backgrounds.

\item A differential graded {\em coalgebra} $(C, \bar B, \bar \Delta)$ 
where the coalgebra  $(C, \bar\Delta)$ is a vector space $C$ with a degree zero coproduct $\bar \Delta:  C \to C \otimes C$ that is coassociative: $ (\textbf{1} \otimes \bar \Delta ) \circ 
\bar \Delta = (\bar\Delta \otimes \textbf{1}) \circ \bar \Delta$.   
Moreover, $(C, \bar B)$ is a differential graded vector space, meaning that 
$\bar B$ is a degree minus one coderivation
with respect to $\bar\Delta$, and $\bar B^2 = 0$.  

\item 
A tensor algebra $( T(V), \mu)$   
with $\mu$ the associative product, 
and the related tensor coalgebra $T^c(V)= (T(V), \bar\Delta)$.
Given a graded vector space $V$ with vectors  $a_i$, called {\em letters,} consider the vector space 
\be
\label{tvdefscoal}
T(V) = \bigoplus_{n=1}^\infty V^ {\otimes n }\,,
\ee  
Elements of $V^{\otimes n}$ are {\em words} written as $a_1 a_2 \cdots a_n$, ommitting the $\otimes$ in between the letters.
In the algebra $T(V)$ the product $m_2$ of two words is defined by concatenation:
\be \mu (a_1 \cdots a_n , b_1\cdots b_m) =  a_1 \cdots a_n b_1 \cdots b_m\,.
\ee
In the tensor coalgebra the coproduct $\bar \Delta$ acts by deconcatenation as follows:
\be \bar\Delta (a_1a_2 \cdots a_n ) = a_1 \otimes' a_2 \cdots a_n \, + a_1 a_2 \otimes' a_3 \cdots a_n \, +
\cdots  \, +  a_1 \cdots a_{n-1} \otimes' a_n \,, \ee
for $n>1$ and $\bar \Delta a_i = 0$.
\end{itemize}

The {\em bar construction} takes a differential graded algebra $(A, m_1, m_2)$ and builds a graded coalgebra $B(A)$ given by
\be B(A)=
(T^c (sA) , \bar B ,\bar \Delta)\,. 
\ee
Here $T^c(sA)$ is the tensor coalgebra over $sA$, with $sA$ the {\em suspension} of the 
graded vector space~$A$. The 
suspension of a vector space is an operation that affects signs 
by setting the degree $k$ subspace of $sA$ equal to the degree $k-1$ subspace of $A$. 
The $m_1, m_2$ operations on $A$  
and on $A\otimes A$
induce, respectively, operations
$b_1$ on $sA$ and  $b_2$ on $sA\otimes sA$, both of degree minus one. 
Moreover, $b_1$ and $b_2$ can be lifted to coderivations $\bar B_1$ and $\bar B_2$
on $T^c(sA)$:
\be
m_1 \ \to \ b_1 \ \to \ \bar B_1  \,, \ \ \ m_2 \ \to\  b_2 \ \to \ \bar B_2\,. 
\ee
Adding $\bar B_1$ and $\bar B_2$ we get a degree minus one coderivation 
\be
\bar B= \bar B_1 + \bar B_2\,,
\ee 
on  $T^c(sA)$. One finds that $\bar B^2=0$.  The operation $\bar \Delta$ is the standard degree-zero coproduct of the tensor coalgebra $T^c(sA)$.

The `cobar' construction is the dual of the bar construction: it takes a differential graded coalgebra
$( C, \bar B, \bar \Delta)$ and produces a differential graded associative algebra 
\be
\Omega (C)=( T(s^{-1}C), D\equiv  B + \Delta, \, \otimes')\, .
\ee
Here the vector space $T(s^{-1} C)$ is the tensor {\em algebra} of the desuspended space $s^{-1}C$, and $\otimes'$ has a prime to denote the product in this algebra. The operations $B$ and $\Delta$ on $s^{-1}C$ 
are induced by the desuspension from $\bar B$ and $\bar \Delta$, respectively. 
Both  $B$ and $\Delta$ are extended 
to $T(s^{-1}C)$ as derivations of the tensor algebra. 
Remarkably, the suspensions result in $\Delta^2 =0$ as well as
$B^2=0$ and  $[ B , \Delta]  = 0$.\footnote{All our commutators are graded: $[a, b] \equiv a b - (-1)^{ab} ba $.} This results in $D^2=0$, as required for a differential graded algebra.

The bar-cobar construction starts with a differential graded associative algebra $(A, m_1, m_2)$,  applies bar to this algebra to get
the coalgebra $B(A)$ and then cobar on $B(A)$ to get an algebra $\Omega B (A)$.  
Thus, the construction begins and ends with a differential graded algebra, the original one on the vector space $A$ and the final algebra on the much larger space 
\be 
\label{fint309}
T (s^{-1} T^c(sA)) = T (s^{-1} C) \,, \ \ \   C \equiv  T^c(V)\,, \ \ V =  sA \,.
\ee
The main claim is that the two differential graded algebras, while living in completely different spaces, are quasi-isomorphic: their homologies are the same, 
and the elements of their Maurer-Cartan 
sets are in one-to-one correspondence.

There is also a bar construction that starts with a differential graded Lie algebra and gives a
symmetric coalgebra, as well as a cobar construction that starts with a symmetric 
coalgebra and gives a differential graded Lie algebra.  Thus, a bar-cobar construction also exists for differential graded Lie algebras. 

For arbitrary field theories and string field theories, the relevant framework, 
as elaborated in~\cite{Hohm:2017pnh}, is that of
$A_\infty$ or $L_\infty$ algebras, rather than 
differential graded associative or Lie algebras. Although these algebras are neither strictly associative 
nor strictly Lie, the bar-cobar
construction remains relevant. In both the $A_\infty$
and $L_\infty$ algebras, the various multilinear products $b_n$ can be extended
to coderivations $\bar B_n$ on a suitable tensor coalgebra with the
sum $\bar B = \sum_n \bar B_n$ being a nilpotent coderivation.
This reformulates  the $\infty$-algebra
as a standard differential graded coalgebra and it is called the bar construction of
the $\infty$-algebra.\footnote{Such a coalgebra, of symmetric type for the Lie case, neatly summarizes the algebraic 
structure of string field theory, as noted in the early formulations~\cite{Zwiebach:1992ie}.} 

Applying the cobar construction one finally gets a differential graded
associative or  Lie algebra, with a quadratic Maurer-Cartan 
equation. In the field theory interpretation, this means that the
resulting field equations are quadratic. 

The Maurer–Cartan equation of motion for the field $\Psi$ in the bar–cobar version of the theory takes the form
\be
\label{33riflk}
(\Delta + B) \Psi + \Psi \otimes' \Psi = 0 \,, \ \ \ \  \Psi\in T(s^{-1} C) , \  C = T^c (V)\,.
\ee
The operations $\Delta$ and $\otimes'$ are universal, defined
in the same way for all field theories.  The information specific to a 
particular field theory resides in the vector space to which $\Psi$ belongs, 
which is constructed from the vector space $V= sA$
of the original
theory, and in $B$, which encodes the multilinear products of the original theory.
Thus the linear term in the equation of motion incorporates the interactions
of the original theory, while the quadratic term of the equation of motion
is universal.

In order to understand this equation, we began by studying in detail
the full set of solutions of the simpler, universal, Maurer-Cartan equation obtained
by setting $B=0$ in~\eqref{33riflk}: 
\be
\label{33riflkx}
\Delta \Phi + \Phi \otimes' \Phi = 0\,,  \ \  \  \  \Phi\in T(s^{-1} C)  , \  C = T^c (V)\,. 
\ee
We calculate the $\Delta$ homology explicitly
 and use this information to show that every solution
of this equation is gauge equivalent 
to a {\em canonical} solution---one in which only the fields in $s^{-1}C$ are nonzero. 
These are type-I multilocal fields, all of which   
turn out to be products of local fields at different points. 

The general knowledge of the solutions of~\eqref{33riflkx} allows us to 
approach the analysis of the physical equations~\eqref{33riflk} using
perturbation theory in the coderivation $B$.
Our analysis  shows that, again, any solution 
is gauge-equivalent to a {\em canonical solution}: a solution
in which only
type-I multilocal fields are nonzero, while still equal to
products of local fields.  
Those are the fields in $s^{-1} C$.
 Such solutions are naturally identified with the solutions
of the original field theory equation of motion. 
Moreover, we 
show how the gauge transformations of the original theory arise from a
class of gauge transformations of the bar-cobar theory.
Reaching these
conclusions is the main technical work in this paper.

\medskip
A few features of the bar-cobar 
construction of quadratic equations  
are seen in the simplified problem of replacing a
quartic interaction of a field $\psi(x)$ with a cubic interaction by including
auxiliary fields. Let $S$ be an action that in addition to terms 
indicated by dots below, contains a rather general quartic 
interaction:
\be
S =  \cdots + \int d^4x_1 d^4 x_2  d^4 x_3 d^4x_4  \ V(x_1, x_2, x_3, x_4) \ \psi(x_1) \psi(x_2) \psi(x_3) \psi(x_4) \,. 
\ee
Here, 
the fully symmetric potential $V(x_1, x_2, x_3, x_4)$ could be nonlocal --
that would happen if $V$ is nonvanishing when the
$x$'s are not coincident. We can replace this action with the action $S'$ that includes
a new bilocal field $\phi (x_1, x_2)$ as follows
\be
\begin{split}
S' =  \cdots \   & - 
\int d^4x_1 d^4 x_2  d^4 x_3 d^4x_4 \  V(x_1, x_2, x_3, x_4) \ \phi (x_1, x_2) \phi (x_3, x_4) \\
&  + 2 \int d^4x_1 d^4 x_2  d^4 x_3 d^4x_4 \  V(x_1, x_2, x_3, x_4) \ \phi (x_1, x_2) \, \psi(x_3) \psi(x_4)\,.
\end{split} 
\ee
The first term is quadratic for the bilocal field and the second term is a cubic interaction of the bilocal field and two original fields. 
The variation of $S'$ 
with respect to the bilocal field gives
\be
\delta_\phi S' =  
\int d^4x_1 d^4 x_2  d^4 x_3 d^4x_4 \  V(x_1, x_2, x_3, x_4) \ \delta\phi (x_1, x_2) 
 \bigl( -2 \phi (x_3, x_4) + 2  \psi(x_3) \psi(x_4)\Bigr) \,. 
\ee
Assuming nondegeneracy of the potential, the $\phi$ equation of motion then sets 
\be
\label{theprodsol}
\phi (x_1, x_2) =  \psi(x_1) \psi(x_2) \,,
\ee
and substitution back in $S'$ gives the original action $S$. 
A rather similar situation
is encountered in the so-called Hubbard-Stratonovich transformation used in condensed matter physics to simplify actions containing interactions of four-electron operators, as reviewed by Weinberg~\cite{Weinberg:1996kr}.  Some of the features above are also observed in the rewriting based on the bar-cobar construction. In particular, we will have multilocal
fields, and often solutions are factorized as in~\eqref{theprodsol}. In the above example, if the interactions were local, one could make the related cubic theory local. 
We want a construction that yields quadratic equations for all theories, including nonlocal ones, and therefore multilocal fields seem unavoidable. The bar-cobar construction is such a universal construction and has multilocal fields.

\bigskip
It is useful to characterize the bar-cobar field $\Psi$ associated to a field theory for
a field $\phi$.  Let~$A$ denote the vector space for the description
of the field theory of $\phi$. In the so-called  $b$-picture, we use the 
space $V = sA$ and
the classical field $\phi$ lies on the degree-zero 
subspace $V_0$ of~$V$:
\be
\label{fieldloc}
\phi = \sum_i \,  \varphi_i \,  v^i  \,, \ \ \   \hbox{with} \ \  v^i \in  
V_0 =  (sA)_0\,.
\ee
Here, $\varphi_i$ is the target space field associated with 
the vector $v^i$.   
As mentioned above, the bar-cobar field  $\Psi \in T(s^{-1} T^c (sA))$, 
and therefore 
\be
\Psi \in T(s^{-1} T^c(V))\,.
\ee   
Moreover the field $\Psi$ must be of degree minus one. 
The outer $T$ in $\Psi$ means a new tensor product $\otimes'$ is used for this tensor
algebra. Thus we write the expansion
\be
\begin{split}
T(s^{-1} T^c(V)) =   & \quad \ \  s^{-1}T^c(V) \\  
& \ \oplus \bigl( s^{-1}T^c(V) \otimes' s^{-1}T^c(V)\bigr) \\
& \ \oplus 
\Bigl( s^{-1}T^c(V) \otimes' s^{-1}T^c(V) \otimes' s^{-1}T^c(V)\Bigr)  \oplus  \cdots \\
 = & \  \bigoplus_{f=1}^\infty   (s^{-1} T^c(V))^{\otimes' f} \,.
\end{split}
\ee
We call $(s^{-1} T^c(V))^{\otimes' f}$ the
subspace of $T(s^{-1} T^c(V))$ of {\em factor number} $f$. 
The field $\Psi$ contains
components $\Psi_f$  at all factor numbers $f=1,2,\ldots\, $. We write 
\be  \Psi = \sum_{f=1}^\infty  \Psi_f \,.
\ee  
To appreciate the field content, consider
first factor number one.  An element of $s^{-1}T^c(V)$ 
is of the form
\be
s^{-1} (v^1 \otimes v^2 \otimes \cdots  \otimes v^\ell) =  s^{-1} (v^1 v^2  \cdots v^\ell)  \in s^{-1}T^c(V)\,,
\ee
where the $\otimes$ is suppressed without possibility of confusion, as noted
below~\eqref{tvdefscoal}. For this
object to be of degree minus one, the sum of the degrees of the various $v$'s must 
add up to zero, since $s^{-1}$ contributes degree minus one.  It follows that not
all $v$'s need lie on $V_0$.
We call $\ell$ above the 
{\em length} of this factor-number-one
element. The length counts the number of $v$'s. For $\Psi_1$ we thus write, summing
over all possible lengths, 
\be\Psi_1  =  \sum_{\ell = 1}^\infty \sum_{i_1, \ldots i_\ell} \varphi_{i_1 \cdots i_\ell}  \, s^{-1} ( v^{i_1} \cdots v^{i_\ell} )\,,
\ee
with multi-indexed component fields $\varphi_{i_1\cdots i_\ell}$
associated to degree minus one states. A field of factor number $f$  is
\be
\Psi_f  =  \sum_{\ell_1, \ldots , \ell_f=1}^\infty
\sum_{\{ i \} } \varphi_{i^{(1)}_1 \cdots i^{(1)}_{\ell_1} |\, \cdots \, \,  | 
i^{(f)}_1 \cdots i^{(f)}_{\ell_f}}  \, \, s^{-1} ( v^{i^{(1)}_1} \cdots v^{i^{(1)}_{\ell_1} } )\otimes'  \cdots 
\otimes' s^{-1} (  v^{i^{(f)}_1} \cdots v^{i^{(f)}_{\ell_f} } )\,.  
\ee
The new $\otimes'$ tensor products imply that the component fields at factor number $f$ 
have $f$ multi-indices, separated by bars. This time, moreover, the various
$s^{-1} T^c(V)$ factors need not have degree minus one; only the sum of their degrees
must be minus one. The length of the above states is $\ell_1 + \cdots + \ell_f$.

The multi-indexed component fields associated
to a local field theory are in fact multilocal fields.  
To see this, begin by writing~\eqref{fieldloc}, separating out a single component. With the sum over $i$ denoting the integral over coordinate space we have 
\be
\phi =  \int{d^Dx}\, \varphi (x) \,  v (x)   + \cdots  \,,
\ee
where $\varphi(x)$ is a local field, and $v(x) \in V_0$ is a basis of position states.
Now consider the factor number
one part of the associated $\Psi$ field.  It would contain the term
\be
\Psi =  \sum_{\ell=1}^\infty  \int {d^Dx_1}\cdots {d^Dx_\ell} \ \, \varphi (x_1, \ldots, x_\ell)
\ s^{-1} \bigl(v(x_1)  \cdots v(x_\ell)\bigr)  + \cdots \,. 
\ee
We immediately see that we
have a multilocal coordinate space field  $\varphi(x_1, \ldots, x_\ell)\,$.
This is a type-I multilocal field. 
For the $\Psi$ components at higher factors 
we have
a more complex form of multilocality. At factor number $f$ the field
depends on $f$ {\em collections} of coordinates, which we separate by the bars included in the notation
\be 
\varphi \bigl(\, x^{(1)}_1 ,\ldots , \ x^{(1)}_{\ell_1} |    \cdots |\, x^{(f)}_1 ,\ldots ,\ x^{(f)}_{\ell_f} \bigr) \,.\ee
This is a type-II multilocal field.

As we discuss in Section~\ref{sec:Discussion}, in string field theory (SFT) the string field can be viewed 
as a vector in the state space ${\cal H}$ of a CFT, a space spanned by
the set of local operators.  A type-I off-shell field of length $\ell$ is an entangled state in ${\cal H}^{\otimes \ell}$. The Riemann surfaces are encoded in the
derivation $B$ through the multilinear products of the SFT, and the type-I entangled state
is inserted across \emph{several} punctures on a Riemann surface. 
In a canonical solution the type-I field is not entangled; it is a product state. 
In general, type-II fields acted by more than one $B$ operator represent disconnected Riemann surfaces.

\medskip
This paper is organized as follows. Section~\ref{dgvsaac} gives the mathematical background on differential graded algebras, coalgebras, suspensions, Maurer-Cartan equations, and other topics while setting the conventions needed to understand the bar and cobar constructions in the following section. Section~\ref{barandcobcon} presents $A_\infty$ algebras in the language of differential graded coalgebras, and proceeds to
describe the bar and cobar constructions.  The section ends with the discussion of the bar-cobar construction
for $L_\infty$ algebras. Section~\ref{thehomofDanddef} gives a detailed analysis of the homology of $\Delta$ 
and the homology of its deformation $\Delta_\Phi$ by a Maurer-Cartan element $\Phi$. Section~\ref{solvtheeqnmc} solves the universal Maurer-Cartan equation \eqref{33riflkx}, defines the canonical solution, and shows that any other solution is gauge equivalent to it. Section~\ref{solveqbplusdelta} then establishes, perturbatively in $B$,  the equivalence of the Maurer-Cartan space of~\eqref{33riflk} with the Maurer-Cartan space of the original theory. 
It also demonstrates how the gauge transformations of the original $A_\infty$ theory arise from a subclass of gauge transformations of the bar-cobar theory. 
Section~\ref{sec:ScalarField} illustrates the bar-cobar construction for a scalar field theory, showing that the bar-cobar equations of motion are equivalent to the original equations of motion. Section~\ref{sec:Discussion} summarizes the results of this paper and discusses open questions. Three appendices collect a set of technical computations.

\section{Differential vector spaces, algebras and coalgebras}\label{dgvsaac}

In this section we review the main ingredients necessary to understand
the bar and cobar construction presented in the following section. 
Most of the material here is standard, although some of it is not well-known
to physicists.
Moreover, the notation and basic properties to be used in the rest of the paper
are discussed here.

In the following, we use the term \emph{morphism}, which is standard in category theory, to refer to a map between spaces. The conditions we put on morphisms will depend on the spaces they are defined on. For example, a morphism of differential graded algebras is always a morphism of differential graded vector spaces, but not vice versa. Similarly, we will also generically use the word isomorphism and, in the context of homological algebra, quasi-isomorphisms and homotopies. As with morphisms, the requirements we put on these maps depend on which spaces they act on.

The section begins by reviewing graded vector spaces, linear maps and suspension
of these vector spaces. We show how to suspend or desuspend several kinds 
of maps. We introduce chain complexes, with a degree minus one differential
that allows one to define homology, chain maps, and quasi-isomorphisms.  
We then turn to differential graded
algebras, discussing in detail the tensor algebra and its universal properties.
We examine Maurer-Cartan equations, the associated Maurer-Cartan set, 
and describe explicitly finite gauge transformations.

The section then turns to differential graded coalgebras, which requires the definition of a  coassociative coproduct,
and a differential that is a coderivation with respect to the coproduct. We explain what a tensor coalgebra $T^c(V)$ associated to a vector space $V$ is, and discuss the lifting property in which a linear map from any coalgebra $C$ to $V$ extends uniquely to a morphism from $C$ to $T^c(V)$.  We study coderivations in $T^c(V)$ and how they 
are defined by linear maps from $T^c(V)$ to $V$.

Some of the following discussion in this section is drawn from the book \cite{Loday:2012operads}, see also \cite{Vallette_AlgebraHomotopy}.

\subsection{Graded vector spaces, suspensions, and chain complexes}

\paragraph{Graded vector spaces.} 
Recall that a graded vector space $V$ is a direct sum $V = \bigoplus_{k \in \mathbb{Z}} V_k$ of vector spaces $V_k$. Given an element $x \in V_k$, we define $\epsilon(x) := k$ and call $\epsilon(x)$ the degree of $x$. Note that an arbitrary element $x \in V$ does not need to have a definite degree. However, any such $x$ is by definition a finite sum of elements of definite degree. If $x$ has a definite degree $k$, we say that $x$ is homogeneous of degree $k$.

Given graded vector spaces $V$ and $W$, we can define new graded vector spaces $V \oplus W$ and $V \otimes W$ by assigning degrees
\be\label{eq:GrSumAndProduct}
(V \oplus W)_{k} := V_k \oplus W_k \, , \qquad (V \otimes W)_k := \bigoplus_{i + j = k} V_i \otimes W_j \, .
\ee
That is, a degree $k$ element $x \in V \oplus W$ is a formal sum $x = v + w$ of a degree $k$ element $v \in V$ and a degree $k$ element $w \in W$. On the other hand, a degree $k$ element $x \in V \otimes W$ is a sum of elements $v \otimes w$, such that $\epsilon(v) + \epsilon(w) = k$. The definition inductively extends to arbitrary sums and products of graded vector spaces.

Given two graded vector spaces $V$ and $W$, we say that a linear map $f: V \rightarrow W$ is of degree~$k$ if
\be 
\epsilon\, (f(v)) = \epsilon(v) + k\, ,
\ee
for any homogeneous element $v \in V$. It follows that for any composable pair of linear maps $f: U \rightarrow V$ and $g: V \rightarrow W$, we have $\epsilon(g\circ f) = \epsilon(g) + \epsilon(f)$. Given any two linear maps $f: V_1 \rightarrow W_1$ and $g: V_2 \rightarrow W_2$, there is the tensor product $f \otimes g: V_1 \otimes V_2 \rightarrow W_1 \otimes W_2$ defined~by
\be\label{eq:TensorProductOnMaps}
(f \otimes g)(v_1 \otimes v_2) = (-1)^{v_1 g}f(v_1) \otimes g(v_2)\, . 
\ee
Here and in the following, we write $(-1)^{v_1 g} := (-1)^{\epsilon(v_1)\epsilon(g)}$. The sign in \eqref{eq:TensorProductOnMaps} can be memorized by noting that by going from the left-hand side to the right-hand side, $v_1$ has to be commuted past $g$ and therefore produces the indicated sign factor.

The degree of the multilinear map $f: V^{\otimes n} \rightarrow W$ is just that of $f$ considered as a linear map from the graded vector space $V^{\otimes n}$ to the graded vector space $W$. According to the definition given in \eqref{eq:GrSumAndProduct}, this means that $f$ is of degree $k$, if for any collection of homogeneous elements 
$v_1,\ldots,v_n$, we have
\be
\epsilon(f(v_1,\ldots,v_n)) = k + \epsilon(v_1) + \cdots + \epsilon(v_n) \, .
\ee 

\paragraph{Suspension of graded vector spaces.} Given any graded vector space $V$, we define a graded vector space $s V$ by assigning the degree
\be\label{eq:VectSuspension}
(s V)_{k} = V_{k-1} \,,  \ \ \ \forall \, k \,.  
\ee
With this definition comes an obvious degree one map
\be
s: V \rightarrow s V \,, \ \ \ \
v \mapsto s v \, ,
\ee
where we use $s$ to denote the map, and $s v$ to denote $v$ as an element in $s V$. 
We call $sV$ the suspension of $V$. The inverse, desuspension map $s^{-1}$ acts
as follows  
\be
\ds: V \rightarrow \ds V \; , \ \ \ \ 
v \mapsto \ds v \, .
\ee

Both the bar and cobar constructions will require us to suspend and desuspend multilinear maps. 
For suspension (desuspension), we pass from  multilinear maps $f_n$ ($\bar f_n$) to new maps $\bar f_n$~($f_n$), as follows:
\be
f_n: V^{\otimes n} \rightarrow W  \qquad \Longleftrightarrow \qquad 
\bar f_n: (sV)^{\otimes n} \rightarrow sW\, , 
\ee
by defining 
\be
\label{cleardepfus} 
\bar f_n = s \circ f_n \circ (s^{-1})^{\otimes n}\, . 
\ee
Note that
\be\label{eq:SuspensionSign}
(s^{-1})^{\otimes n}(sv_1 \otimes \cdots \otimes sv_n) = (-1)^{\sum_{i = 1}^n (n-i) (v_i+1)} v_1 \otimes \cdots \otimes v_n \, ,
\ee
which follows from \eqref{eq:TensorProductOnMaps}. From this we see that $\bar f_n$ is given by
\be
\label{3fje3iu}
\bar f_n(sv_1,\ldots,sv_n) = (-1)^{\sum_{i = 1}^n (n-i) (v_i+1)} s f_n(v_1,\ldots,v_n) \, ,
\ee
which is the convention used, for example, 
in~\cite{GetzlerJones1990AInfinity}.\footnote{Some authors
(e.g.,~\cite{keller2001Amodules}) choose the sign of $\bar f_n$ by demanding that
$\bar f_n \circ  s^{\otimes n}  = s \circ f_n$, which differs from~\eqref{cleardepfus} because $s^{\otimes n}$ and $(s^{-1})^{\otimes n}$ are only inverses up to a sign.  In this different convention,
\be
\bar f_n(sv_1,\ldots,sv_n) = (-1)^{\sum_{i = 1}^n (n-i)v_i} s f_n(v_1,\ldots,v_n) \, .
\ee
This differs from~\eqref{3fje3iu} by a factor of $ (-1)^{\sum_{i = 1}^n (n-i)} = (-1)^{\frac{n(n-1)}{2}}$.
} Since $s$ has degree one, we find that  
\be\epsilon(\bar f_n) = \epsilon(f_n) - n + 1\, .
\ee

As is the case with coproducts and their iterated action, we will also encounter maps taking a vector in $V$ to a vector in $V^{\otimes n}$.
This time we need to relate
\be
g_n: \ds V \rightarrow (\ds W)^{\otimes n}\qquad \Longleftrightarrow \qquad 
\bar g_n: V \rightarrow W^{\otimes n}\,.
\ee
We do this by  defining 
\be 
\label{gsuspensio30}
\bar g_n = s^{\otimes n}\circ  g_n \circ \ds \,.
\ee
It is quick to check, for example, that 
\be
\label{30rdfjhe}
 g_2 = - (s^{-1}\otimes s^{-1})  \circ  \bar g_2 \, \circ  s \,. 
\ee

\paragraph{Chain complexes} Graded vector spaces are the underlying vector spaces of differential graded vector spaces $(V,\df)$. This is a graded vector space $V$, equipped with a linear map $\df: V \rightarrow V$ of degree $\epsilon(\df) = -1$, satisfying $\df^2 = 0$. Associated to any differential graded vector space $(V,\df)$ is the homology $H(\df)$, defined as the space of vectors $v \in V$ satisfying $\text d v = 0$ with the identification of vectors of the same degree $v_1 \sim v_2$ if
\be
v_2 = v_1 + \text d w\, ,
\ee
for some element $w$ of degree $\epsilon(w) = \epsilon(v) + 1$. As a graded vector space, the homology is therefore given by the quotient
\be 
H(\df) := \frac{\ker \df}{\text{im}\; \df} \, .
\ee

\bigskip  
A morphism  $f: (V,\df_V) \rightarrow (W,\df_W)$ between differential graded vector spaces is called a chain map. This is a morphism of graded vector spaces, namely, a degree zero linear map $f: V \to W$ that commutes with the differential in the sense
\be 
f \df_V = \df_W f \, .
\ee 
This chain condition guarantees that $f$ is well-defined on homology:
any chain map $f: (V,\df_V) \rightarrow (W,\df_W)$ induces a map 
\be 
\label{maponhomology}
H(f): \ H(\df_V) \rightarrow H(\df_W)\,,  
\ee
defined by 
\be
\label{explitictymaponH}
H(f)([v]) := [f(v)]\,.  
\ee
A chain map is a {\em quasi-isomorphism} if it is an isomorphism on homology.

It can happen that different chain maps $f: V \rightarrow W$ and $g: V \rightarrow W$ induce the same map on homology. When working with differential graded vector spaces, this is equivalent to $f$ and $g$ being homotopic. By this we mean that there is a degree one linear map $h: V \rightarrow W$, called a homotopy, such that
\be\label{eq:DefofHomotopy}
g - f = \df_W h + h \df_V \, .
\ee
Notice that when $a \in \ker \df_V$, the homotopy relation implies that
\be
g(a) = f(a) + \df_W h(a) \, .
\ee
So the existence of a homotopy tells us that $f$ and $g$ agree on homology: the images of any element $a \in \ker \df_V$ under $f$ and $g$  differ by an exact term. 

A special case of a homotopy is a contraction
to a subspace of a given differential graded vector space $V$. Let $\boldsymbol{\pi}: V \rightarrow V$ be a projecting chain map, that is, it satisfies $\boldsymbol{\pi}^2 = \boldsymbol{\pi}$. 
We call it a contraction if it is homotopic to the identity, meaning there is a homotopy $h: V \rightarrow V$,
such that
\be\label{eq:ContractingHomotopy}
    [\df,h] := \df h  + h \df = 1 - \boldsymbol{\pi} \, .
\ee
Here, $[-,-]$ denotes the graded commutator of linear maps, defined as
\begin{equation}\label{gradedcomm}
    [f,g] = f g - (-1)^{fg}g f\,,
\end{equation}
for linear maps $f$ and $g$.
 Since both $h$ and $\df$ are odd, there is a relative plus sign in \eqref{eq:ContractingHomotopy}. This relation implies that $\boldsymbol{\pi}$ becomes the identity on homology, so $\boldsymbol{\pi}$ is a quasi-isomorphism. If also $\df \boldsymbol{\pi} = \boldsymbol{\pi}\df = 0$, then  $\boldsymbol{\pi}$ projects to a subspace isomorphic to $H(\df)$.

The direct sum $(V\oplus W,\df_{V\oplus W})$ and tensor product $(V\otimes W,\df_{V\otimes W})$ of two differential graded vector spaces $(V,\df_V)$ and $(W,\df_W)$ can also be defined. We have
\be\label{eq:SumAndProdOfDifferential}
\df_{V\oplus W} := \df_V \oplus \df_W \, , \qquad \df_{V\otimes W} := \df_V \otimes 1_W + 1_V \otimes \df_{W} \, .
\ee
It is straightforward to check that these differentials also square to zero. The definition of a differential on larger sums and products follows inductively. Moreover, given a linear map $f: (V,\df_V) \rightarrow (W,\df_W)$, we can define a differential acting on linear maps $f$:  
\be\label{eq:DifferentialOnMaps}
\df_{(V,W)} f := \df_W \circ f - (-1)^f f\circ \df_V \, . 
\ee
Note that $f$ is a chain map if it is of degree zero and closed with respect to $\df_{(V,W)}$. Furthermore, two chain maps $f$ and $g$ are homotopic, if their difference is $\df_{(V,W)}$-exact. Indeed, since $h$ is of degree one, the homotopy
relation~\eqref{eq:DefofHomotopy} implies $g-f= \df_{(V,W)} h$.  

As a combination of the above two definitions, we shall write
\be\label{eq:CommutatorDifferential}
[\df,f_k] := \df_{(V^{\otimes k},V)} f_k = \df_V \circ f_k - (-1)^{f_k} f_k \circ (\sum_{i = 1}^k 1^{\otimes {(i-1)}} \otimes \df_V \otimes  1^{\otimes {(k-i)}}) \,,
\ee
for a given multilinear map $f_k: V^{\otimes k} \rightarrow V$ on a differential graded vector space $(V,\df_V)$. Note that this is a generalization of the graded commutator given in \eqref{eq:DifferentialOnMaps}.

\subsection{Differential graded algebras}

A differential graded algebra combines the concept of a differential graded vector space with that of an algebra. A differential graded algebra $(A,m_1,m_2)$ is a differential graded vector space $(A,m_1)$, equipped with a map $m_2$ of degree zero
\be
\begin{split}
m_2 : A \otimes A &\rightarrow A \, , \\
a\otimes b &\mapsto m_2(a, b)\,, 
\end{split}
\ee
satisfying the associativity condition
\be
m_2(m_2(a, b), c) = m_2(a, m_2(b, c))\,, 
\ee
and such that the differential   
$m_1$ is a derivation with respect to the product. A derivation of $m_2$ is a linear map $D: A \rightarrow A$ satisfying the Leibniz rule
\be
D (m_2(a, b)) = m_2(D a, b) + (-1)^{aD}m_2(a, D b) \, .
\ee

A morphism $f: (A,m_1^A,m_2^A) \rightarrow (B,m_1^B,m_2^B)$ of differential graded algebras is a chain map that is also an algebra morphism, so it satisfies
\be
f(m_2^A(a, b)) = m_2^B(f(a), f(b)) \, .
\ee

\medskip\noindent\textbf{ Example: Tensor algebra.} 
 Given any graded vector space $V$, we can associate to it the 
 tensor algebra 
 \be   (T(V),\mu),   
 \ee
 which as a graded vector space is given by
\be\label{eq:CompletedTensorAlgebra}
T (V) = \widehat{\bigoplus}_{n \ge 1} V^{\otimes n} \, .
\ee 
The hat symbol in \eqref{eq:CompletedTensorAlgebra} is there to indicate that we allow for infinite sums.\footnote{In math literature, one often finds $\prod$ instead of $\widehat \bigoplus$ to indicate the inclusion of infinite formal sums. The tensor algebra with infinite sums is often called the completed tensor algebra.}
To keep notation simple, we write
\be
 a_1 \otimes a_2 \otimes \cdots \otimes a_n \equiv a_1a_2 \cdots a_n\,,
\ee
for the elements in $V^{\otimes n}$. In this notation, an element of $V^{\otimes n}$ is a {\em word} consisting of $n$ {\em letters} $a_i \in V$. 
An element in $T(V)$ is then a potentially infinite formal sum of such words, where each individual word can have arbitrary length. 
The product on $T(V)$ 
is given by concatenation of words, i.e.
\be\label{eq:DefTensAlgProduct}
\begin{split}
\mu: T (V)  \otimes' T (V) &\rightarrow T (V) \; , \\ 
a_1 \cdots a_m \otimes'
b_1 \cdots b_n &\mapsto a_1 \cdots a_m b_1 \cdots b_n \; .
\end{split}
\ee
This clearly defines an associative product. We use $\otimes'$ to distinguish from the $\otimes$ used to construct words from letters.

The main property of the tensor algebra is that given an algebra $(A,m_2)$, a linear map $f: V \rightarrow A$ uniquely lifts to a morphism $F: (T(V), \mu) \rightarrow (A,m_2)$ of graded algebras, such that
\be
\label{diagrtenalgmorph}
\begin{tikzcd}
 T(V) \arrow[dr,"F"]& \\
    V \arrow[r,"f"] \arrow[u,"\iota_1"] & A
\end{tikzcd}
\ee
commutes. Here $\iota_1: V \rightarrow T(V)$ is the inclusion. 
The map $F$ is given by 
\be
F(a_1a_2 \cdots a_n) = f(a_1) \cdot f(a_2) \cdots f(a_n) \; ,
\ee
where the right-hand side denotes the product of elements $f(a_i)$ in $A$. Equivalently, regarded as a map,
\be
F = \sum_{n \ge 1} m^{(n)} \circ f^{\otimes n}\, ,
\ee
where $m^{(n)}$ is inductively defined by $m^{(1)} = \text{id}_A$ and $m^{(n+1)} = m_2 \circ (m^{(n)} \otimes \text{id}_A)$.

Any derivation $D_X$ on $T(V)$ is determined by a linear map 
$X: V \rightarrow T(V)$  
as
\be \label{eq:DerLift}
D_X(a_1 \cdots a_n) = \sum_{i = 1}^n (-1)^{(a_1 + \cdots + a_{i-1})X} a_1 \cdots a_{i-1}X(a_i)a_{i+1} \cdots a_n
\ee
such that $D_X \circ \iota_V = X$. In particular, if $(T(V),B,\mu)$ is a differential graded algebra, the differential $B$ is a degree minus one  
derivation of the product and is therefore determined by a linear 
map $b$ in the sense that $b= B|_V$.  Then, 
$B$ acts as in~\eqref{eq:DerLift}.
\be \label{eq:DerLiftb}
B(a_1 \cdots a_n) = \sum_{i = 1}^n (-1)^{a_1 + \cdots + a_{i-1}} a_1 \cdots a_{i-1}b(a_i)a_{i+1} \cdots a_n \; .
\ee

\paragraph{Maurer-Cartan equation}

Given a differential graded algebra $(A,m_1,m_2)$ and an element $a\in A_{-1}$, define a degree minus-one $m_1^a$ as
\be
m_1^a b := m_1 b + m_2(a,b)-(-1)^{b}m_2(b,a) \, .
\ee
Here, $m_1^a$ should be regarded as a candidate deformed differential: it is a genuine differential only if $(m_1^a)^2=0$.  Equivalently, using the graded commutator
\be 
[a,b] := m_2(a,b)-(-1)^{ab}m_2(b,a) \, ,
\ee
we may write
\be
m_1^a b = m_1 b + [a,b] \, .
\ee
One can verify that the operator $[a,-]$ is a graded derivation of the product $m_2$. Since $m_1$ is already a derivation of $m_2$, we see that $m_1^a$ is also a derivation of $m_2$. Therefore, the only remaining condition for $(A,m_1^a,m_2)$ to be a differential graded algebra is the square-zero condition
$(m_1^a)^2=0$. A straightforward computation gives
\be
\label{eq:flat}
(m_1^a)^2 b = [m_1 a + m_2(a,a), b] \, .
\ee
Thus the square of the deformed operator is inner, generated by the element
\be
F(a):=m_1 a + m_2(a,a)\, ,
\ee
called the field strength of $a$. Consequently,
$(m_1^a)^2=0$
holds precisely when $F(a)$ graded-commutes with every element of $A$. In particular, a sufficient condition is
\be
F(a)=0 \, .
\ee
When this stronger condition holds,  $(A,m_1^a,m_2)$ is again a differential graded algebra.
The equation
\be
m_1 a + m_2(a,a)=0\, ,
\ee
is called the Maurer--Cartan equation of the differential graded algebra $(A,m_1,m_2)$. Thus Maurer--Cartan elements $a\in A_{-1}$ are precisely the elements for which the above deformation has vanishing field strength, and hence gives a new differential graded algebra structure with the same product $m_2$.

We view  $(A,m_1^a,m_2)$ as a {\em deformation} of $(A,m_1,m_2)$. These two
algebras are not in general quasi-isomorphic: the homologies of $m_1$ and $m_1^a$
need not agree. 
We call the
set of all deformations the \emph{Maurer-Cartan space} $\mathcal{MC}(A)$
\be
\mathcal{MC}(A) = \{a \in A_{-1} \; | \; F(a) = 0\}\,.
\ee
We declare
two deformations to be equivalent if they differ by a gauge transformation.
The equivalence means that the algebras generated by the two deformations
are isomorphic, as we will demonstrate
now.  For this we need to give the gauge transformations. 

Infinitesimal gauge transformations generated by a gauge parameter $\lambda\in A_{0}$ are given by
\be\label{eq:InfinitesimalGT}
\delta_\lambda a = m_1 \lambda + [a,\lambda]\,.
\ee
The Maurer-Cartan space is invariant under this transformation, since
\be
\delta_\lambda F(a) = [F(a),\lambda] \; .
\ee
The finite version of the gauge transformation can be obtained by introducing a
parameter $t$ and solving the differential equation
\be 
\dot a(t) = \delta_\lambda a(t) = m_1 \lambda + [a(t),\lambda]\,,
\ee
with initial condition $a(0) = a$.  The solution is  
\be\label{eq:FiniteGTx}
a(t)  = e^{-t\lambda}(m_1 e^{t\lambda}) + e^{-t\lambda} a\,  e^{t\lambda} \; ,
\ee
where
\be\label{eq:Exponential}
e^{\lambda} = \sum_{n \ge 0} \frac{1}{n!} \lambda^n \; .
\ee
Here, $\lambda^n = m_2(\lambda,\lambda^{n-1})$, i.e.~it is the $n$th power of $\lambda$ under $m_2$. We then define $a' = a(t=1)$ as the finite gauge transformation with parameter $\lambda$:
\be\label{eq:FiniteGT}
a' = e^{-\lambda}(m_1 e^{\lambda}) + e^{-\lambda} a \, e^{\lambda} \; .
\ee
 We assume that \eqref{eq:Exponential} converges. 

We now show that gauge transformations define isomorphisms of deformations. 
Explicitly, 
the isomorphism is an invertible map $\Phi: A \to A$ given by:   
\be
\Phi(x) = e^{-\lambda} \, x\,  e^{\lambda} \,,  \ \    x \in A  \,.  
\ee
Writing the product $m_2 (x, y) = xy$, we see that the isomorphism
preserves it:   
\be
m_2(\Phi(x),\Phi(y)) = e^{-\lambda} x  e^{\lambda}  e^{-\lambda} y  e^{\lambda} = e^{-\lambda} x y e^{\lambda} = \Phi(m_2(x,y)) \; .
\ee
Furthermore, we have that
\be
\begin{split}
\Phi (m_1^a  
\Phi^{-1} (x)) &= \Phi 
( m_1   
\, ( e^\lambda x e^{-\lambda})) +  \Phi (a e^{\lambda} x e^{-\lambda}) - (-1)^x  \Phi(e^{\lambda} x e^{-\lambda} a) \\
&= e^{-\lambda} m_1( e^\lambda x e^{-\lambda})e^{\lambda} + e^{-\lambda} a e^{\lambda} x - (-1)^x x e^{-\lambda} a e^{\lambda} \\
&= \big(e^{-\lambda} (m_1 e^{\lambda}) + e^{-\lambda} a e^{\lambda}\big) x
+ m_1 x - (-1)^x x\big(-(m_1 e^{-\lambda})e^{\lambda} +  e^{-\lambda} a e^{\lambda}\big) \\
&= m_1 x + \big(e^{-\lambda} (m_1 e^{\lambda}) + e^{-\lambda} a e^{\lambda}\big) x - (-1)^x x\big(e^{-\lambda}(m_1 e^{\lambda}) +  e^{-\lambda} a e^{\lambda}\big) \\
&= m_1^{a'} x \, .
\end{split}
\ee
This equation implies the relation
$\Phi m_1^a  =  m_1^{a'} \Phi$ and therefore 
 $\Phi: (A,m^a_1,m_2) \rightarrow (A,m_1^{a'},m_2)$ is an {\em isomorphism} of differential graded algebras. From this observation, it makes sense to define
\be
MC(A) = \mathcal{MC}(A)/(\text{gauge transformations}) \, ,
\ee
where the quotient means that
we identify solutions that differ by a gauge transformation. We call $MC(A)$ the \emph{Maurer-Cartan set} of $A$. It contains the true deformations of~$A$.

A morphism $f: (A,m_1^A,m_2^A) \rightarrow (B,m_1^B,m_2^B)$ of differential graded algebras induces a well-defined map between Maurer-Cartan sets. If $a \in A_{-1}$ is a Maurer-Cartan element,  
so is $f(a)$, since
\be\label{eq:MorphismsPreserveMC}
m_1^B f(a) + m^B_2(f(a),f(a)) = f(m_1^A a + m_2^A(a,a)) \; .
\ee
Furthermore, if $a' = e^{-\lambda}(m_1^A e^{\lambda}) + e^{-\lambda} a e^{\lambda}$, we clearly have
\be
f(a') = f(e^{-\lambda})(m_1^B f(e^{\lambda})) + f(e^{-\lambda}) f(a) f(e^{\lambda}) \, .
\ee
Using $f(e^\lambda)= e^{f(\lambda)}$, we see that $f(a)$ and $f(a')$ are gauge equivalent via $e^{f(\lambda)}$. It follows that any morphism of differential graded algebras induces a well-defined map on the respective Maurer-Cartan sets.

There is also a convenient rewriting of the gauge relation \eqref{eq:FiniteGT}. Let us define $u := e^\lambda-1$. In that case, the gauge relation can be equivalently stated as
\be\label{eq:AltFiniteGT}
a' - a = m_1(u) + a u - u a' \, .
\ee
Note that $u$ is not infinitesimal, it is a finite gauge parameter. 
In this form, the gauge relation has a straightforward generalization to the homotopy associative case we will discuss in the next section.

\paragraph{Example: Chern-Simons theory}

A standard example of a differential graded algebra in physics arises in Chern-Simons theory. Let $M$ be a three-dimensional manifold and $G$ a Lie group with Lie algebra $(\mathfrak{g},[-,-]_\mathfrak{g})$.  

Let $\Omega^\bullet(M,\mathfrak{g})$ be the space of $\mathfrak{g}$-valued forms. In order to match our grading conventions, we say that $\Omega^k(M,\mathfrak{g})$ has degree $-k$. Then $\Omega^k(M,\mathfrak{g})$ has a differential of degree minus one
\be
\begin{split}
\df: \Omega^\bullet(M,\mathfrak{g}) &  \rightarrow \Omega^\bullet(M,\mathfrak{g})\, , \\
\omega \,  T   &  \mapsto \df \omega \, T  \, , 
\end{split}
\ee
where $\omega \in \Omega^k (M), T \in \mathfrak{g}$.
There is also a Lie bracket of degree zero
\be
\begin{split}
[-,-]: \Omega^\bullet(M,\mathfrak{g}) \otimes \Omega^\bullet(M,\mathfrak{g}) &\rightarrow \Omega^\bullet(M,\mathfrak{g}) \, , \\
\omega_1 T_1 \otimes \omega_2 T_2 &\mapsto (\omega_1 \wedge \omega_2) [T_1,T_2]_\mathfrak{g}\,,
\end{split}
\ee
with $\omega_{1}\in  \Omega^{k_1}(M)$,  $\omega_2 \in  \Omega^{k_2}(M)$,  $T_1,T_2 \in \mathfrak{g}$. 
The differential graded Lie algebra is then $(\Omega^\bullet(M,\mathfrak{g}), \df, [ - , - ] ) $.

If $\mathfrak{g}$ admits an invariant form $\langle-,-\rangle$, we can define the Chern-Simons action for $A \in \Omega^1(M,\mathfrak{g})$:
\be
S_{CS}(A)= \int_M \langle A,\tfrac{1}{2}\text d A + \tfrac{1}{6}[A,A]\rangle\, .
\ee
 The equations of motion are given by
\be
\label{dfghliis}
\text d A + \tfrac{1}{2}[A,A] = 0 \, .
\ee
Assume now that we have a finite-dimensional representation of $\mathfrak{g}$ on a vector space $V$, namely, a map
\be
\rho: \mathfrak{g} \rightarrow \operatorname{End}(V)\,,
\ee
that is a Lie algebra homomorphism: $\rho([T_1,T_2]_\mathfrak{g}) = [\rho(T_1),\rho(T_2)]_{\operatorname{End} (V)}$, 
with $[ - , - ]_{\operatorname{End} (V)}$ the commutator bracket. The representation  defines a map $\Omega^\bullet(M,\mathfrak{g}) \rightarrow \Omega^\bullet(M,\operatorname{End}(V))$. The elements of $\Omega^\bullet(M,\operatorname{End}(V))$ then form a differential graded algebra
with product
\be (\omega_1  R_1 ) (\omega_2  R_2)  =   \omega_1 \wedge \omega_2  \, R_1 \circ R_2\,, \ \ \ \hbox{with} \ \  R_1, R_2 \in \operatorname{End} (V) \,. 
\ee
With $\mathcal{A} \in \Omega^1(M,\operatorname{End}(V))$, the Maurer-Cartan equation for this differential graded associative algebra is
\be
\label{csmc}
\text d \mathcal{A} + \mathcal{A}^2 = 0 \, .
\ee
This equation also arises from the differential graded Lie algebra equation~\eqref{dfghliis}
by replacing $A \to \mathcal{A}$ and noting that 
$[\mathcal{A},\mathcal{A}]_{\Omega^\bullet(M,\operatorname{End}(V))} = 2\mathcal A^2$. 
Equation~\eqref{csmc} is covariant
under the standard gauge transformations 
\be
\mathcal A' = e^{-\lambda} (\df e^\lambda) + e^{-\lambda} \mathcal  A e^\lambda \, ,
\ee
with gauge parameter $\lambda \in  \Omega^0(M,\operatorname{End}_V)$.

\subsection{Differential graded coalgebras}

The dual notion of a differential graded algebra is that of a differential graded coalgebra $(C,\bar B, \bar \Delta)$. 
Here $(C,\bar B)$  is a differential graded vector space, so $\bar B$ squares
to zero and has degree minus one:
\be
\epsilon ( \bar B) = -1\,. 
\ee
Moreover, we have a  
degree zero coproduct $\bar \Delta$:
\be
\epsilon (\Delta) = 0 \,. 
\ee
The coproduct is a map 
\be
\bar \Delta: C \rightarrow C \otimes C\,,
\ee
satisfying the coassociativity condition
\be
(\bar \Delta \otimes 1) \circ \bar \Delta = (1 \otimes \bar \Delta) \circ \bar \Delta \, .
\ee
This is just like the associativity condition $m_2 \circ (m_2 \otimes 1) = m_2 \circ (1 \otimes m_2)$, but with the order of operations reversed. Moreover, we demand that $\bar B$ is a coderivation on $C$. In general, a coderivation on $C$ is a linear map $D: C \rightarrow C$ that satisfies the co-Leibniz rule
\be
\bar \Delta \circ D = (D \otimes 1 + 1 \otimes D) \circ \bar \Delta \, .
\ee
This is just the Leibniz rule $D \circ m_2 = m_2 \circ (D \otimes 1 + 1 \otimes D)$ with the order of operations reversed. It is straightforward to check that the space of coderivations closes with respect to the graded commutator $[D_1,D_2] = D_1 \circ D_2 - (-1)^{D_1D_2} D_2 \circ D_1$. In particular, if $D$ is of odd degree, then $D^2 = \frac{1}{2}[D,D]$ and hence $D^2$ is also a coderivation.

\medskip\noindent
\textbf{Coalgebra morphisms.}
A morphism $F$ of coalgebras,  
\be F: (C_1,\bar B_1,\bar \Delta_{1}) \rightarrow (C_2,\bar B_{2},\bar \Delta_{2})\ee 
is a chain map $F: C_1 \rightarrow C_2$, meaning,
\be   F \bar B_{1} =  \bar B_{2}  F \,,\ee
that in addition satisfies
\be
\bar \Delta_{2} \circ F = (F \otimes F) \circ \bar \Delta_1 \, .
\ee
The morphism $F$ is in fact a {\em quasi-isomorphism} if the induced map $H(F)$ of
homologies:
\be
H(F) : H (\bar B_1) \to H(\bar B_2)
\ee
is invertible -- that is, when $F$ is an isomorphism on homology.  The map $H(F)$ is defined as in~\eqref{explitictymaponH}:  for a cycle $u \in C_1$ and with $[u]$ denoting its homology class in $H(\bar B_1)$ we have
$H(F)([u]) =  [ F(u)]$.

\medskip
Given a coalgebra morphism $F$,  a map $D_F: C_1 \to C_2$ is said to be an $F$-coderivation if
\be
\label{fcoderivation} 
\bar\Delta_2  D_F = \bigl(  D_F \otimes F + F \otimes D_F ) \bar\Delta_1 \,. 
\ee
One can quickly verify that given coderivations $D_1$ and $D_2$ on the coalgebras $C_1, C_2$, respectively, then $F D_1$ and $D_2 F$ are both $F$-coderivations:  morphisms composed with coderivations are $F$-coderivations. 

\medskip
\noindent\textbf{Example: Tensor coalgebra.}
An important example of a coalgebra is the tensor coalgebra. To any graded vector space $V$, we associate the graded coalgebra $T^c(V)$, whose underlying vector space is
\be
T^c(V) = \widehat \bigoplus_{n \ge 1} V^{\otimes n} \, ,
\ee
which is the same as that of the tensor algebra $T(V)$. We also use concatenation of letters to denote elements of $T^c(V)$, which are words. We have a coproduct $\bar\Delta :  T^c(V) \to T^c(V) \otimes' T^c(V)$, where
we use $\otimes'$ in order to distinguish it from the tensor product $\otimes$ in $T^c (V)$.
The coproduct is given by the deconcatenation operation
\be
\bar \Delta(a) = 0 \, , \qquad \bar \Delta(a_1 \cdots a_n) = \sum_{i= 1}^{n-1} a_1 \cdots a_i \otimes' a_{i+1} \cdots a_n \, , n \ge 2 \, .
\ee
It is straightforward to show that $\bar \Delta$ satisfies the coassociativity condition. 

The tensor coalgebra has a lifting property similar to that of the tensor algebra. Given any coalgebra $(C,\bar \Delta_C)$ and a vector space $V$, any degree zero linear map $f: C \rightarrow V$ lifts uniquely to a coalgebra morphism $F: C \rightarrow T^c(V)$, such that
\be
\label{morphcoalgdiag}
\begin{tikzcd}
& T^c (V) \arrow[d,"
\pi_1"] \\
C  \arrow[ur,"F"] \arrow[r,"f"] & V
\end{tikzcd}
\ee
commutes. Here, $\pi_1: T^c (V) \rightarrow V$ is the projection to the tensor-degree-one component. The lift $F$ is given by the formula
\be\label{eq:CoHomLift}
F = \sum_{n \ge 1} \iota_n \circ f^{\otimes n} \circ \bar \Delta_C^n \; ,
\ee
where the maps $\iota_n$ are the embeddings $\iota_n: V^{\otimes n} \rightarrow T^c (V)$ and
\be
\bar \Delta_C^n: C \rightarrow C^{\otimes n}\,,
\ee
is the linear map inductively defined by $\bar \Delta_C^1 = 1$ and $\bar \Delta_C^n = (\bar \Delta_C^{n-1} \otimes 1) \circ \bar \Delta_C$ for $n \ge 2$.

As a special case of the above lifting property, consider the case where
the coalgebra $C$ is $T^c(V)$ and  the above diagram is
\be\label{liftingcoalgebra}
\begin{tikzcd}
& T^c (W) \arrow[d,"
\pi_1"] \\
T^c(V)  \arrow[ur,"F"] \arrow[r,"f"] & W
\end{tikzcd}
\ee
with the degree zero linear map $f: T^c (V) \rightarrow W$, which we can decompose as
\be
\label{Fmorphf}
f = \sum_{k \ge 1} f_k \;  \qquad  \ f_k := f|_{V^{\otimes k}}\,.
\ee
On an element $v_1 \cdots v_n \in V^{\otimes n}$, the coalgebra morphism $F$ in~\eqref{eq:CoHomLift} 
is given by
\be\label{eq:TCoHomLift}
\begin{split}
F(v_1 \cdots v_n) = & \ \sum_{k = 1}^n\, \sum_{i_1 + \cdots + i_k = n} 
\hskip-8pt f_{i_1}(v_1,\ldots,v_{i_1}) f_{i_2}(v_{i_1 + 1},\ldots,v_{i_1 + i_2}) \cdots f_{i_k}(v_{i_1 + \cdots + i_{k-1} + 1},\ldots , v_n) \\[0.5ex]
=& \ f_n(v_1,\ldots,v_n) + \cdots +  f_1(v_1)f_1(v_2) \cdots f_1(v_n)\,,
\end{split}
\ee
where the last equality shows the $k=1$ and $k=n$ terms in the sum. 
The image of the word $v_1\cdots v_n$ under $F$ is a collection of words
in $T^c(W)$ whose lengths range from one to $n$.

\bigskip
Along the same lines, any linear map $X: T^c (V) \rightarrow V$ uniquely determines a coderivation $D_X$ on $T^c( V)$. The map $X$ can be expressed as a sum of multilinear maps $X_k : V^{\otimes k} \to V$ as follows  
\be\label{derivationlin}
X = \sum_{k=1}^\infty X_k \,, \ \ \   X_k := X|_{V^{\otimes k}} \,.\ee
With this notation, the action of $D_X$ on an element $v_1 \cdots v_n$ is given by
\be\label{eq:CoderLift}
D_X(v_1 \cdots v_n) = \sum_{k = 1}^n \sum_{i = 0}^{n-k} (-1)^{X_k(v_1 + \cdots + v_i)} v_1 \cdots v_{i} X_k(v_{i+1},\ldots,v_{i+k}) v_{i+k+1} \cdots v_n \; .
\ee
The right-hand side is a collection of words in $T^c(V)$ whose lengths range from $1$ to $n$. 
Note that $D_{X_k} (v_1 \cdots v_n) = 0$ for $k> n$.
This extension of a linear map to a coderivation will be used repeatedly to encode the set  of multilinear products of an $A_\infty$ algebra in a coderivation.

Consider now a coalgebra morphism $F: (T^c(V_1), \bar\Delta_1)  \to (T^c(V_2), \bar\Delta_2)$
as in~\eqref{Fmorphf}, 
and let $D_F$ be an $F$-coderivation (see~\eqref{fcoderivation}).  Then $D_F$ is 
determined by the map $d$:
\be  d =  \pi_1^{(2)}  D_F :  \ T^c(V_1) \to V_2 \, \ \ \ \hbox{with} \ \ 
d = \sum_{k=1}^\infty d_k\,, \ \ \ d_k : V_1^{\otimes k} \to V_2 \,. 
\ee
We then have
for $D_F$
\be 
\label{dfdetbyd}
D_F = \sum_{m\geq 1} \sum_{\ell=1}^m  \Bigl( f^{\otimes (\ell-1)} \otimes d \otimes f^{\otimes (m-\ell)} \Bigr) \bar \Delta_1 ^{(m)} \,.
\ee
This means that on the right-hand side each term must have a $d_k$ and the remaining terms
have $f_k$'s.  For example, for an odd coderivation $D_F$,
\be
D_F (v_1 v_2) = d_2 (v_1, v_2) + d_1 (v_1) f_1(v_2) + (-1)^{v_1} f_1 (v_1) d_1 (v_2) \,. \ee

\section{Bar and cobar constructions}\label{barandcobcon}

This section begins with the presentation of the bar construction,
that given a differential graded algebra $A$ provides an associated differential
graded coalgebra $B(A)$. We then turn to $A_\infty$ algebras, a natural extension of differential graded algebras, showing that
a variant of the bar construction associates to the $A_\infty$ algebra
a differential graded coalgebra $B_\infty (A)$. There is a Maurer-Cartan 
equation for $A_\infty$ algebras, 
and morphisms of $A_\infty$ algebras
are defined as morphisms of the associated bar coalgebras. 

We then turn to the dual cobar construction that starting with a differential
graded coalgebra $C$ gives us a differential graded algebra $\Omega (C)$. 
In this way, we can have a bar-cobar construction in which starting with
an $A_\infty$ algebra, we apply bar followed by cobar to get the differential graded algebra $\Omega(B_\infty(A))$.
The equivalence of $A_\infty$ algebras and their bar-cobar construction in terms of a quasi-isomorphism is explained, taking for granted some hard technical results in the mathematics literature. In fact, the following sections provide a physicist-accessible proof of the claimed quasi-isomorphism.

The section concludes with a discussion of the bar-cobar construction for $L_\infty$ algebras.

Again, some of the material in this section is drawn from~\cite{Loday:2012operads}, see also \cite{Vallette_AlgebraHomotopy}. 
 $A_\infty$ algebras first appeared in \cite{Stasheff1,Stasheff2}.  In the context of bosonic string field theory, definitions of $A_\infty$ algebras can be found, for example, in \cite{Gaberdiel:1997ia,Gaberdiel:1997mg,Kajiura:2003ax}.  While the notion of homotopy Lie algebras was clear to exist after the case of $A_\infty$, the detailed presentation of the 
 $L_\infty$ axioms and a nontrivial example in closed string field theory was given in~\cite{Zwiebach:1992ie}, a discussion expanded shortly thereafter in~\cite{Lada:1992wc}.

\subsection{Bar construction}

Given a differential graded algebra $(A,m_1,m_2)$, we can define a differential graded coalgebra $B(A)$, called the bar construction of $A$, and given as follows.
\be
\label{cobardefexp}
B(A) =   ( T^c(sA),  \bar B , \bar \Delta ) \, .
\ee
As stated, the underlying  graded vector space is $T^c(sA)$ -- a tensor coalgebra over the suspension of $A$ defined as in~\eqref{eq:VectSuspension}. Moreover, as usual, $\bar \Delta$ is deconcatenation on
$T^c(sA)$. 

Here $\bar B$ is a differential and a coderivation of degree minus one.  Its construction begins
from the operations of the differential graded algebra. 
The differential $m_1: A \rightarrow A$ and the product $m_2: A \otimes A \rightarrow A$ after suspension become linear maps
\be\label{eq:SuspendendedBarMaps}
b_1: sA \rightarrow sA \; , \qquad b_2: sA \otimes sA \rightarrow sA\,, 
\ee
that, following~\eqref{cleardepfus} take the form $b_n = s \circ m_n \circ (s^{-1})^{\otimes n}$
for $n=1,2$
and thus give
\be\begin{split}
b_1(sa) =& \  sm_1 (a)\,,\\
b_2(sa_1,sa_2) = &   (-1)^{a_1+1}s m_2(a_1,a_2)\,. 
\end{split}
\ee
using the sign rule~\eqref{eq:SuspensionSign}.
Note that both $b_1$ and $b_2$ are of degree minus one.    
As linear maps $T^c(sA) \rightarrow sA$,  $b_1$ and $b_2$ 
can be lifted to coderivations $\bar B_1, \bar B_2$ on~$T^c(sA)$ 
\be
\bar B_i: T^c(s A) \rightarrow T^c(sA)\,,
\ee
by following the rule stated in~\eqref{eq:CoderLift}. 
Now define the coderivation $\bar B$ as the sum, 
\be\bar B = \bar B_1 + \bar B_2\,. \ee 
The condition $\bar B^2 = 0$ is equivalent to the defining relations of a differential graded algebra $(A,m_1,m_2)$, as we  will now verify. 

Given that  $\bar B$ is of odd degree, we have that $\bar B^2 = \frac{1}{2}[\bar B,\bar B]$ is also a coderivation. Since coderivations are fully determined by their image in $sA$, we have $\bar B^2 = 0$ if and only if $\pi_1 \circ \bar B^2 = 0$, with $\pi_1: T^c(sA) \to sA$ the projection.   
By explicit calculation we find that, acting on a single vector, 
\be
\label{ononeentry}
\begin{split}
\pi_1 \circ \bar B^2(sa) = & \  \pi_1 \circ \bar B  \circ \bar B_1 (sa)  = \pi_1 \circ \bar B  \circ b_1 (sa) =  \pi_1 \circ \bar B  \circ  s m_1 (a) \\[0.5ex]
= & \  \bar B_1   s m_1 (a)  = b_1   s m_1 (a)  =  s m_1 (m_1(a))  =  s m_1^2  a \,. 
\end{split}
\ee
Acting on two vectors we get 
\be
\begin{split}\label{ontwoentries}
\pi_1 \circ \bar B^2(sa \otimes sb) &= 
\pi_1 \circ \bar B \circ  \Bigl(\bar B_1 (s a \otimes s b) + \bar B_2 (sa \otimes s b) \Bigr) \\
&= \pi_1 \circ \bar B \circ  \Bigl( b_1 (s a) \otimes s b + (-1)^{a+1}  s a \otimes b_1 (s b)
+  b_2 (s a \otimes s b) \Bigr) \\ 
&= \pi_1 \circ \bar B \circ  \Bigl( s m_1(a) \otimes s b + (-1)^{a+1}  s a \otimes s m_1(b) 
+  (-1)^{a+1} s\, m_2 (a, b) \Bigr) \\ 
&= b_2\Bigl( sm_1 (a) \otimes s b + (-1)^{a+1} sa \otimes s m_1 (b) \Bigr)  + (-1)^{a+1} b_1(sm_2(a,b)) \\
&= (-1)^{a} s m_2(m_1 a,b) + s m_2(a,m_1 b) + (-1)^{a+1} s m_1 m_2(a,b) \\
&= (-1)^{a+1}\, s \Bigl(  m_1 m_2(a,b) - m_2(m_1 (a),b) - (-1)^a m_2(a,m_1( b))\Bigr) \; . \\
\end{split}
\ee
Acting on three arguments, the $\pi_1$ forces both $\bar B$'s to become $\bar B_2$:
\be
\begin{split}\label{onthreeentries}
\pi_1 \circ \bar B^2(sa \otimes sb \otimes sc) 
& =  \pi_1 \circ \bar B_2 \circ \bar B_2 (sa \otimes sb \otimes sc)  \\
& =  \pi_1 \circ \bar B_2 \Bigl(  b_2 (sa , sb)  \otimes sc + (-1)^{a+1} sa \otimes b_2 (sb, sc) \Bigr)  \\
& =   b_2 \Bigl(  (-1)^{a+1} s m_2 (a , b)  \otimes sc + (-1)^{a+b} sa \otimes sm_2 (b, c) \Bigr)  \\
&= (-1)^{b}s  \Bigl( m_2(m_2(a,b),c) -  m_2(a,m_2(b,c)) \Bigr) \; .
\end{split}
\ee
Since $\pi_1 \circ \bar B^2 = 0$ automatically 
on $(sA)^{\otimes k}$   
for $k > 3$, the condition $\pi_1 \circ \bar B^2 = 0$ is equivalent to the vanishing of the three expressions given in \eqref{ononeentry},~\eqref{ontwoentries}, and~\eqref{onthreeentries}. Therefore, the defining relations of the
differential graded algebra imply that the coderivation $\bar B$ is a differential
\be
m_1^2 = 0\; , \ m_1 m_2 = m_2(m_1 \otimes 1 + 1 \otimes m_1) \; , \ m_2(m_2 \otimes 1) =  
m_2 (1 \otimes m_2) \quad \Rightarrow \quad \bar B^2 = 0 \,\; . 
\ee 
This confirms that, as constructed, $\bar B$ in \eqref{cobardefexp} is a differential and a coderivation of degree minus one.  Thus, indeed, $B(A)$ is a differential graded coalgebra. 
Note that the above arrow can be reversed: $\bar B$ being a differential implies
the three conditions listed for the $m$'s.

Given a morphism $f_1: (A_1,m_1^{A_1},m_2^{A_1}) \rightarrow (A_2,m_1^{A_2},m_2^{A_2})$ of differential graded algebras, there is a morphism $F: (B(A_1),\bar B_1,\bar \Delta) \rightarrow (B(A_2),\bar B_2,\bar \Delta)$ of differential graded coalgebras. It is given~by
\begin{equation}
F(sa_1\otimes \cdots \otimes sa_n) = s f_1(a_1) \otimes \cdots \otimes s f_1(a_n) \; .
\end{equation}
Note that this is a special case of the lift~\eqref{eq:TCoHomLift}  of a map $f: T^c(sA_1) \rightarrow sA_2$, which only has a linear term $f(sa) = s f_1(a)$. 

\subsection{\texorpdfstring{$A_\infty$}{A-infinity} algebras}
The differential $\bar B$  
of the  bar construction is a coderivation   
constructed out of the linear map $b_1$ and the bilinear map $b_2$ given in \eqref{eq:SuspendendedBarMaps}, themselves  
defined through the differential and the associative product of the original differential graded algebra. As discussed before, 
a coderivation 
on $T^c(sA)$ in general has components
$b_k: (sA)^{\otimes k} \rightarrow sA$, 
acting as given in \eqref{eq:CoderLift}.  An $A_\infty$ algebra (or homotopy associative algebra) is defined by allowing for these higher contributions.

More concretely, an $A_\infty$ algebra consists of a graded vector space $A$, together with a collection of multilinear maps $m_k: A^{\otimes k} \rightarrow A$ of degree $\epsilon(m_k) = k-2$ for all $k \ge 1$.  
We denote the data of an $A_\infty$ algebra as follows:
\be
(A,\{m_k\}_{k \ge 1})\,.
\ee
Differential graded algebras are $A_\infty$ algebras with $m_{k \ge 3} = 0$. The constraints on the maps $m_k$ follow by introducing the multilinear maps $b_k: (sA)^{\otimes k} \to sA$ given by suspension following~\eqref{cleardepfus},    
\be 
\label{bnsusp}
b_k  =  s \circ m_k \circ (s^{-1})^{\otimes k}  \,.
\ee
The $b_k$ define coderivations $\bar B_k:  T^c(sA) \to T^c (sA)$, and then we form
the coderivation $\bar B$ on 
$T^c(sA)$ by writing
\be
\bar B = \sum_{k=1}^\infty \bar B_k \,. 
\ee
The constraints on $m_k$ follow by requiring $\bar B$ to be a differential, namely, $\bar B ^2 =0$.
We thus have a differential graded coalgebra 
\be 
\label{barofainfty}
B_\infty(A) =  (T^c(sA),\bar B, \bar \Delta)\,, 
\ee
that we call the {\em bar construction} of a given $A_\infty$ algebra $(A,\{m_k\}_{k \ge 1})$.

\bigskip
We call $(A,\{m_k\}_{k \ge 1})$ the $m$-picture of an $A_\infty$ algebra. Letting $V:= sA$, we call $(V,\{b_k\}_{k \ge 1})$ the $b$-picture of an $A_\infty$ algebra.

The maps $m_{k \ge 3}$ will modify the relations satisfied by a differential graded algebra. The first modification will appear in the associativity condition. In the $b$-picture, the square-zero condition $\bar B^2 = 0$ is satisfied if $\pi_1 \circ \bar B^2 = 0$ which implies
\be \label{eq:BPictureAinfinity}
\begin{split}
0 &= b_1^2(u) \; , \\
0 &= [b_1,b_2](u, v) \; , \\
0 &= [b_1,b_3](u,v,w) + b_2(b_2(u,v),w) + (-1)^{u} b_2(u,b_2(v,w))\, 
\end{split}
\ee 
on at most three inputs. Here, the definition of $[b_1,b_i]$ is 
given in \eqref{eq:CommutatorDifferential}. In the $m$-picture, the first two lines give the square-zero condition and the derivation property identical to those of differential graded algebras. The first modifications appear for three inputs. In terms of $m_i$, the third line in \eqref{eq:BPictureAinfinity} gives
\be
[m_1,m_3](a,b,c) = m_2(m_2(a,b),c) - m_2(a,m_2(b,c)) \; .
\ee
This condition says that $m_2$ is no longer an associative product. Rather, the associator $m_2 \circ (m_2 \otimes 1) - m_2 \circ (1 \otimes m_2 )$ is zero only up to a homotopy $m_3$, in the sense given in \eqref{eq:DefofHomotopy}. So $m_3$ defines a homotopy from the map $m_2 \circ (1 \otimes m_2)$ to $m_2 \circ  (m_2 \otimes 1)$. In particular, $m_2$ is not associative on $A$, but only on the homology $H(A)$. The higher $m_k$ induce an infinite set of relations, which will not be given explicitly. They can always be rederived from the single condition $\bar B^2 = 0$.

\medskip\noindent
{\bf A$_\infty$ morphisms.} One  defines 
a morphism $(A_1,\{m_k^{(1)}\}) \rightarrow (A_2,\{m_k^{(2)}\})$ 
of $A_\infty$ algebras as a morphism 
$\bar F$ of the associated differential graded coalgebras 
\be \bar F : \ \ \ (T^c(V_1),\bar B_1,\bar \Delta_1) \ \to \ (T^c(V_2),\bar B_2, \bar \Delta_2)\,, \ \ 
V_i \equiv sA_i \,, \ i = 1,2. 
\ee
Recall from \eqref{eq:TCoHomLift} that such a morphism is a map $\bar F: T^c(V_1) \rightarrow T^c(V_2)$ determined by a collection of $n$-linear maps 
\be\label{eq:AinftyMorphism}
\bar f_n: (V_1)^{\otimes n} \rightarrow V_2
\ee
of degree zero. The lifting formula \eqref{eq:TCoHomLift} then guarantees that $\bar F$ preserves the coproduct. The nontrivial condition is the chain map condition 
\be 
\bar F \circ \bar B_1 = \bar B_2 \circ \bar F \,.
\ee
Recall that both the left-hand side and the right-hand side are $\bar F$-coderivations
and so is their difference $ D_B \equiv \bar F \circ \bar B_1 - \bar B_2 \circ \bar F$.   To check that the $\bar F$-coderivation $D_B$ vanishes, we just need to check that the projection $\pi_1^{(2)}$ 
of its image vanishes (recall the discussion around~\eqref{dfdetbyd}).  Thus it suffices to check that
\be
\label{realeqntoc}
\pi_1^{(2)} \circ \bar F \circ \bar B_1 =  \pi_1^{(2)} \circ \bar B_2 \circ \bar F \,.
\ee
We expand the coderivations as $\bar B_i = \sum_{k\geq 1} \bar B^{(i)}_k$ and then find (in the $b$-picture) for the action on one and on two arguments
\be 
\label{tworelbse}
\begin{split}
\bar f_1 b_1^{(1)} &= b_1^{(2)} \bar f_1 \; , \hskip84pt  \hbox{on} \ \ V_1 \,,  \\[1.0ex]
\bar f_1  b_2^{(1)}  + \bar f_2 (b_1^{(1)} \otimes 1 +  1 \otimes b_1^{(1)})  &= b_1^{(2)} \bar f_2 + b_2^{(2)}(  \bar f_1 \otimes \bar f_1) \,, \hskip10pt \hbox{on} \ \  V_1 \otimes V_1 \,.  
\end{split}
\ee
More generally, the condition~\eqref{realeqntoc} acting on $(V_1)^{\otimes n}$ gives:
\be
\label{generalbpiccond}
\sum_{\ell=1}^n  \bar f_\ell \circ \bar B^{(1)}_{n+1-\ell} \  = \  \sum_{\ell=1}^n  b^{(2)}_\ell \sum_{i_1 + \cdots + i_\ell = n}  ( \bar f_{i_1} \otimes \cdots \otimes \bar f_{i_\ell} ) \,,  \ \ \ \hbox{acting on} \ \ V_1^{\otimes n} \,.
\ee
Both sides of the equation yield a vector in $V_2$. The $\bar B$ on the left-hand side has
a bar because it should be treated as a coderivation. The structure of the left-hand side
is clear recalling that $\pi_1^{(2)}\circ \bar F=  \bar f :  T^c(V_1) \to V_2$.
The structure of the right-hand side
is clear noting that $\pi_1^{(2)}\circ \bar B_2 =  \sum_{\ell\geq 1}  b^{(2)}_\ell \circ \pi^{(2)}_\ell$, with $\pi^{(2)}_\ell:  \, T^c(V_2) \to (V_2)^{\otimes \ell}$.

In the $m$-picture we use analogous notation, and the $f_k$'s have no bars. 
Using the suspension formulae relating the $b_n$'s and the $\bar f_n$ to
the $m_n$'s and $f_n$'s, respectively, equations~\eqref{tworelbse} give, with a little rearrangement,
\be
\begin{split}
f_1 m_1^{(1)} &=\  m_1^{(2)}f_1\; , \hskip134pt  \hbox{on} \ \ A_1 \,,  \\[1.0ex]
f_1 \,m_2^{(1)} - m_2^{(2)}(f_1 \otimes f_1) &= \ 
m_1^{(2)} f_2 + f_2 \bigl(\,  m_1^{(1)}\otimes 1  + 1 \otimes m_1^{(1)} )  \hskip10pt \hbox{on} \ \  A_1 \otimes A_1\,.  
\end{split}
\ee 
The first equation is the chain map condition for $f_1$, while the second equation tells us that $f_1$ is an algebra morphism up to homotopy $f_2$.

\medskip\noindent 
{\bf Maurer-Cartan for A$_{\infty}$.}
The Maurer-Cartan equation also generalizes in the $A_\infty$ setting. In the $b$-picture, the Maurer-Cartan equation is
\be\label{eq:bpictureMC}
0 = \sum_{k \ge 1} b_k(v,\ldots,v)\,, 
\ee
for some $v \in V$ of degree zero.
In fact, the Maurer-Cartan equation in the $m$-picture is equally simple.
To see this, act on the above equation with $s^{-1}$, set $v = sa$ for some degree minus one $a\in A_{-1}$, and use~\eqref{bnsusp} to find
\be\label{eq:bpicturexx}
0 =   \sum_{k \ge 1} s^{-1} b_k(sa,\ldots,sa)  = 
\sum_{k \ge 1} m_k \circ (s^{-1} )^{\otimes k} (sa,\ldots,sa)\,, 
\ee
Since the $sa$'s are all of degree zero, there is no sign from letting the
$s^{-1}$ desuspensions act and we find the $m$-picture Maurer-Cartan equation:
\be\label{mpicMC}
0 = \sum_{k \ge 1} m_k(a,\ldots, a) \; .\ee

The Maurer-Cartan equation defines the Maurer-Cartan space $\mathcal{MC}(A)$ of deformations of the $A_\infty$ algebra. The deformed products $m_n^a$ then satisfy the $A_\infty$ algebra relations. 
The deformations are expressed most conveniently in the $b$-picture, where the deformed products are given by
\be  b_{n}^v(w_1,\ldots,w_n) = \sum_{k_0,\ldots,k_n \ge 0} b_{n + k_0 + \cdots + k_n}(v^{k_0},w_1,v^{k_1},w_2,v^{k_2},\ldots,v^{k_{n-1}},w_n,v^{k_n}) \;,\ee
for $v$ a solution to~\eqref{eq:bpictureMC}. In the $m$-picture, the deformed product
\be
m_{n}^a(x_1,\ldots,x_n) = \sum_{k_0,\ldots,k_n \ge 0}(-1)^{\sigma} m_{n + k_0 + \cdots + k_n}(a^{k_0},x_1,a^{k_1},x_2,a^{k_2},\ldots,a^{k_{n-1}},x_n,a^{k_n}) \; ,\ee
has a sign given by $\sigma = \sum_{j = 1}^n k_j(j + \sum_{i = 1}^j\epsilon(x_i))$.
The above two formulae are related by suspension, letting
$v= sa$ and $w_i = sx_i$. The infinitesimal gauge transformation preserving the Maurer-Cartan equation in the $b$-picture reads
\be\label{ainftygauge} \delta_\eta v = b_1^v(\eta) = \sum_{k_0,k_1 \ge 0} 
b_{k_0 + k_1 + 1}(v^{k_0},\eta,v^{k_1}) \,, \ee
for some gauge parameter $\eta \in V_1$. 
In the $m$-picture, we have
\be
\delta_\lambda a = m_1^a(\lambda) = \sum_{k_0,k_1 \ge 0} (-1)^{k_1} m_{k_0 + k_1 + 1}(a^{k_0},\lambda,a^{k_1}) \, ,
\ee
where $\lambda \in A_0$. 
Finite gauge transformations are most conveniently described in a form that generalizes \eqref{eq:AltFiniteGT}. For $A_\infty$ algebras, two solutions $a,a' \in \mathcal{MC}(A)$ are gauge equivalent, if they satisfy
\be
a' - a = \sum_{k,l \ge 0} (-1)^l m_{k+l+1}(a^k,u,{a'}^l)\,
\ee
for some $u \in A_0$, representing the finite gauge parameter. In the $b$-picture, this becomes
\be
v' - v = \sum_{k,l \ge 0} b_{k+l+1}(v^k,z,{v'}^l) \, ,
\ee 
with gauge parameter $z \in V_1$. This form of gauge equivalence can be found in \cite{MILHAM2023384}.  

From equation \eqref{eq:MorphismsPreserveMC} we were able to deduce that morphisms $f: (A,m_1^A,m_2^A) \rightarrow (B,m_1^B,m_2^B)$ between differential graded algebras preserve Maurer-Cartan elements. This fact generalizes to $A_\infty$ morphisms as follows. Let $\bar F: \  (T^c(V_1), \bar B_1, \bar\Delta_1 ) \rightarrow (T^c(V_2),\bar B_2 , \bar \Delta_2)$ be a morphism of the associated bar constructions of $A_\infty$ algebras. Then $\bar F$ defines a map
\be\label{eq:MorphismOnMC}
\begin{split}
f_{\mathcal{MC}}: \mathcal{MC}(A_1) &\rightarrow \mathcal{MC}(A_2) \, , \\
 v_1 &\mapsto f_{\mathcal{MC}}(v_1) :=  \sum_{k \ge 1} \bar f_k (v_1,\ldots,v_1) \, ,
\end{split}
\ee
where the $\bar f_k$ are related to $\bar F$ via
\be
\bar f = \pi_1 \circ \bar F = \sum_{k\geq 1}  \bar f_k \circ \pi_k \, .
\ee
Observe that the map $f_{\mathcal{MC}} $ in~\eqref{eq:MorphismOnMC} indeed generalizes the differential graded algebra case by including contributions $\bar f_k$ for $k \ge 2$. But note that for a general $A_\infty$ morphism, $f_{\mathcal{MC}}(v_1)$ depends nonlinearly on $v_1$.

To establish \eqref{eq:MorphismOnMC}, we need to prove that given $v_1 \in (V_1)_{0}$ a solution,
\be
\label{v2ansatz}
v_2  = f_{\mathcal{MC}}(v_1) = \sum_{k \ge 1} \bar f_k (v_1,\ldots,v_1)\,,
\ee
with $v_2 \in (V_2)_{0}$, is a solution to the Maurer-Cartan equation of the algebra. Proving this efficiently requires a bit of notation.  For the coderivations $\bar B_i$ we write
\be
\label{defineBbFf}
\bar B_i  = \sum_{k\geq 1} \bar B^{(i)}_k  \,,  \ \ \ \ \qquad 
b_{(i)} \equiv  \pi_1 \circ \bar B_i = \sum_{k\geq 1}  b_k^{(i)} \pi_k \, \quad i =1,2\,.
\ee
We also recall from \eqref{eq:CoHomLift} that $\bar F$ is related to $\bar f$ through
\be
\bar F =  \sum_{n\geq 1} \iota_n \circ \bar f^{\otimes n} \circ \bar \Delta_1^{n}  \, . 
\ee
We then define the object $E(v_1) \in T^c(V_1)$ for $v_1 \in V_1$: 
\be\label{eq:DefGroupLikeElement}
E(v_1) :=  \sum_{n\geq 1}  (v_1)^{\otimes n} =  v_1 + v_1 \otimes v_1  + \cdots \,.
\ee
The Maurer-Cartan equation for $v_1$  is 
\be
0 = b_{(1)} E(v_1) =  \sum_{k\geq 1}  b_k^{(1)} (v_1, \ldots, v_1) \,.  
\ee
With these definitions one can prove the following relations
\be
\label{3ordlk4}
\begin{split}
\bar\Delta E(v_1) = & \   E(v_1) \otimes' E(v_1) \,, \\
\bar F ( E (v_1) ) =  & \   E \bigl( \bar f (E (v_1))\bigr) \,,\\ 
\bar B_1 E(v_1) = & \  \bigl( 1 + E(v_1)\bigr) b_{(1)} ( E(v_1) )  
\bigl( 1 + E(v_1) \bigr)\,. 
\end{split} 
\ee
We note that~\eqref{v2ansatz}, giving the image of the Maurer-Cartan element on the
second algebra, can be rewritten as
\be\label{eq:RewritingMCImage}
v_2 = f_{\cal MC} (v_1) =  \bar f  (E(v_1))  = \sum_{k\geq1} \bar f_k \circ \pi_k \,  E(v_1) \,.
\ee
Now, finally, we note that, by the third equation in~\eqref{3ordlk4}, if the Maurer-Cartan equation for~$v_1$ holds, we have $\bar B_1 E(v_1) = 0$, and therefore $\bar f$ acting on this is zero:
\be
0 = \bar f (\bar B_1 E(v_1)) = \pi_1^{(2)}  \circ \bar F \bar B_1  E(v_1)
=  \pi_1^{(2)}  \circ \bar B_2 \bar F  E(v_1) = b_{(2)}   E(\bar f(E(v_1)))= b_{(2)}   E(v_2)\,, 
\ee 
using the chain map property.  This shows that $v_2 = f_{\mathcal{MC}}(v_1)$ solves the Maurer-Cartan equation for the second algebra, as claimed.

The map $f_{\mathcal{MC}}: \mathcal{MC}(A_1) \rightarrow \mathcal{MC}(A_2)$ of Maurer-Cartan spaces is also well-defined as a map of Maurer-Cartan sets, that is, it induces a map
\be\label{eq:MapOnMCSet}
\begin{split}
    f_{MC}:  {MC}(A_1) &\rightarrow MC(A_2) \, , \\
[v_1] &\mapsto [v_2] = [f_{\mathcal{MC}}(v_1)] \, ,
\end{split}
\ee
where $[v_1]$, resp.~$[v_2]$, denote the gauge equivalence classes in $\mathcal{MC}(A_1)$, resp.~$\mathcal{MC}(A_2)$. To see that this map is well-defined, we need to show that if $[v_1] = [v_1']$, then $ [v_2] = [v_2']$. Assuming that $[v_1] = [v_1']$ means that $v_1$ and $v_1'$ satisfy
\be
v'_1 - v_1 = \sum_{k,l \ge 0} b^{(1)}_{k+l+1}(v^k_1,z_1,{v'_1}^l)
\ee
for some $z_1 \in (V_{1})_1$. One can then show that $v_2 = f_{\mathcal{MC}}(v_1)$ and $v_2' = f_{\mathcal{MC}}(v'_1)$ are related by a gauge transformation
\be
v'_2 - v_2 = \sum_{k,l \ge 0} b^{(2)}_{k+l+1}(v^k_2,z_2,{v'_2}^l)
\ee
with
\be
z_2 = \sum_{k,l\ge 0} \bar f_{k+l+1}(v_1^k,z_1,{v'_1}^l) \, .
\ee
Therefore, $[v_2] = [v_2']$. This shows that the map \eqref{eq:MapOnMCSet} is indeed well-defined. 

It is proven in \cite{MILHAM2023384} that if a morphism $\bar F: (T^c(V_1),\bar B_1,\bar \Delta_1) \rightarrow (T^c(V_2),\bar B_2,\bar \Delta_2)$ of $A_\infty$ algebras is a quasi-isomorphism, that is, the associated map $H(\bar F): H(\bar B_1) \rightarrow H(\bar B_2)$ between homologies is invertible, then also the associated map \eqref{eq:MapOnMCSet} between Maurer-Cartan sets is invertible.

\subsection{Cobar construction} \label{sec:CobarConstruction}

The cobar construction is the dual of the bar construction: it associates to a differential
graded {\em coalgebra} $C$ a differential graded {\em algebra} $\Omega(C)$. 
Explicitly, given a differential graded coalgebra $(C,\bar B, \bar \Delta)$, there is a differential graded algebra $\Omega(C)$
given by
\be
\label{thecobarstate}
\Omega(C) = (T(s^{-1}C), B + \Delta,\mu)\,. 
\ee
The graded algebra structure is given by the tensor algebra $(T(s^{-1}C),\mu)$. The differential $B + \Delta$ is constructed out of the desuspensions of $\bar B$ and $\bar \Delta$, as we will explain below.

To relate $B$ and $\bar B$ we follow~\eqref{cleardepfus} and
define $ \bar B = s\circ  B \circ s^{-1}$, so that 
\be\label{bsuspensiondef}
B(s^{-1}c) = s^{-1}\bar B(c)\,,  \ \ \  c \in C \,. \ee
The operator $B$ on $s^{-1} C$ is then extended to $T(\ds C)$ as a degree minus one derivation:
\be  
\epsilon(B) = -1 \,, \ee
with the rule
\be\label{eq:CobarLiftB}
B(\ds c_1 \otimes' \cdots \otimes' \ds c_n) = \sum_{i = 1}^n (-1)^{(c_1-1) +  \cdots + (c_{i-1}-1)} s^{-1}c_1 \otimes' \cdots \otimes' B (s^{-1} c_i) \otimes' \cdots \otimes' 
s^{-1} c_n \; .\ee 
Since  $\bar \Delta$ is a map from $C$ to $C \otimes'C$ we follow~\eqref{30rdfjhe} to set
$\Delta= - (s^{-1} \otimes's^{-1}) \circ \bar \Delta \circ s$. Equivalently,
\be
\label{desuspbardelta}
\Delta \circ s^{-1} = - (s^{-1} \otimes's^{-1}) \circ \bar \Delta \,. 
\ee
We see that $\Delta$ has degree minus one:
\be
\epsilon (\Delta ) = -1 \,. 
\ee
Acting on vectors,~\refb{desuspbardelta} implies that  
\be
\bar \Delta(c) = \sum_i c_i \otimes' c^i  \quad \Rightarrow \quad 
\Delta (s^{-1}c) 
= \sum_i (-1)^{c_i+1} s^{-1}c_i 
\otimes' 
s^{-1}c^i \; .
\ee
From $\bar B^2 = 0$ and $B = s^{-1}\circ \bar B  \circ s$ we see that
\be
B^2 = 0 \,.
\ee
This holds both on $s^{-1} C$ and on $T(s^{-1} C)$.

Consider now the equation that states that $\bar B$ is a $\bar \Delta$ coderivation. Letting $s^{-1}$ act on the right,
\be
\begin{split}
   \Delta  B s^{-1}  = \Delta s^{-1} \bar B  = -  (s^{-1}\otimes's^{-1})\bar\Delta \bar B = & \ -(s^{-1}\otimes's^{-1})\bigl( \bar B \otimes' \textbf{1}  + \textbf{1}  \otimes'  \bar B\bigr)  \bar \Delta\\
   = &\ \ \bigl( s^{-1} \bar B \otimes's^{-1}  - s^{-1}  \otimes's^{-1}  \bar B\bigr)  \bar \Delta   \\
   = & \ \ \bigl(  B \otimes'\textbf{1} + \textbf{1} \otimes' B\bigr)(s^{-1}\otimes's^{-1})\bar\Delta   \\ 
   = & \  -\bigl(  B \otimes'\textbf{1}  + \textbf{1} \otimes' B\bigr) \Delta s^{-1} 
  \,.
\end{split}
\ee
We thus learn that $\Delta$ and $B$ commute: 
\be
\Delta B + ( B \otimes'\textbf{1}  + \textbf{1}  \otimes'  B)  \Delta = 0   \quad \Rightarrow \quad \Delta B + B \Delta = 0 \quad \Rightarrow \quad  [\Delta, B] = 0 \,. 
\ee
Coassociativity of $\bar \Delta$ also gives a new relation for $\Delta$. Letting $s^{-1}\otimes's^{-1}\otimes's^{-1}$ act on the coassociativity relation, we find
that
\be
\begin{split}
 s^{-1} \otimes'(s^{-1} \otimes's^{-1}) (\textbf{1} \otimes'\bar\Delta) \bar\Delta \  = & \  \ (s^{-1} \otimes's^{-1}) \otimes's^{-1}  (  \bar\Delta  \otimes'\textbf{1}) \bar\Delta \,, \\
 -(s^{-1} \otimes' \Delta s^{-1} )\bar\Delta\  = & \  \ - ( \Delta s^{-1}  \otimes's^{-1}) \bar \Delta\,,\\
 (\textbf{1} \otimes' \Delta)(s^{-1}\otimes's^{-1})\bar\Delta\  = & \  \ - (\Delta \otimes'\textbf{1}) (s^{-1}\otimes's^{-1})\bar\Delta \,,\\
-(\textbf{1} \otimes' \Delta)\Delta \, s^{-1}\  = & \  \  ( \Delta \otimes'\textbf{1})\Delta s^{-1}\,.
\end{split}
\ee
We thus learn that the desuspended $\Delta$ is now a differential:
\be
\bigl( \textbf{1} \otimes'\Delta +  
\Delta  \otimes'\textbf{1} \bigr) \Delta = 0  \quad \Rightarrow \quad   \Delta^2 = 0. 
\ee
Having shown that 
$B^2 = \Delta^2 = [ B , \Delta ] = 0$  we now conclude that, as claimed in the statement
of~\eqref{thecobarstate}, the sum $B+ \Delta$ is a differential on $T(s^{-1} C)$: 
\be
(B + \Delta)^2 =  B^2 + \Delta^2  + B \Delta + \Delta B = 0 \, .
\ee
Thus, $\Omega(C)$ is  a differential graded algebra. 

Just like in the bar construction, morphisms carry over in the cobar construction. One can show that a morphism of differential graded coalgebras $\bar f: C_1 \rightarrow C_2$ defines a morphism $F: \Omega(C_1) \rightarrow \Omega(C_2)$. Explicitly,
\be
F(s^{-1}c_1 \otimes' \cdots \otimes' s^{-1}c_n) = s^{-1} \bar f(c_1) \otimes' \cdots \otimes' s^{-1}\bar f(c_n) \; .
\ee
This is the special case of the lift~\eqref{diagrtenalgmorph} of the map $f= s^{-1}\circ\bar f\circ s :s^{-1} C_1 \to T(s^{-1} C_2)$, which has image only in $s^{-1} C_2 \subseteq T(s^{-1} C_2)$:
\be
\label{diagrtenalgmorphXS}
\begin{tikzcd}
 T(s^{-1} C_1) \arrow[dr,"F"]& \\
    s^{-1} C_1 \arrow[r,"f"] \arrow[u,"\iota_1"] & T(s^{-1} C_2) 
\end{tikzcd}
\ee

The bar and cobar constructions are related as follows. There is a bijective correspondence between a morphism $F: C \rightarrow B(A)$ and an \emph{adjoint} morphism $F^+: \Omega(C) \rightarrow A$. We find it convenient to display this fact in the following diagram\footnote{This is not to be understood as a commutative diagram, since $bar$ and $cobar$ are not morphisms.}
\be\label{eq:BarCobarAdjunction}
\begin{tikzcd}
A \arrow[r,"bar"] & B(A)  \\ 
\Omega(C) \arrow[u,"F^+"] & \arrow[l,"cobar"] C\arrow[u,"F"]
\end{tikzcd} 
\ee
The relation works as follows. Recall that any morphism $F: C \rightarrow B (A)= T^c(sA)$ of coalgebras is determined by a degree zero linear map~\eqref{morphcoalgdiag},
\be f = \pi_1 \circ F: C \rightarrow sA.
\ee
We can view the same map as a degree zero linear map 
\be
f^+: \ds C \rightarrow A\,. 
\ee
As explained in~\eqref{diagrtenalgmorph}, this induces a morphism of algebras $F^+: \Omega(C) \rightarrow A$. 
The map $F^+: \Omega (C) \rightarrow A$ is a morphism of differential graded algebras if and only if 
$F: C \rightarrow B (A)$
is a morphism of differential graded coalgebras (Theorem 2.2.6 in \cite{Loday:2012operads}). Furthermore, $F$ is a quasi-isomorphism if and only if $F^+$ is a quasi-isomorphism (Theorem 2.3.1 in \cite{Loday:2012operads}). Proving this
relation between quasi-isomorphisms requires sophisticated mathematical techniques that we will not discuss here.

\subsection{Bar-cobar construction of \texorpdfstring{$A_\infty$}{A-infinity} algebras}\label{sec:BarCobarAinfty}

Given an $A_\infty$ algebra $(A,\{m_k\}_{k \ge 1})$ in terms of a coalgebra $C = (T^c(sA),\bar B,\bar \Delta)$ via the bar construction,
one can ask whether there is an equivalent differential graded algebra. The cobar construction applied to $C$ provides such a differential graded algebra:
\be\label{eq:DefBarCobarComplex}
\Omega (C) = (T(\ds T^c(sA)),B + \Delta, \mu) \; .
\ee
This tensor algebra has a very simple product structure $\mu$ given by taking the tensor product, denoted as~$\otimes'$. 
All the information about the original $A_\infty$ algebra is stored in the differential $B$. The price we have to pay for the simplification in the product structure is that the underlying graded vector space $T(s^{-1} T^c(sA))$ is much larger than the original space $A$.

Before discussing the equivalence of $(A,\{m_k\}_{k \ge 1})$ with $\Omega (C)$, let us first explore the algebra $\Omega (C)$ in more detail. We define 
\be
V := sA\,,
\ee
so that the vector space of $\Omega(C)$ is $T(s^{-1} T^c(V))$. We then use the concatenation notation
\be
A_\ell := v_1 \cdots v_\ell \in T^c(V) \; , 
\ee
where $v_i \in V$. We say that $\ell$ is the \emph{length} of $A_\ell$. An element in $\Omega(C)$ is then given by
\be
R_{f,\ell} := \ds A_{\ell_1} \otimes' \ds A_{\ell_2} \otimes' \cdots \otimes' \ds 
A_{\ell_f} \; ,
\ee
where $A_{\ell_i} \in T^c(V)$ is an element of length $\ell_i$. We define $\ell = \sum_i {\ell_i}$ to be the length of $R_{f,\ell}$. Moreover, we say that $f$ is the \emph{factor number} of $R_{f,\ell}$. We necessarily have $f \le \ell$. We write 
\be\label{eq:factorlengthsubspace}
\Omega (C)_{f,\ell}\,,
\ee
to denote the subspace of factor number $f$ and length $\ell$.

Following~\eqref{desuspbardelta} 
on an element in $\ds T^c(V)$, the differential $\Delta$ acts as
\be
\label{keyformulaforDelta}
\Delta \, \ds (v_1 \cdots v_n)
= -\sum_{i = 1}^{n-1}(-1)^{v_1 + \cdots + v_i} \ds (v_1 \cdots v_i) \otimes' \ds (v_{i+1} \cdots v_n)  \, ,
\ee
and it is extended over $\mu$ via Leibniz rule. This implies that $\Delta$ increases factor number by one and preserves the length of an element. The product $\mu$, however, is a binary operation: it maps inputs of factor numbers $f_1$ and $f_2$ to factor number $f_1+f_2$. Additionally, it maps inputs of lengths $\ell_1$ and $\ell_2$ to length $\ell_1+\ell_2$. On the other hand,  we write
\be B = \sum_{k \ge 1} B_k\,, \ee  
and applying~\eqref{bsuspensiondef} we note that 
\be
B_k s^{-1} = s^{-1} \bar B_k, 
\ee 
and the coderivation
$\bar B_k$ acts via the multilinear maps (also denoted by) $b_k$, so that 
\be
B_k\, \ds (v_1 \cdots v_n)
=  \sum_{i = 0}^{n-k} (-1)^{v_1 + \cdots + v_i}\ds
(v_1 \cdots v_i \,b_k(v_{i+1},\ldots, v_{i+k}) v_{i+k+1} \cdots v_n) \; .
\ee
This is extended as an odd-derivation over $\mu$. Unlike $\Delta$, the operator $B_k$
does not produce an~$\otimes'$. As a result, $B_k$   
preserves factor number while reducing the length by $k-1$.

We now wish to identify the claimed quasi-isomorphism
\be
\label{searchedformorph}
(A,\{m_k\}_{k \ge 1}) \ \ \to  \ \ 
\Omega(C) = (T (\ds C),B + \Delta,\mu)\,, 
\ee
where the target is a differential graded algebra that is itself a particular case of an $A_\infty$ algebra. 
Here $C= T^c(sA)$ is the vector space  of the coalgebra
$B_\infty(A) = ( C, \bar B, \bar \Delta)$ 
 associated with the $A_\infty$ algebra $(A,\{m_k\}_{k \ge 1})$.
We recall now from the discussion around equation~\eqref{eq:AinftyMorphism} that a morphism of $A_\infty$ algebras is to be viewed as a morphism of the associated coalgebras obtained by the bar
construction. 
Thus the morphism above 
is the same as a morphism of 
differential graded coalgebras 
\be\label{eq:UniversalInclusion}
I: \ \  B_\infty(A) = C \rightarrow B(\Omega (C))\,. 
\ee
Fitting this morphism $I$ as the right vertical one in diagram~\eqref{eq:BarCobarAdjunction} 
we complete the diagram as follows
\be\label{eq:BarCobarAdjunction1}
\begin{tikzcd}
\Omega(C) \arrow[r,"bar"] & B(\Omega(C))  \\ 
\Omega(C) \arrow[u,"I^+"] & \arrow[l,"cobar"] C\arrow[u,"I"]
\end{tikzcd} 
\ee
We now see that the coalgebra morphism $I$ is equivalent to an adjoint algebra morphism $I^+$ mapping
$\Omega(C)$ to itself:
\be  I^+: \Omega (C) \rightarrow \Omega (C)\,.
\ee
We can choose 
\be
I^+ = \text{id}_{\Omega (C)}\,. 
\ee
Since $I^+$ is trivially a quasi-isomorphism, it is guaranteed that $I$ is also a quasi-isomorphism. Interestingly, while $I^+$ is in fact an isomorphism, $I$ is merely a quasi-isomorphism.

Explicitly, as a morphism from the tensor algebra to itself 
\be 
I^+:T(s^{-1} C) \rightarrow T(s^{-1} C)\,,
\ee
the following diagram, a version of~\eqref{diagrtenalgmorph}, tells us what is the map $i^+: s^{-1} C \to T(s^{-1} C)$ whose lift
gives $I^+$:
\be
\label{diagrtenalgmorphcobar}
\begin{tikzcd}
 T(s^{-1} C) \arrow[dr,"I^+"]& \\
    s^{-1} C \arrow[r,"i^+"] \arrow[u,"\iota_{s^{-1}C}"] & T(s^{-1} C)
\end{tikzcd}
\ee
We see that 
\be
i^+ = I^+ \circ \iota_{s^{-1}C} = \iota_{s^{-1}C}\,,  \ \ \
\ee
since $I^+$ is the identity map.  Here, $\iota_{s^{-1}C}$ is the inclusion map
\be \  \iota_{s^{-1}C}: s^{-1} C \rightarrow T(s^{-1} C) \,. \ee
By suspending $i^+$ we obtain the inclusion map $i$
\be 
i: C \rightarrow sT(s^{-1}C)\,, 
\ee
which determines a morphism of coalgebras 
\be
I: C \rightarrow T^c(s T(s^{-1}C))\,,
\ee
via the
diagram~\eqref{morphcoalgdiag} applied to this case:
\be
\label{morphcoalgdiagforbarcobar}
\begin{tikzcd}
& T^c (sT(s^{-1}C)) \arrow[d,"
\pi_1"] \\
C  \arrow[ur,"I"] \arrow[r,"i"] & sT(s^{-1} C) 
\end{tikzcd}
\ee

In our discussion $C$ is the coalgebra $B_\infty(A)$ 
representing the 
$A_\infty$ algebra, therefore $C = T^c(sA) = T^c(V)$.
The map $i$ above is thus:
\be i: T^c(V) \rightarrow s T(s^{-1}T^c(V))\,,
\ee
and maps elements $v_1 \cdots v_k\in T^c(V)$ to the same element in $sT(s^{-1} T^c(V))$ in factor number one. The $k$-linear component of $i$ is therefore
\be
i_k(v_1,\ldots,v_k) = s\big(s^{-1}(v_1 \cdots v_k)\big)\ \in \ s s^{-1}V^{\otimes k} \subseteq s s^{-1}T^c(V) \subseteq s T(s^{-1} T^c(V)) \, .
\ee

Since $i$ is a morphism of $A_\infty$ algebras, it induces a map $i_{\mathcal{MC}}$ between Maurer-Cartan spaces, see \eqref{eq:MorphismOnMC}. We now want to verify this fact in this particular case, in order to explicitly see how Maurer-Cartan solutions of $(A,\{m_k\})$ are embedded into Maurer-Cartan solutions in
$(\Omega (B_\infty(A)),B + \Delta,\mu)$.
As before, we write $V = sA$ and let $v \in V_0$
be a solution to the Maurer-Cartan equation associated with $A$. So $v$ satisfies
\be\label{eq:MConV}
0 = \sum_{k \ge 1} b_k(v,\ldots,v) = \pi_1 \circ \bar B(E(v)) \, ,
\ee
where $E(v)$ is defined in \eqref{eq:DefGroupLikeElement}
and $\bar B$ is defined in~\eqref{defineBbFf}. Following \eqref{eq:RewritingMCImage}, 
we can write the image of $v$ under  $i_{\mathcal{MC}}$ as the element $sa$ with $a$ the
degree minus one Maurer-Cartan element in $\Omega (B_\infty(A))$
\be
sa := i_{\mathcal{MC}}(v)
= i (E(v)) =  E(v) = s(s^{-1} E(v)) \in s s^{-1}T^c(V) \subseteq s T(s^{-1}T^c(V)).
\ee
Note that the final set is $s \Omega (B_\infty(A))$, the extra suspension appears because
$i$ is defined as a morphism in $b$-picture. As an element in the $m$-picture, $a$
is therefore given by
\be
a = s^{-1} E(v)  \,. 
\ee
We check that $ a$
satisfies the Maurer-Cartan equation
\be\label{eq:v2MC}
\Delta a + B a + a \otimes' a = 0
\ee
of $(\Omega(B_\infty(A)),B + \Delta,\mu)$. Recall from \eqref{3ordlk4} that $E(v)$ satisfies
\be
\bar \Delta E(v) = E(v) \otimes' E(v) \, .  
\ee
Together with the desuspension given in~\eqref{desuspbardelta}, this implies that
\be\label{eq:MCwoB}
\Delta a = \Delta s^{-1} E(v) = - s^{-1}E(v) \otimes' s^{-1}E(v) = - a \otimes' a \, . 
\ee
Using this in Maurer-Cartan equation \eqref{eq:v2MC} we find that the Maurer-Cartan equation reduces to $B a = 0$.  
But then
\be
B a = B s^{-1} E(v) = s^{-1}\bar B E(v)\,.
\ee
The right-hand side vanishes when $v$ satisfies the Maurer-Cartan equation. 
So the solutions to \eqref{eq:MConV} are parametrized by $s^{-1}E(v)$ so that $v$ satisfies the original Maurer-Cartan equation on $V = sA$.

The map $i_{\mathcal{MC}}$ identifies the solutions to the original Maurer-Cartan equation with an element in $\Omega (B_\infty (A))$ of factor number one. In general, there will also be solutions in other (in general inhomogeneous) factor numbers. Since $I: C \rightarrow  T^c(s T(\ds C))$ is a quasi-isomorphism, we know that the map $i_{MC}$ on Maurer-Cartan sets, as defined in \eqref{eq:MapOnMCSet}, is invertible on Maurer-Cartan sets.

To conclude that $i_{MC}: MC(A) \rightarrow MC(\Omega(B_\infty(A)))$ is invertible, we relied on two nontrivial mathematical facts:
\begin{enumerate}
    \item The quasi-isomorphism property is preserved under taking the adjoint.
    \item A quasi-isomorphism induces an isomorphism of Maurer-Cartan sets.
\end{enumerate}
These facts rely on certain assumptions on the structure of the original algebra $A$. Sections \ref{thehomofDanddef}-\ref{solveqbplusdelta} are dedicated to studying the nontrivial equivalence of $MC(A)$ and $MC(\Omega(B_\infty(A)))$. We will explicitly solve the Maurer-Cartan equation of $\Omega B_\infty(A)$ and find the following:
\begin{enumerate}
    \item Any solution to the Maurer-Cartan equation in $\Omega B_\infty(A)$ is a gauge transformation of a solution in factor number one only. We call them \emph{canonical solutions} and show that they are necessarily of the form $i_{\mathcal{MC}}(v)$.
    \item Gauge transformations in factor number one preserving canonical solutions can be identified with gauge transformations in $A$.
\end{enumerate}
We will do so by considering solutions perturbatively in powers of $B$.

\subsection{Bar-cobar construction of \texorpdfstring{$L_\infty$}{L-infinity} algebras} 

Perturbative field theories can often be encoded by $L_\infty$ algebras, namely, homotopy Lie algebras. They are the Lie algebra analogue of $A_\infty$ algebras. The description of $L_\infty$ algebras follows closely the description of $A_\infty$ algebras. 

We first recall that a differential graded Lie algebra $(A,l_1,l_2)$ is a differential 
graded vector space $(A,l_1)$, meaning that $l_1$ is of degree minus one and
$l_1l_1=0$, together with a degree zero Lie bracket $l_2$,
\be
l_2: A \otimes A \rightarrow A\,,
\ee
so that $l_1$ satisfies the Leibniz rule
\be
l_1l_2(a,b) = l_2(l_1 a,b) + (-1)^a l_2(a,l_1 b) \; .
\ee
As a Lie bracket, $l_2$ is graded antisymmetric
\be 
l_2 (a, b)= -(-1)^{ab} \, l_2 (b, a)\,,   
\ee
and satisfies a graded Jacobi identity
\be
(-1)^{ac}\, l_2 \bigl( a, l_2 (b,c)\bigr)
+
(-1)^{ba} \, l_2\bigl( b, l_2 (c,a)\bigr)
+
(-1)^{cb}\, l_2 \bigl( c, l_2 (a,b)\bigr)
=0.
\ee
This is often written in derivation form using a bracket to represent $l_2: [a,b] =l_2 (a,b)$ 
\be
[ a, [b,c]] = [ [a,b],c] +(-1)^{ab} [ b, [a,c]]\,.
\ee

Differential graded Lie algebras have a Maurer-Cartan equation associated to them. Let $a \in A_{-1}$ be an element of degree minus one. Its Maurer-Cartan equation is given by
\be
F(a) = l_1(a) +   \tfrac{1}{2}  l_2( a , a ) = 0 \; . 
\ee
The Maurer-Cartan equation transforms covariantly
under the infinitesimal gauge transformations with gauge parameter $\lambda$ of degree zero:
\be
\delta_\lambda a = l_1 (\lambda)  
+ l_2( a , \lambda ) \,,  \ \ \lambda\in A_0\,. 
\ee
Indeed, one finds that
\be
\delta_\lambda F(a) = l_2( F(a), \lambda)  \, . 
\ee
As in the associative case, the finite gauge transformations are defined as the integrated infinitesimal gauge transformation and the Maurer-Cartan set is given by the solutions modulo gauge transformations. 

\medskip\noindent
\textbf{Bar construction.} This construction associates to a  differential graded Lie algebra $(A,l_1,l_2)$ the 
{\em symmetric} coalgebra $B(A)$ given as 
\be B(A)  = \bigl(\, S^c(sA), \bar B , \bar \Delta_c ) \, .
\ee
Let us discuss the various ingredients in here.
The vector space in the coalgebra $B(A)$ is $S^c(sA)$, the {\em symmetric
tensor coalgebra} of the suspended space $sA$.
In fact, for any graded vector space $V$, we define 
the symmetric tensor coalgebra
\be
S^c(V) = \bigoplus_{n \ge 1} V^{\wedge n}\, 
\ee
where $V^{\wedge n}$ is $V^{\otimes n}$, but with elements identified that differ by a graded permutation. For example, on $V^{\wedge 3}$ we have
\be
v_1 \wedge v_2 \wedge v_3 = (-1)^{v_1v_2}v_2 \wedge v_1 \wedge v_3 = (-1)^{v_1(v_2+ v_3)} v_2 \wedge v_3 \wedge v_1 = \ldots \; .
\ee 
The coalgebra $S^c(V)$ has a coproduct  $\bar\Delta_c$ defined as follows
\be
\label{coprodsymtencoalg}
\bar \Delta_c(v_1 \wedge \cdots \wedge v_n) = \sum_{i = 1}^{n-1}
\sum_{\sigma \in uSh(i,n-i)} 
\epsilon(\sigma,\{v\}) 
v_{\sigma(1)} \wedge \cdots \wedge v_{\sigma(i)} 
\otimes' v_{\sigma(i+1)} \wedge \cdots \wedge v_{\sigma(n)} \; .
\ee
The second sum runs over `unshuffles', 
all permutations satisfying $\sigma(1) < \ldots < \sigma(i)$ and $\sigma(i+1) < \ldots < \sigma(n)$-- equivalently, in the unshuffle one chooses $i$ of the $n$ inputs to go to the left tensor factor with the remaining $n-i$ going to the  right tensor factor, without reordering within either group. The factor $\epsilon(\sigma,\{v\})$ is the sign obtained from the graded permutation of the $v_i$'s  by~$\sigma$. 
As a simple example, we have 
\be  
\label{symtcoalgex} 
\bar \Delta_c  (v_1 \wedge v_2) =  v_1 \otimes' v_2  + (-1)^{v_1v_2} v_2 \otimes' v_1 \,.
\ee 
The coproduct $\bar \Delta_c$ can be verified to be coassociative, meaning that it satisfies
\be
(1 \otimes \bar \Delta_c)\circ \bar \Delta_c = (\bar \Delta_c \otimes 1) \circ \bar \Delta_c \; .
\ee
Moreover, as defined, the coproduct is cocommutative.
This means that
\be\label{eq:CoCom}
\bar \tau \circ \bar \Delta_c = \bar \Delta_c\,, 
\ee
where $\bar\tau$ is the graded exchange operator defined by
\be
\bar\tau  ( x \otimes' y ) = (-1)^{xy} \, y \otimes' x \,, \ \   
x, y \in S^c(V)\,. 
\ee
It is immediate to check that indeed, the right-hand side of~\eqref{symtcoalgex}
is invariant under the action of $\bar\tau$.  It is not hard to see that the definition~\eqref{coprodsymtencoalg} of $\bar\Delta_c$ makes it cocommutative.

It remains to explain what $\bar B$ is. The story is analogous to the one for associative algebras. The maps $l_1$ and $l_2$ induce maps $b_1$ and $b_2$ using the suspension formula~\eqref{eq:SuspensionSign},  
\be
b_1(sa) = sl_1(a) \; , \qquad b_2(sa,sb) = (-1)^{a+1} sl_2(a,b) \,.
\ee
Both $b_1$ and $b_2$ have degree minus one. These maps in turn can be lifted to coderivations $\bar B_1$ and $\bar B_2$ of $\bar \Delta_c$ using the rule
\be\label{eq:LinftyLift} 
\bar B_k(v_1\wedge \ldots \wedge v_n) = \sum_{\sigma \in uSh(k,n-k)} \epsilon(\sigma,\{v\})\  b_k(v_{\sigma(1)},\ldots,v_{\sigma(k)}) \wedge v_{\sigma(k+1)} \wedge \cdots \wedge v_{\sigma(n)} \; ,
\ee 
where the $v_i \in V := sA $. This sum also runs over unshuffles as defined above, and the 
sign factor $\epsilon (\sigma, \{ v\} )$  again arises from the graded permutation of the $v_i$'s. 
Defining now the coderivation 
$\bar B = \bar B_1 + \bar B_2$, one can verify that the properties that make $l_1$ and $l_2$ define a differential graded Lie algebra make $\bar B$ a differential on $S^c(sA)$. 

\medskip\noindent
\textbf{L$_\infty$ algebras.} These homotopy Lie algebras are best described with a bar construction, as a differential graded symmetric coalgebra whose differential (and coderivation) $\bar B$  includes higher contributions from coderivations $\bar B_k$ with $k \geq 3$.  These $\bar B_k$ coderivations arise from multilinear maps $b_k: V^{\wedge k} \to V$, with $V=sA$ 
using~\eqref{eq:LinftyLift}.

The square-zero condition $\bar B^2 = 0$ then produces an infinite set of relations among the $b_k$'s, implying by desuspending
an infinite set of
relations for the $l_k$, now including $k\geq 3$. 
In particular, if $[b_1,b_3]$ is nonzero, then $l_2$ satisfies the Jacobi identity only up to homotopy. Morphisms of $L_\infty$ algebras are the morphisms of the respective differential graded symmetric coalgebras. 
Following the associative case, we write
\be
\label{binftyLie}
B_\infty(A) = (S^c(s A),\bar B ,\bar \Delta_c)\,, 
\ee
for the bar construction of an $L_\infty$ algebra $(A,\{l_k\}_{k \ge 1})$. 

The $L_\infty$ algebras also have a Maurer-Cartan equation. 
For an element $v \in sA = V$ of degree zero we have
\be
\sum_{n \ge 1} \frac{1}{n!} b_n(v,\ldots,v) = 0 \; .
\ee
On the other hand, the infinitesimal gauge transformations with parameter $\Lambda$ read
\be
\delta_\Lambda v = \sum_{n \ge 1} \frac{1}{(n-1)!} b_n(v,\ldots,v,\Lambda)\,, \ \ \Lambda \in V_1  \; .
\ee

In order to go into the cobar construction it is useful to
introduce a Lie algebra that has properties that are similar
to those of the tensor algebra $T(V)$.  

\medskip\noindent
\textbf{The Lie algebra $\mathbf{(L(V),[-,-])}$.} 
Given a graded vector space $V$, we have the tensor algebra $T(V)$ with product $\mu$ given in \eqref{eq:DefTensAlgProduct}. Since $\mu$ is an associative and in general noncommutative product, it has an associated Lie bracket given by the $\mu$-commutator
\be
[x,y] := \mu(x,y) - (-1)^{x y} \mu(y,x) = x\otimes' y - (-1)^{x y} y \otimes' x \, .
\ee
In this way, $(T(V),[-,-])$ is a graded Lie algebra.

As an important aside, note that $\bar\tau$ acting on the Lie bracket changes its sign:
\be
\label{tauonbracket}
\bar \tau [ x, y] = - [x, y] \,.
\ee
Moreover, the converse holds: any $R= \sum_i C_i \otimes' D_i$ that is odd under exchange is the sum of brackets.  Indeed, $\bar \tau R = - R$ means that 
\be
 \sum_i  (-1)^{C_i D_i} D_i \otimes' C_i  = - \sum_i C_i \otimes' D_i\,.
\ee
Then, as claimed, we see that
\be
R = \tfrac{1}{2} \sum_i  \Bigl( C_i \otimes' D_i - (-1)^{C_i D_i} D_i \otimes' C_i \Bigr)  = \tfrac{1}{2}  \sum_{i} [ C_i , D_i] \,. 
\ee

Returning to the Lie algebra, the next step is to identify $L(V)$ as a Lie subalgebra of $(T(V),[-,-])$. We can define 
\be L(V) = \widehat \bigoplus_{n \ge 1} L^n(V)\,,
\ee
where $L^n(V)\subset V^{\otimes n}$ is defined inductively as follows. First,
\be
L^1(V):=V.
\ee
Assuming that $L^k(V)$ has been defined for all $k<n$, we define $L^n(V)$ to be the linear span of all brackets
\be
[x,y],
\qquad
x\in L^{k_1}(V),\quad y\in L^{k_2}(V),\quad k_1+k_2=n.
\ee
Equivalently, $L^n(V)$ is spanned by all iterated commutators of length $n$ in elements of $V$. For example, $L^2(V)$ is spanned by elements of the form
\be
[v_1,v_2],
\qquad v_1,v_2\in V,
\ee
while $L^3(V)$ is spanned by elements of the form
\be
[[v_1,v_2],v_3],
\qquad
[v_1,[v_2,v_3]],
\qquad v_1,v_2,v_3\in V.
\ee
Thus $L(V)$ is closed under the commutator bracket, and $(L(V),[-,-])$ is a graded Lie subalgebra of $(T(V),[-,-])$.

The graded Lie algebra $(L(V),[-,-])$ has properties very similar to those of the tensor algebra $T(V)$. 
In particular, any graded derivation $D_X: L(V) \rightarrow L(V)$ of $[-,-]$ is uniquely determined by a linear map $X: V \rightarrow L(V)$, such that
$D_X(v) = X(v)$ for $v \in V$.
We then use the 
derivation property to define the action on commutators recursively:
\be
D_X [u, w] :=  [D_X u, w]  + (-1)^{X u} [u, D_X w]\,,
\ee
for homogeneous elements $u, w \in L(V)$. This rule is applied first
for $[v_1, v_2]$ and then can be applied for $[[v_1, v_2], v_3]$, and so on and so forth.

\medskip
\noindent 
\textbf{Cobar construction.}  Given any cocommutative differential graded coalgebra $(C,\bar B,\bar \Delta_c)$ there is the cobar construction, producing a differential graded Lie algebra. 
\be\label{eq:LieCobar}
\Omega( C)  =  \Bigl(\,  L(s^{-1} C) ,\,  B + \Delta_c, \, [- \,, - ] \Bigr)\,. 
\ee
The underlying graded vector space is $L(\ds C)$, the bracket $[-,-]$ is the Lie bracket of $L(\ds C)$. Moreover, the differential $\bar B: C \rightarrow C$ defines a degree minus one differential $B: s^{-1} C \to s^{-1} C$:
\be
B(\ds x) = \ds \bar B(x) 
\,,  \ \ x \in C \,,  
\ee 
following \eqref{cleardepfus}. Since we have a linear map $B: s^{-1} C\to s^{-1} C \subset L(s^{-1} C)$, we can extend $B$
to a derivation in $L(\ds C)$ as explained above. 
For simplicity, the derivation is still called $B$.

Given  the degree zero map $\bar \Delta_c : C \to C \otimes' C$, we define the desuspended version $\Delta_c$ just like in \eqref{desuspbardelta}
\be
\label{desltdesuseo3ir}
\Delta_c \circ s^{-1} = - (s^{-1} \otimes' s^{-1}) \circ \bar \Delta_c\,.
\ee
$\Delta_c$ is of degree minus one. 
In the desuspended complex we have a $\tau$ that acts as follows
\be
\tau (s^{-1} x \otimes' s^{-1} y) = (-1)^{(x +1)(y+1)}  
s^{-1} y \otimes' s^{-1} x\,.
\ee
Writing, on the right-hand side $s^{-1} y \otimes' s^{-1} x= (-1)^{y+ xy}  (s^{-1} \otimes' s^{-1})\circ \bar \tau (x\otimes' y)$
one quickly finds that
\be\label{taudesusljr93}
\tau \circ (s^{-1} \otimes' s^{-1}) =  -  (s^{-1} \otimes' s^{-1})\circ \bar \tau\,.
\ee
We can now use these identities to see what happens to cocommutativity after suspension. Using~\eqref{desltdesuseo3ir} and~\eqref{taudesusljr93} 
and $\bar \tau \circ \bar \Delta_c = \bar\Delta_c$ we have
\be\begin{split} 
\tau \circ \Delta_c \circ s^{-1} = & \  - \tau \circ (s^{-1} \otimes' s^{-1}) \circ \bar \Delta_c\ = (s^{-1} \otimes' s^{-1}) \circ \bar\tau \circ \bar \Delta_c \\[0.5ex]
= & \ (s^{-1} \otimes' s^{-1}) \circ \bar \Delta_c = \  - \Delta_c \circ s^{-1} \,,
\end{split}
\ee
from which we conclude that 
\be  
\tau\circ \Delta_c \ = \ - \Delta_c\,.
\ee
Recalling the discussion following~\eqref{tauonbracket} we see that
$\Delta_c$ maps $s^{-1}C$ into the linear space of 
Lie {\em brackets} of two elements:
\be
\Delta_c: \ds C \rightarrow L^2(\ds C) \, .
\ee
Since $L^2 (\ds C) \subseteq L(\ds C)$, $\Delta_c$ also defines a map
\be
\Delta_c: \ds C \rightarrow L(\ds C) \, .
\ee
As such, it can be lifted to a derivation $\Delta_c: L(\ds C) \rightarrow L(\ds C)$ of $[-,-]$. Note also that as shown in Section~\ref{sec:CobarConstruction}, desuspension makes $\Delta_c$ a differential, and  $[B, \Delta_c]=0$  follows from $\bar B$
being a coderivation of $\bar\Delta_c$.  The sum $B + \Delta_c$ of anticommuting differentials (and derivations) $B$ and $\Delta_c$ then defines a differential and derivation on $L(\ds C)$:
\be\label{eq:CobarLiesqzero}
(B + \Delta_c)^2 = 0 \, .
\ee
With this we have completed the verification that the cobar construction defining 
$\Omega(C)$ in~\eqref{eq:LieCobar} is indeed 
a differential graded Lie algebra.

\medskip\noindent
\textbf{Bar-cobar construction.} The bar construction associates to a differential graded Lie algebra $(A,l_1,l_2)$ a differential graded cocommutative coalgebra $B(A)$. This construction extends to the $L_\infty$ case, associating to an $L_\infty$ algebra $(A,\{l_k\}_{k \ge 1})$ the coalgebra $B_\infty(A)$ in~\eqref{binftyLie}. The cobar construction associates to any differential graded cocommutative coalgebra $(C,\bar B,\bar \Delta_c)$ the differential graded Lie algebra $\Omega(C)$. Combining these constructions allows us to associate to any $L_\infty$ algebra an equivalent differential graded Lie algebra that can be viewed as a 
strict model for the original $L_\infty$ algebra.

As in the $A_\infty$ algebra case, the differential graded Lie algebra $\Omega (B_\infty(A))$ is {\em quasi-isomorphic} to the $L_\infty$ algebra $(A,\{l_k\}_{k \ge 1})$ via a morphism $I$ of the associated coalgebras:
\be
I : B_\infty(A) \rightarrow B(\Omega (B_\infty(A))) \, .
\ee
This is the $L_\infty$ analog of \eqref{eq:UniversalInclusion}. 
Under some assumptions on the structure of the $L_\infty$ algebra $(A,\{l_k\}_{k \ge 1})$, it can be proven that such a quasi-isomorphism defines an equivalence of the Maurer-Cartan sets of $L_\infty$ algebras (see, for example, \S 3 of \cite{DOLGUSHEV2015260}, or theorem 4.8 and proposition 4.9 of \cite{GetzlerNilpotent} for proofs of this fact under certain conditions.) 

\section{The homology of \texorpdfstring{$\Delta$}{Delta} and its deformation \texorpdfstring{$\Delta_\Phi$}{Delta-Phi}}\label{thehomofDanddef}

 In this section, starting from a coalgebra $(C, \bar\Delta)$, 
 where $C= T^c(V)$ and $V= sA$, 
 we work in the cobar complex $\Omega (C) = (T(s^{-1} C), \Delta, \mu)$, 
with $T(s^{-1} C)$ the tensor algebra of the desuspended version of $C$. This is the differential graded algebra \eqref{eq:DefBarCobarComplex} of the bar-cobar construction, but with $B = 0$. 
As we have seen, the desuspended coproduct $\Delta$ squares to zero and
can be defined to be a derivation of the tensor algebra. 
Since $\Delta$ is a differential, we can speak of the  homology $H(\Delta)$.  Understanding
this homology will be key to finding the full solution set of the
Maurer-Cartan equation $\Delta \Phi + \Phi \otimes' \Phi = 0$.

We find that the homology $H(\Delta)$ is isomorphic to the space 
$\Omega (C)_{1,1}$ 
consisting of elements of factor number one and length one -- factor number and length as defined at the beginning of Section~\ref{sec:BarCobarAinfty}. 
To show this, we will introduce a degree one  
contracting homotopy $h$ such that
\be 
[\Delta , h] = 1 - \boldsymbol{\pi}_{1,1} \,   \ \ \  \epsilon(h) = +1 ,  
\ee
where $\boldsymbol{\pi}_{1,1}$ is the projector to the space $\Omega (C)_{1,1}$ and $[-,-]$ denotes the graded commutator as in~\eqref{eq:ContractingHomotopy}.  
This homotopy 
will allow us to find the complete set of factor-number-one solutions
of the Maurer-Cartan equation $\Delta \Phi_1 + \Phi_1 \otimes' \Phi_1=0$ in Section~\ref{solvtheeqnmc}.

We then turn to the calculation of the $\Delta_\Phi :=  \Delta + \operatorname{ad}_\Phi$ homology $H(\Delta_\Phi)$ in $\Omega (C)$. 
Here, `$\operatorname{ad}$' represents adjoint action, as will be defined below. 
The operator $\Delta_\Phi$ is nilpotent when $\Phi$ satisfies the
Maurer-Cartan equation $\Delta \Phi + \Phi \otimes' \Phi=0$.  We prove that
there is a chain isomorphism relating $\Delta_\Phi$ and $\Delta$ and thus
the homologies are isomorphic, with
\be
{1\over 1 + h\operatorname{ad}_\Phi}:\   H(\Delta) \to H(\Delta_\Phi)\,.
\ee
When $\Phi = \Phi_1$ is a state of factor number one, 
the homology $H(\Delta_\Phi)$ vanishes at all factor numbers greater or equal to two.
The homology is concentrated at factor number one and we
determine a useful explicit form for  the representatives.
The understanding of $\Delta_{\Phi_1}$ developed here will be needed
to find the general solution of the Maurer-Cartan equation 
$\Delta \Phi + \Phi \otimes' \Phi= 0$ in Section~\ref{solvtheeqnmc}
and to solve the Maurer-Cartan equation
$(\Delta + B) \Psi + \Psi\otimes'\Psi = 0$
in Section~\ref{solveqbplusdelta}.

\subsection{\texorpdfstring{$\Delta$}{Delta} homology on \texorpdfstring{$\Omega(C)$}{Omega-of-C}}

\medskip
\noindent
\textbf{Defining a contracting homotopy.}  
We now define a degree-one operator $\tilde h$ that essentially converts a $\otimes'$, one at a time, 
into a $\otimes$. 
Therefore, it reduces the factor number of a vector by one unit while preserving its length.  We will eventually construct the homotopy $h$ from $\tilde h$. We declare that
 \be
 \tilde h:   \Omega (C)_{f,\ell}  \rightarrow  \Omega (C)_{f-1,\ell} \,.
 \ee
 We can also say, focusing on factor number only, that  
 \be
 \tilde h:  (s^{-1} C)^{\otimes' f}  \   \to \   (s^{-1} C)^{\otimes' ( f-1) }  \,.  
 \ee
 Since there are no vectors 
 with factor number 
 less than one, we define for any $\ell \geq 1$ 
 \be
 \label{tildehkillsfone}
 \tilde h:   \Omega (C)_{1,\ell}  \rightarrow  0\,, \ \ \hbox{so that} \ \ 
 \tilde h (s^{-1} A) = 0,  \ \  \forall A \in C\,. 
 \ee
We define $\tilde h:  s^{-1} C\otimes's^{-1} C \rightarrow s^{-1} C$ as
\begin{equation}
\label{htilontwof}
	\tilde h( s^{-1}A_1\otimes's^{-1} A_2) :=  (-1)^{A_1+1} s^{-1}(A_1 A_2)\,,
    \ \ \ A_1, A_2 \in C = T^c (V) \,,  
\end{equation}
and extend the action of $\tilde h$ to the algebra $T(s^{-1} C)$ by defining its 
action on vectors 
of factor number $f>2$: 
\begin{equation}
\label{hhiszero}
\begin{split}
	\tilde h (s^{-1}A_1\otimes'\cdots\otimes's^{-1} A_f )  &   = \  \sum_{i=1}^{f-1} 
	(-1)^{\sum_{k=1}^{i-1} (1+ A_k)} 
	 \, s^{-1} A_1\otimes'\cdots\otimes' \tilde h(s^{-1}A_i\otimes's^{-1}A_{i+1})\otimes'\cdots\otimes's^{-1} A_f \\
	 &= \ \sum_{i=1}^{f-1} 
	 (-1)^{\sum_{k=1}^{i} (1+ A_k)} 
	 \, s^{-1} A_1\otimes'\cdots\otimes' s^{-1}(A_i A_{i+1})\otimes'\cdots\otimes' s^{-1} A_f \, .\\
\end{split}
\end{equation}
We prove in Appendix~\ref{appA} that the operator $\tilde h$ satisfies $\tilde{h}^2 = 0$.

It is convenient to introduce the following notations.
Consider $A\in C$ of the form
\be A =  v_1\cdots v_n\in C. 
\ee
We define for $1 < i < n$, 
\be
\label{handynot}  \  \check A_i :=  v_1 \cdots v_i\,, \ \ \   \check A^i :=  v_{i+1}  
\cdots v_n, 
\ee
so that 
\be A =  \check A_i\otimes \check A^i \equiv \check A_i \check A^i .
\ee
We also use the vertical bar
$|$ as a shorthand 
for $\otimes'$ as
\be
s^{-1}A_1|\cdots|s^{-1} A_n \equiv s^{-1}A_1\otimes'\cdots\otimes's^{-1} A_n\, .
\ee
 We can then easily write the action of 
 $\Delta$ following~\eqref{keyformulaforDelta}
\begin{equation}
\label{deltaonCs} 
\Delta 
(s^{-1}A)\ = \  -\sum_ {i=1}^{n-1} (-1)^{\check A_i } s^{-1}
\check A_i\otimes's^{-1}\check A^i \equiv \  -\sum_ {i=1}^{n-1} (-1)^{\check A_i } s^{-1}
\check A_i|s^{-1}\check A^i\,. 
\end{equation}

We now claim that the following general relation holds for the action of the graded 
commutator $[ \Delta , \tilde h ]$ on elements of $T(s^{-1} C)$
\begin{equation}\label{4tidn}
    [\Delta, \tilde h] \, (s^{-1}A_1|\cdots|s^{-1} A_f) \, = \,  (\abs{A}-1)\, \bigl(s^{-1}A_1|\cdots| s^{-1} A_f \bigr) \,, \,  \ \ \   |A| \equiv
		\sum_{i=1}^f  \abs{A_i}\, , 
\end{equation}
where $|A_i|$ denotes the length of $A_i$ and $|A|$ denotes the length of $A$.
The proof is given in Appendix~\ref{appA}. To give some insight we consider here the case when the graded commutator 
acts on vectors of factor number one. 

\medskip
\noindent
For that case we consider $s^{-1} A$,  with 
$A =  \check A_i  \check A^i$ and 
$A= v_1  \cdots  v_m$,
such that  $|A|= m$. Recalling that $\tilde h (s^{-1} A)=0$, 
we have 
\begin{equation}
	[\Delta, \tilde h] \, s^{-1}A =  \, \tilde h \, \Delta s^{-1} A =  -\sum_{i=1}^{|A|-1}(-1)^{\check A_i}\tilde h ( s^{-1}\check A_i|s^{-1}\check  A^i) = \sum_{i=1}^{|A|-1}
	 s^{-1}(\check A_i \check A^i) \,.
\end{equation}
Since $\check A_i \check A^i = A$ for all $i$ from one to $|A|-1$, 
the sum on the right-hand side gives $ (|A|-1) s^{-1} A$ so that 
\begin{equation}
\label{simplestcase} 
    [ \Delta, \tilde h] \, s^{-1}A =   (|A|-1) s^{-1}A \,, \,\, \ \ \  \  |A| \geq 1\,, 
	 \end{equation}
	 confirming the validity of~\eqref{4tidn} for factor number one.
     Note that the equation was derived for $|A|> 1$ but it actually holds
	 for $|A|=1$, since in that case the left-hand side is manifestly
	 zero: both $\Delta$ and $\tilde h$ 
     vanish on a length one element.  
Clearly, vectors of length one belong to the homology $H(\Delta)$: they 
are annihilated by $\Delta$ but are not in the image of $\Delta$ because that image
contains vectors of length greater than one.

\medskip
\noindent
We now show that the homology of $\Delta$ is in fact given by those elements of length one:  
\be
\label{cohdelbequal0}
H (\Delta) \cong  \Omega (C)_{1,1}\,. 
\ee
To see this, we can work in vector spaces spanned by elements with a 
 fixed number of factors~$f$ and fixed length greater than one.
 Let $U$ be a fixed-factor-number vector of fixed length $\abs{U}>1$
 such that $\Delta U = 0$. Equation~\eqref{4tidn} then shows that this
 vector is $\Delta$-exact 
 \be
U =  {1\over |U|-1} \Delta \, (\tilde h\, U ) \,,
\ee
confirming that $\Delta$ homology vanishes for length greater than one,  and establishing
the claim in~\eqref{cohdelbequal0}. 

\bigskip
We can recast ~\eqref{4tidn}  using $W_{f,\ell}$ to denote
a general vector 
with factor number $f$ and length $\ell$; any such term, of course,
having $\ell \geq f$.  Equation~\eqref{4tidn} implies that  
\be
\label{keyexpli}
 \ [\Delta, \tilde  h] \   W_{f,\ell}   =  (\ell-1)  W_{f,\ell} \,, 
\ee
which holds for all $W_{f,\ell}$. 
If $\ell =1$ this implies
$f=1$ and both $\tilde h$ and $\Delta$ vanish on
$W_{1,1}$, that is, both the left-hand side and the right-hand side correctly vanish. 

We can use the projectors  $\boldsymbol{\pi}_{f,\ell}$ to the
subspaces of $\Omega (C)$ 
with factor number $f$ and length $\ell$
to define the operator
$h$ that will provide a contracting homotopy for $\Delta$
\be
\label{Hdefinition} 
\ h  \ \equiv  \ 
 \sum_{f=2}^\infty \sum_{\ell=f}^\infty {1\over \ell-1} \,  \tilde h\,  \boldsymbol{\pi}_{f,\ell}\,. 
\ee
The definition of $h$ above means that 
\be
\label{Hdef}
\begin{split}
h \,  W_{1,\ell}  = & \ 0 \,, \ \  \ell\geq 1\,,\\
h \,  W_{f,\ell}  = & \  {1\over \ell-1} \, \tilde  h\,  W_{f,\ell} \,, \ \    f\geq 2\,, \ \ell\geq f\,.   \\
\end{split}
\ee
The first equation holds because all projectors in the definition of $h$ kill $W_{1,\ell}$. 
This was to be anticipated, since $h$ is a rescaled version of $\tilde h$ and
we noted that the factor-number-one space $\Omega (C)_1$ 
is in the kernel of $\tilde h$ (see~\eqref{tildehkillsfone}).
Note also that the relations in~\eqref{Hdef} imply
that $h^2 =0$ on account of $\tilde h^2= 0$.

We can now write the commutator $[ h , \Delta]$ 
on all subspaces of $\Omega (C)$. 
It is the identity operator except on
$(1,1)$ spaces, where it is zero.  
Thus we have
\be
\label{keycommutator}
[ \Delta \,, \, h ] =  1 -  \boldsymbol{\pi}_{1,1}  \,. 
\ee
Thus, $\boldsymbol{\pi}_{1,1}$ is a contraction as introduced in \eqref{eq:ContractingHomotopy}. This implies that $H(\Delta)$ is contained in 
$\Omega(C)_{1,1}$. 
Since also $\Delta \boldsymbol{\pi}_{1,1} = \boldsymbol{\pi}_{1,1} \Delta = 0$, we actually have \eqref{cohdelbequal0}. 

\subsection{\texorpdfstring{$\Delta_\Phi$}{Delta-Phi} homology on \texorpdfstring{$\Omega(C)$}{Omega-of-C}}  \label{delphohomonomc}

We now focus on the properties of the operator
\be
\Delta_\Phi \equiv  \Delta + \operatorname{ad}_\Phi \,,
\ee
where $\Phi$ is a degree minus one element 
that satisfies the Maurer-Cartan equation
	\be
    \label{MCforfi}
	\Delta \Phi + \Phi \otimes' \Phi = 0\,.
	\ee
The adjoint action `ad' is defined by the graded commutator 
\be
\operatorname{ad}_U V \equiv U \otimes' V - (-1)^{UV}  V \otimes' U =  [U, V] \,.
\ee
From \eqref{eq:flat} we know that on account of the Maurer-Cartan equation~\eqref{MCforfi}
\be
\Delta_\Phi\Delta_\Phi= 0\, .
\ee 
We also have
\be
\Delta_\Phi \operatorname{ad}_\Phi= - \operatorname{ad}_\Phi\Delta \, . 
\ee

\medskip\noindent
\textbf{$\Delta_\Phi$ homology.}
We compute this homology by exhibiting a chain isomorphism relating $\Delta_\Phi$ to $\Delta$.
To see this, first consider the commutator
\be\label{commuatator}
\begin{split} 
[  \Delta  ,  h \, \operatorname{ad}_\Phi] = & \   \Delta\,  h\,  \operatorname{ad}_\Phi- h \, \operatorname{ad}_\Phi\Delta \\
= & \  [ \Delta,  h ]  \operatorname{ad}_\Phi-  h \, (\Delta \,  \operatorname{ad}_\Phi+  \operatorname{ad}_\Phi\Delta)  \\
= & \    \operatorname{ad}_\Phi- h\operatorname{ad}_{\Delta \Phi}  
= \    \operatorname{ad}_\Phi+ h\operatorname{ad}_{\Phi \Phi} \\ 
= & \  \operatorname{ad}_\Phi+  h \operatorname{ad}_\Phi\operatorname{ad}_\Phi = (1 + h\operatorname{ad}_\Phi) \operatorname{ad}_\Phi \,.
\end{split}
\ee
Here we used $[ \Delta , h] = 1 - \boldsymbol{\pi}_{1,1} $, the Maurer-Cartan equation for $\Phi$, 
and  $\boldsymbol{\pi}_{1,1} \operatorname{ad}_U = 0$, valid for arbitrary $U$. 
We then claim that the chain isomorphism relating the differentials is
\be
\label{chainmxx} 
(1 + h\operatorname{ad}_\Phi)\Delta_\Phi =   \,  \Delta (1 + h\operatorname{ad}_\Phi)\,, 
\ee
which follows directly from~\eqref{commuatator}.
It follows from~\eqref{chainmxx} that we have a map of homology classes
\be
1 + h\operatorname{ad}_\Phi:\  H(\Delta_\Phi) \to  H(\Delta)\,,
\ee
and its inverse
\be
{1\over 1 + h\operatorname{ad}_\Phi}:\   H(\Delta) \to H(\Delta_\Phi)\,.
\ee
For arbitrary length-one vectors $s^{-1} w\in 
H(\Delta)$, with $w\in V$,
\be
\label{rephdelphi}
 {1\over 1 + h\, \operatorname{ad}_\Phi} s^{-1} w
 \in  H(\Delta_\Phi) \,.
\ee
Note that if $\Phi$ is a vector of factor number one, the homology class of $\Delta_\Phi$ can be
represented by a factor-number-one vector with components of various lengths.

Of particular interest are $H(\Delta_\Phi)$ representatives when $\Phi= \Phi_1$ is of factor number one.
Such a $\Phi_1$ takes the form 
\be
\label{phi1solguessed} \Phi_1 = s^{-1}E(v) \,, 
\ \ \  E (v)  =  {v\over 1 - v} = v + vv + vvv  + \cdots  \,.  
\ee
The vector $E(v)$ was defined in~\eqref{eq:DefGroupLikeElement}, $v\in V$ is of degree zero, and the products here are with~$\otimes$. This $\Phi_1$ satisfies the Maurer-Cartan equation 
\be
     \Delta \Phi_1 + \Phi_1 \otimes'  \Phi_1 = 0 
    \ee
as we saw in \eqref{eq:MCwoB}, where $\Phi_1 \equiv a$. 
Using~\eqref{rephdelphi}, a representative $W_1\in H(\Delta_{\Phi_1})$ 
of factor number one parameterized by an arbitrary vector $w\in V$ takes the explicit form
\be
\label{detx1coh} 
\begin{split}
W_1=  {1\over 1 + h\, \operatorname{ad}_{\Phi_1}} s^{-1} w
=  & \   s^{-1} \bigl[  (1 + E(v)) \, w\, 
(1+ E(v)) \bigr] 
=\ s^{-1}  \Bigl[  {1\over 1 - v } \, w\, 
{1\over 1-v } \Bigr] 
\  \,.
\end{split}\ee
Here, $W_1$ is a vector of factor number one, built as a sum of vectors of lengths going from one to infinity. 
The expressions
in terms of $v$ were found by expanding 
$(1 + h \, \operatorname{ad}_{\Phi_1})^{-1}$ in terms of 
powers of $h\,\operatorname{ad}_{\Phi_1}$, using the formula for $h$ in~\eqref{Hdefinition} and 
the expression for $\Phi_1$ in~\eqref{phi1solguessed}.
While the algebra involved in proving the first
equality  in~\eqref{detx1coh} is nontrivial, it is easier to verify that 
$\Delta_{\Phi_1} W_1 = 0$.
Indeed, one can check quickly that  
\be\Delta_{\Phi_1} s^{-1}   = -s^{-1} \otimes' s^{-1}  \bar \Delta_{E(v)} \,, \ \ \  \hbox{with} \ \  \ 
\bar \Delta_{E(v)} =  \bar \Delta  +\overline{\operatorname{ad}}_{E(v)} \,,
\ee 
with $\overline{\operatorname{ad}}_U V := (-1)^{U+1} (U\otimes' V + (-1)^{UV} V\otimes' U)$. 
It then follows that 
\be
\label{intermo3odfkj}
\Delta_{\Phi_1} W_1 = -s^{-1} \otimes' s^{-1}  \bar \Delta_{E(v)} \bigl[  (1 + E(v)) \, w \, 
(1+ E(v)) \bigr]\,.   \ee
Moreover, another short computation shows that for arbitrary vector ${\cal O}\in C$
\be
\bar \Delta_{E(v)} \bigl[  (1 + E(v)) {\cal O} (1+ E(v)) \bigr] =  \bigl[  (1 + E(v)) (\bar \Delta {\cal O}) (1+E(v)) \bigr]\,.
\ee
Since $\bar \Delta \, w = 0$, 
this confirms that $\bar\Delta_{E(v)}[  (1 - v)^{-1} \, w\, 
(1-v)^{-1}] = 0$, 
and back in~\eqref{intermo3odfkj} we conclude that $\Delta_{\Phi_1} W_1 = 0$, as we wanted to confirm.

In summary, the homology $H(\Delta_{\Phi_1})$ on $\Omega (C) = T(s^{-1} C)$, with
$C = T^c (V)$ is given by the following representatives all at factor number one, 
and parameterized by arbitrary vectors $w \in V$:
\be
H(\Delta_{\Phi_1} ) \cong \Bigl\{  s^{-1}  \Bigl[  {1\over 1 - v } \, w\, 
{1\over 1-v } \Bigr] \,, \   w \in V  \Bigr\} \,, \ \ \hbox{with} \ \ 
\Phi_1 = s^{-1} E(v), \  v \in V \,.  
\ee
The homology $H(\Delta_{\Phi_1})$ vanishes for any factor number greater than or equal to two. By this we mean that if $Z$ is such that $\boldsymbol{\pi}_1(Z) = 0$ with $\boldsymbol{\pi}_1: \Omega(C) \rightarrow \Omega(C)_1$ the projection to the 
factor-number-one subspace, we have
\be
\Delta_{\Phi_1} (Z) = 0 \ \Rightarrow \ Z = \Delta_{\Phi_1}(Z') \, .
\ee

\bigskip
\noindent\textbf{A contracting homotopy for $\Delta_\Phi$.}
Recalling that $[ \Delta , h] = 1 - \boldsymbol{\pi}_{1,1}$ we expect an analogous 
result for the operator $\Delta_\Phi$.  We claim that if $\Phi$ satisfies the Maurer-Cartan equation, there is
an operator $h_\Phi$ satisfying 
\be
\label{conjconhom} 
[ \Delta_\Phi\,, \,  h_\Phi ] =  \ 1 -  {1\over 1+ h \, \operatorname{ad}_\Phi} \, 
  \boldsymbol{\pi}_{1,1}\,, \ \ \hbox{with} \ \ \   h_\Phi\equiv   {1\over 1 + h\, \operatorname{ad}_\Phi}  \, h  \, = \,  h  \, {1\over 1 + \operatorname{ad}_\Phi\, h}  \,. 
\ee
The last equality follows from 
\be
h(1 + \operatorname{ad}_\Phi \, h) = h + h \,\operatorname{ad}_\Phi \, h = (1 + h\,\operatorname{ad}_\Phi)h \, ,
\ee
and implies that
\be
\label{w823} 
h =  ( 1 + h\, \operatorname{ad}_\Phi) h_\Phi  = h_\Phi (1 + \operatorname{ad}_\Phi \, h ) \, . 
\ee 
We have seen that $h^2 = 0$, and therefore, using the two alternative forms
of $h_\Phi$ and forming the product of the first version to the left of the second version, we 
see that 
\be
h_\Phi^2 = 0 \,.
\ee
The proof of~\eqref{conjconhom} is given in Appendix~\ref{appB}.

\medskip
\noindent
\textbf{Solving the key equation.}
Consider now the equation 
\be
\label{modeleqn}
 \ \Delta_\Phi \,  W = Z\,, 
\ee
whose consistency requires
\be
\label{consistcond}
\Delta_\Phi \, Z = 0 \,, \ \ \hbox{and} \ \ \   \boldsymbol{\pi}_1 Z = 0\,, 
\ee
with
$\boldsymbol{\pi}_1$ the projection to $\Omega (C)_1$.  The first consistency condition
is required because $\Delta_\Phi$ is nilpotent and
the second because the 
image of $\Delta_\Phi$ lies at factor number two or higher. 

Consider the gauge-fixing condition
\be
h_\Phi W = 0\,.
\ee
Then, acting with $h_\Phi$ on equation~\eqref{modeleqn} we have
\be
h_\Phi \Delta_\Phi  \, W =  [ h_\Phi, \Delta_\Phi] W  =  h_\Phi \, Z  
\ee
Using the contracting homotopy this gives
\be
\Bigl(  1 \   -  {1\over 1+ h \, \operatorname{ad}_\Phi} \, \boldsymbol{\pi}_{1,1} \, \Bigr) W =  h_\Phi \, Z \,. 
\ee
This gives a solution
\be
\label{sold1}
W  =   {1\over 1+ h \,\operatorname{ad}_\Phi} \,  W_{1,1}  +   h_\Phi \, Z   \,. 
\ee
We can see that $h_\Phi W =0$ as expected. For this, the first term requires 
a little bit of thought, but follows from $h^2= 0$ and $h \, W_{1,1}=0$:
\be
h_\Phi  {1\over 1+ h \,\operatorname{ad}_\Phi} \,  W_{1,1} =  {1\over 1+ h \,\operatorname{ad}_\Phi} \,  h \, {1\over 1+ h \,\operatorname{ad}_\Phi} \,  W_{1,1} =  {1\over 1+ h \,\operatorname{ad}_\Phi} \,  h  \,  W_{1,1} = 0 \,. 
\ee
We can add an arbitrary solution by including an arbitrary $\Delta_\Phi$ exact term 
\be
\label{sold1x} 
W  =   {1\over 1+ h \,\operatorname{ad}_\Phi} \,  W_{1,1}  +   h_\Phi \, Z    + \Delta_\Phi S\,. 
\ee
The original gauge condition $h_\Phi W=0$  will no longer be satisfied, of course. We check that this is indeed a solution by acting with $\Delta_\Phi$.  The first term is annihilated, being a homology representative, and so is the last term.  We are left with  
\be
\label{sold1xx}
\Delta_\Phi \, W  =     \Delta_\Phi h_\Phi \, Z  = [ \Delta_\Phi\,,  h_\Phi ] \, Z    = Z \,, 
\ee
since $\Delta_\Phi$ kills $Z$
and   $\boldsymbol{\pi}_{1,1} Z = 0$. This shows that equation~\eqref{modeleqn} is satisfied.

We note that the right-hand side of \eqref{sold1x} has the form of a Hodge decomposition. Using the properties of $h_\Phi$, one can show that the complex $\Omega(C)$ splits into
\be
\Omega(C) = H(\Delta_\Phi) \oplus \text{Im} \, h_\Phi  \oplus \text{Im} \, \Delta_\Phi \, .
\ee
Any $W \in \Omega(C)$ can therefore be written in the form given in \eqref{sold1x}.

\section{Solving the equation \texorpdfstring{$\Delta \Phi+ \Phi\otimes'\Phi=0$}{MC-equation}}\label{solvtheeqnmc}
In this section, we solve the Maurer-Cartan equation on $\Omega(C)$, which as a vector space is $T(s^{-1} C)$, with $C = T^c (V)$.
As we decompose the field $\Phi$ into a sum of factor numbers, 
we show that the factor-number-one field $\Phi_1$ is determined uniquely in terms of the length one element $\Phi_{1,1}$, in turn parameterized by a vector in $V$. Setting all higher-factor components of $\Phi$ to zero gives what we call the {\em canonical}
solution of the Maurer-Cartan equation. 
 We use
finite gauge transformations and the homology $H(\Delta_{\Phi_1})$
to demonstrate that the general solution of the Maurer-Cartan equation is a finite gauge transformation of the 
canonical solution.  That general solution has components in all factor numbers. 

\subsection{Exploring the Maurer-Cartan equation}

We aim to understand the Maurer-Cartan equation
\be
{\cal E} := \Delta \Phi + \Phi \otimes' \Phi  = 0 \,.
\ee
Here $\Delta$ and $\Phi$ are both of degree minus one,
while $\otimes'$ carries no degree. 
From \eqref{eq:InfinitesimalGT}
we know that this equation is supplemented with a gauge transformation
\be
\label{smgadlje}
\delta_\Lambda \Phi = \Delta \Lambda + \Phi\otimes' \Lambda - \Lambda \otimes' \Phi \, ,
\ee
where the gauge parameter $\Lambda$ is of degree zero.  Under this transformation 
the Maurer-Cartan equation of motion 
transforms covariantly:
\be
\delta_\Lambda {\cal E} =   {\cal E} \otimes' \Lambda  - \Lambda \otimes' {\cal E} \, . 
\ee
One quickly checks by acting with $\Delta$ on this equation of motion that it consistently results in $\Delta ( \Phi \otimes' \Phi ) = 0$.

Recall from Section \ref{sec:BarCobarAinfty} the decomposition of vectors into factor number and length. We write
\be
\Phi= \sum_{f=1}^\infty    \Phi_{f}\,, \  \ \
\Phi_f= \sum_{\ell = f}^\infty  \Phi_{f, \ell} \,, \ \ 
\ \Phi= \sum_{f=1}^\infty  \sum_{\ell= f}^\infty  \Phi_{f,\ell}  \,,
\ee
where $f$ is the factor number and $\ell$ is the length.
Thus we have
\be
s^{-1} A_1  \otimes' \cdots  \otimes'  s^{-1} A_f   \in \Omega (C)_{f,\ell} \,, \ \ 
\hbox{with} \ \ \ell = \sum_{i=1}^f |A_i| \,. 
\ee
 Note that in $\Phi_{f,\ell}$ we must have $\ell \geq f$ since each factor must
 contain at least one vector.
 Moreover, the field $\Phi_{f,\ell}$ stands for many component fields.  
 Being of length $\ell$ we have $\ell$ vectors that must be ordered.
 This can be done in $\ell!$ ways, assuming for the time being
 that all the vectors are different.  Then the possible splittings 
  into $f$ factors can be counted as the number of ways one can put
  $f-1$ bars in $\ell-1$ positions since the first bar cannot appear
  before the first vector and the last bar cannot appear after the last vector. 
  Such a splitting can be done in $\bigl( {\ell-1\atop f-1} \bigr)$ ways, giving
  the total number of component fields equal to $\ell!  \bigl( {\ell-1\atop f-1} \bigr)$.
  When the vectors are not all different, the result must be divided by
  $r_1! r_2! \cdots$  when we have sets of $r_1, r_2, \ldots $ equal vectors. 
So $\Phi$ can be organized as
 \be
 \label{fieldarr}
\begin{split} 
\Phi = \ & \   \Phi_{1,1}  \  + \Phi_{1,2}  \ + \Phi_{1,3}  \ + \Phi_{1,4} \ + \cdots \\
\ & \  \phantom{\Phi_{1,1}\  } + \Phi_{2,2}  \ + \Phi_{2,3}  \ + \Phi_{2,4} + \cdots \\
\ & \  \phantom{\Phi_{1,1}\ + \Phi_{2,2}  \ }  + \Phi_{3,3}  \ + \Phi_{3,4} + \cdots  \\
\ & \  \phantom{\Phi_{1,1}\ + \Phi_{2,2} + \Phi_{3,3}  \ \ }   + \Phi_{4,4} + \cdots \, ,
\end{split}
\ee
where length increases from left to right and factor number increases from top to bottom.

The infinitesimal gauge transformation of the field
was given in~\eqref{smgadlje}. The  corresponding finite gauge transformation can be written directly
using~\eqref{eq:FiniteGT} and noting that $m_1$ is here $\Delta$ and the product is defined with $\otimes'$:
\be
	\Phi' = e^{-\Lambda} \otimes' \Delta e^{\Lambda} + e^{-\Lambda} \otimes' \Phi \otimes'e^{\Lambda}\,. 
\ee
The exponential here is defined via the formal power series $e^{\Lambda} = 1 + \Lambda + \frac{1}{2!}\Lambda \otimes' \Lambda + \cdots$. 

It is often convenient to rewrite this expression by adding and subtracting $e^{\Lambda} \otimes' \Phi$:
\be
\Phi' = e^{-\Lambda} \otimes' \bigl( \Delta e^{\Lambda} + \Phi  \otimes' \, e^{\Lambda} \bigr)
=  e^{-\Lambda} \otimes' \bigl( \Delta e^{\Lambda} + \Phi  \otimes' \, e^{\Lambda} 
-   e^{\Lambda} \otimes' \Phi  + e^{\Lambda} \otimes' \Phi \bigr)
\ee
which elegantly simplifies to:
\be
\label{welrkjeofhe}
\Phi' =  \Phi +   e^{-\Lambda} \otimes' \bigl( \Delta +  \operatorname{ad}_\Phi \bigr)  e^{\Lambda} 
 =  \Phi +   e^{-\Lambda} \otimes' \Delta_\Phi e^{\Lambda}\, . 
\ee
Under this finite transformation, the equation of motion transforms covariantly:
\be
\Delta \Phi' + \Phi' \otimes' \Phi'  =  e^{-\Lambda} \otimes' \bigl( \Delta \Phi + \Phi \otimes'\Phi \bigr) \otimes'e^{\Lambda} \,,
\ee
demonstrating that if $\Phi$ satisfies the equation of motion, then $\Phi'$ does as well. 

\subsection{Solution to the Maurer-Cartan equation} \label{30dlkr0sd}

In this subsection, we will solve for the general solution 
of 
\be
\Delta \Phi + \Phi \otimes' \Phi = 0 \, ,
\ee
by expanding the field in factor number $\Phi = \sum_{f=1}^\infty \Phi_f$.
This expansion gives the following equations
\begin{equation}
\label{3edkoe38}
    \begin{split}
        & \Delta \Phi_1 + \Phi_1\otimes'\Phi_1 = 0
        \, ,\\
        & \Delta_{\Phi_1}\Phi_f + \sum_{i=2}^{f-1}\Phi_i\otimes'\Phi_{f+1-i} = 0, \quad 
        f \geq  2\, .
    \end{split}
\end{equation}
In the second equation, the sum vanishes for $f=2$. 

We start by showing that the solution $\Phi_1$ to the first equation is {\em unique} once we specify $\Phi_{1,1}$, and in fact coincides with the solution noted
 in~\eqref{phi1solguessed}. Expanding the top equation in terms of length, 
\be
\Delta \Phi_{1,\ell} + \sum_{p=1}^{\ell-1}  \Phi_{1,p} \otimes' \Phi_{1, \ell-p}  = 0,  \,\quad \ell\geq 2.
\ee
Applying the homotopy operator $h$, we get
\be
h\Delta \Phi_{1,\ell} =  -\sum_{p=1}^{\ell-1} h \bigl(  \Phi_{1,p} \otimes' \Phi_{1, \ell-p} \bigr) \,, \quad \ell\geq2\,.
\ee
On the left-hand side, $h\Delta$ is equal to the anticommutator as $h$ kills 
factor-number-one elements. Using \eqref{keycommutator} and \eqref{Hdef}, we get
\be
\Phi_{1,\ell} =  -\frac{1}{\ell-1}\sum_{p=1}^{\ell-1} \tilde h \bigl(  \Phi_{1,p} \otimes' \Phi_{1, \ell-p} \bigr) \,, \quad \ell\geq2\,.
\ee
We use the notation $\Phi_{1,\ell} = s^{-1} v_\ell$ to solve for the $v_\ell$'s in terms of
$v_1\in V$ appearing in 
\be
\Phi_{1,1} = s^{-1} v_1\,.
\ee
The equation then becomes
\be
 s^{-1} v_\ell =  -{1\over \ell-1}  \sum_{p=1}^{\ell-1} \tilde h \bigl(   s^{-1} v_p \otimes' s^{-1} v_{\ell-p}  \bigr)= \, {1\over \ell-1}  \sum_{p=1}^{\ell-1} s^{-1} (v_p \otimes v_{\ell-p}) \,,
\ee
using~\eqref{htilontwof}, so that we learn that
\be
\label{musthavecon} 
 v_\ell =\, \, {1\over \ell-1}  \sum_{p=1}^{\ell-1} \, v_p \otimes v_{\ell-p}  \,.
\ee
This gives   $v_2 = v_1 \otimes v_1$,  $v_3 = v_1\otimes v_1\otimes v_1$, and in
general,
\be
v_\ell = v_1^{\otimes \ell} \,. 
\ee
One immediately confirms consistency with~\eqref{musthavecon}.
We have thus shown that
\be\label{higher1}
\Phi_{1,\ell} =  s^{-1} (v_1^{\otimes \ell}) \,, \ \ \  \ell\geq 1 \quad \to \quad 
 \Phi_1 = \sum_{\ell =1}^\infty s^{-1} (v_1^{\otimes \ell}) = s^{-1} E(v_1)=s^{-1}{v_1\over 1-v_1}\,.
\ee
This is indeed the answer~\eqref{phi1solguessed}, which we already checked satisfies the Maurer-Cartan equation. Now we know this is the unique solution. 
The field $\Phi_1$ is gauge invariant as gauge transformation produces factor number two  and higher contributions. Supplemented with $\Phi_{f\geq 2} = 0$, we
still have a solution of the full Maurer-Cartan equation, since the second equation in~\eqref{3edkoe38} is also satisfied -- the product of fields never has two $\Phi_1$'s. 
We thus have that
\be
\Phi = \Phi_1  = s^{-1}E(v_1)  \, , 
\ee
is, in fact, the {\em canonical solution} of $\Delta \Phi + \Phi\otimes' \Phi = 0$,
and holds for any $v_1 \in V_0$.

\medskip The general solution retains $\Phi_1$ from the canonical solution but includes nonzero higher-factor fields. We begin by solving the equation at the next factor number: 
\be
\Delta_{\Phi_1}\Phi_2 = 0\,.
\ee
Since the homology of $\Delta_{\Phi_1}$ vanishes at any factor number greater than one, we can write the general form of $\Phi_2$ as:
\begin{equation}\label{guess2} 
\Phi_2 = \Delta_{\Phi_1}\tilde S_1\,,
\end{equation}
with $\tilde S_1$ an arbitrary vector of degree zero and factor number one. 
The equation at factor number three is:
\be
\Delta_{\Phi_1}\Phi_3 = -\Phi_2\otimes'\Phi_2 = -\Delta_{\Phi_1}(\tilde S_1\otimes'\Delta_{\Phi_1} \tilde S_1)\,.
\ee
Again, utilizing the triviality of the homology, the general solution for $\Phi_3$ is:
\begin{equation}\label{guess3} 
\Phi_3= \Delta_{\Phi_1}\tilde S_2 - \tilde S_1\otimes'\Delta_{\Phi_1} \tilde S_1\,,
\end{equation}
with $\tilde S_2$ an arbitrary vector of degree zero and factor number two. 
So far this gives 
\begin{equation}
\label{firstfew}
    \Phi = \Phi_1 + \Phi_2 +\Phi_3  + \cdots =  \, 
    \Phi_1 \   - \tilde S_1 \otimes'  \Delta_{\Phi_1} \tilde S_1  + \Delta_{\Phi_1} (\tilde S_1 + \tilde S_2) + \cdots  .
\end{equation}
Defining a degree zero element $\tilde S$  as the sum of $\tilde S_f$ for all values of $f\geq 1$ we write
\be
\label{ssumdef}
\tilde S = \sum_{f\geq 1}  \tilde S_f  = \tilde S_1 + \tilde S_2 + \cdots \, .
\ee
We now claim that the {\em general} solution $\Phi$ is given as follows:
\be
\label{welrkjeofhezz} 
\ \Phi =  \Phi_1 +   {1\over 1+\tilde S} \otimes' \Delta_{\Phi_1} \tilde S \, .
\ee
Expanding the above expression in factor number using~\eqref{ssumdef}, we recover the
terms indicated in~\eqref{firstfew} -- a simple consistency check.  Comparing with~\eqref{welrkjeofhe}, we see that 
$\Phi$ is just the gauge transformation of $\Phi_1$ with a gauge parameter $\Lambda$ related to $\tilde S$ as $\tilde S = e^{\Lambda} -1$ (note that $\Delta_{\Phi_1} 1 = 0$).

In order to prove~\eqref{welrkjeofhezz} we first note a couple of simple
facts. First, $\Delta_{\Phi_1}$ is a derivation with respect to the $\otimes'$ product. 
Second, letting $(\cdots)_f$ denote the factor number $f$ projection of $(\cdots)$,
we note that
\be
\label{identia3a5}
\begin{split}
(A \otimes' C)_f = & \  \sum_{i=1}^{f-1}  A_i \otimes' C_{f-i}\,, \\ 
(\Delta_{\Phi_1}A)_{f+1} = & \ \Delta_{\Phi_1} A_f \,.  
\end{split}
\ee

Our induction hypothesis is that formula~\eqref{welrkjeofhezz} gives the correct 
expression for the field at factor number $f$ and we will prove that then
it gives the correct expression for factor number $f+1$.  Having shown that~\eqref{welrkjeofhezz} works for $f=2$, this will prove our claim.
So, we write
\be
\label{indhyp}
\Phi_f = \Bigl(  {1\over 1+\tilde S} \otimes' \Delta_{\Phi_1} \tilde S 
\Bigr)_f\,.
\ee
Note that in the above formula only $\tilde S_1, \ldots , \tilde S_{f-1}$ contribute to $\Phi_f$.
Thus, defining
\be 
\label{deftildesf}
\tilde S (f) := \tilde S_1 + \cdots + \tilde S_{f-1} \,,
\ee
we can refine the induction hypothesis to read
\be
\label{indhypex}
\Phi_f = \Bigl(  {1\over 1+\tilde S(f)} \otimes' \Delta_{\Phi_1} \tilde S(f)  \Bigr)_f\,.
\ee
Consistent with this hypothesis, the above formula can be used for any
$2\leq i \leq f$:
\be
\label{indhypexx}
\Phi_i = \Bigl(  {1\over 1+\tilde S(f)} \otimes' \Delta_{\Phi_1} \tilde S(f)  \Bigr)_i\,, \ \ \   2 \leq i \leq f \,. 
\ee
Now consider the equation of motion for $\Phi_{f+1}$. From~\eqref{3edkoe38} it reads
\be
\begin{split}
\Delta_{\Phi_1} \Phi_{f+1} = & \ - \sum_{i=2}^f  \Phi_i \otimes'\Phi_{f+2-i} \\
= & \ - \sum_{i=2}^f \Bigl( {1\over 1+\tilde S(f)} \otimes' \Delta_{\Phi_1} \tilde S(f) \Bigr)_i  \otimes'  \Bigl({1\over 1+\tilde S(f)} \otimes' \Delta_{\Phi_1} \tilde S(f)  \Bigr)_{f+2-i}\\
= & \ - \sum_{i=1}^{f+1} \Bigl( {1\over 1+\tilde S(f)} \otimes' \Delta_{\Phi_1} \tilde S(f) \Bigr)_i  \otimes'  \Bigl({1\over 1+\tilde S(f)} \otimes' \Delta_{\Phi_1} \tilde S(f)  \Bigr)_{f+2-i}
\end{split}
\ee
where in the last step we extended the range of the sum to include $i=1$, where the
left factor vanishes, as well as $i = f+1$, where the second factor vanishes.  Using now property~\eqref{identia3a5} we see that   
\be
\begin{split}
\Delta_{\Phi_1} \Phi_{f+1} 
= & \ -  \Bigl( {1\over 1+\tilde S(f)} \otimes' \Delta_{\Phi_1} \tilde S(f)   \ \otimes' \ {1\over 1+\tilde S(f)} \otimes' \Delta_{\Phi_1} \tilde S(f)  \Bigr)_{f+2}\,. 
\end{split}
\ee
Recalling that $\Delta_{\Phi_1}$
is a derivation over $\otimes'$ and squares to zero, we can rewrite
the expression within parentheses on the right-hand side as 
\be
\Delta_{\Phi_1} \Phi_{f+1} 
=  \  \Bigl( \Delta_{\Phi_1} \Bigl[ 
 {1\over 1+\tilde S(f)} \otimes' \Delta_{\Phi_1} \tilde S(f)\Bigr] \Bigr)_{f+2}
 =  \Delta_{\Phi_1}  \Bigl( 
 {1\over 1+\tilde S(f)} \otimes' \Delta_{\Phi_1} \tilde S(f) \Bigr)_{f+1}\,,
\ee
where we used the second line of~\eqref{identia3a5} in passing to the
last expression.  We then have 
\be
\Delta_{\Phi_1} \Bigl[ \Phi_{f+1} 
- \Bigl( 
 {1\over 1+\tilde S(f)} \otimes' \Delta_{\Phi_1} \tilde S(f)\Bigr)_{f+1}\Bigr]
 =  0\, .
\ee
Again, the absence of nontrivial $\Delta_{\Phi_1}$ homology at factor number 
greater than one implies that there is an $\tilde S_f$ such that 
\be
 \Phi_{f+1} =  \Bigl( 
 {1\over 1+\tilde S(f)} \otimes' \Delta_{\Phi_1} \tilde S(f)\Bigr)_{f+1}
 + \Delta_{\Phi_1} \tilde S_f  \,.
\ee
Given the definition~\eqref{deftildesf}, $\tilde S (f+1) = \tilde S(f) + \tilde S_f$
and the $\Delta_{\Phi_1} \tilde S_f $ term is the only one generated
when $\tilde S(f)$ on the first term is replaced by $\tilde S(f+1)$, resulting in
\be
 \Phi_{f+1} =  \Bigl( 
 {1\over 1+\tilde S(f+1)} \otimes' \Delta_{\Phi_1} \tilde S(f+1)\Bigr)_{f+1}
  \,.
\ee
This is precisely the statement of the induction hypothesis~\eqref{indhypex}
for $f+1$, completing the inductive step and proving that 
the general solution $\Phi$ is exactly the canonical solution $\Phi= \Phi_1$ acted on by a gauge transformation generated by $\tilde S = \sum_{k=1}^\infty \tilde S_k$, as expressed in~\eqref{welrkjeofhezz}.

\section{Solving the equation 
\texorpdfstring{$(\Delta + B) \Psi+ \Psi\otimes'\Psi=0$}{MC-equation-with-B}}\label{solveqbplusdelta}

In this section we solve the full equation of motion of nontrivial theories in their bar-cobar reformulation. This is the Maurer-Cartan equation of the bar-cobar theory~\eqref{eq:DefBarCobarComplex} where the differential
is $\Delta + B$ and the product is $\otimes'$:
\be
\label{3ouhrfdk}
(\Delta + B) \Psi+ \Psi\otimes' \Psi= 0\,.
\ee 
 The most general solution to~\eqref{3ouhrfdk}, we will show, is
 a finite gauge transformation of a {\em canonical solution}: a purely 
  factor-one solution $\Psi= \Psi_1$, now subject to the subsidiary condition $B\Psi_1 = 0$. This canonical solution is in fact fully determined by the
 choice of $\Psi_{1,1}$ subject to a constraint that is in fact the original $A_\infty$ equation of motion for $\Psi_{1,1}$. 

 The infinitesimal gauge transformation of the field
is  given as
\be
\label{infgaugbbb}
\delta \Psi = (\Delta + B)\Lambda + \Psi\otimes'\Lambda - \Lambda \otimes' \Psi \, .
\ee
The  corresponding finite gauge transformation can be written directly
using~\eqref{eq:FiniteGT}
and noting that $m_1$ is here $\Delta + B$ and the product is defined with $\otimes'$:
\be\label{gaugetansnffull}
	\Psi' = e^{-\Lambda} \otimes' (\Delta + B) e^{\Lambda} + e^{-\Lambda} \otimes' \Psi \otimes'e^{\Lambda}\,. 
\ee
The exponential is defined via the formal power series $e^{\Lambda} = 1 + \Lambda + \frac{1}{2!}\Lambda \otimes' \Lambda + \cdots$. We suppress the $\otimes'$ from now on.

 We also consider gauge transformations of the bar-cobar theory that preserve canonical solutions: acting on a canonical solution $\Psi_1$, the transformed solution is still a factor equal one element. 
 While the higher factor elements remain zero, $\Psi_1$ is changed. 
 The transformations of $\Psi_1$  are none other than those induced by
  the $A_\infty$ gauge transformations of the field $\Psi_{1,1}$.

 In the analysis that follows, we set up a perturbative expansion of the Maurer-Cartan equation
 in terms of $B$. This results in  equations tractable with the knowledge of
 $\Delta_{\Phi}$ 
 homology developed in Section~\ref{delphohomonomc}.

\subsection{Useful identities} 
We begin by establishing some identities that will help our analysis. 
We have in~\eqref{welrkjeofhezz} an expression for the general solution $\Phi$ of the Maurer-Cartan equation 
$\Delta\Phi + \Phi  \Phi = 0$ 
as a gauge transformation of the canonical solution $\Phi_1$~\eqref{higher1}. 
Introducing a new
parameter $S$ related to $\tilde S$ as follows, $e^S = 1 + \tilde S$, the expression for $\Phi$ becomes
\be
\label{gtofx11} 
\Phi =  \Phi_1 +  e^{-S}   \Delta_{\Phi_1} e^S \,. 
\ee
We have relations that express the action of $\Delta_\Phi$ and that of $\Delta_{\Phi_1} $ in terms of each other.  One can check that
\begin{equation}
\label{pairofdeltas}
\begin{split}
	\Delta_{\Phi}f =& \  e^{-S}\{\Delta_{\Phi_{1}}(e^{S}fe^{-S})\}e^{S}\, ,  \\[1.0ex]
	\Delta_{\Phi_1}f =& \  e^{S}\, \{\Delta_{\Phi}(e^{-S}fe^{S})\}\, e^{-S} \,. 
	\end{split}
\end{equation}
These equations allow us to invert~\eqref{gtofx11} and write $\Phi_1$ in terms of $\Phi$:
\be
\label{gtofx2} 
\Phi_1 =  \Phi +  e^{S}  \Delta_{\Phi} e^{-S} \,,
\ee
consistent with the fact that the inverse gauge transformation maps $\Phi$ back to $\Phi_1$. 
Moreover, from~\eqref{pairofdeltas} we see that
\be
\label{frXX1} 
\Delta_{\Phi}f = 0  \ \ \ \Rightarrow \ \   \  \Delta_{\Phi_{1}}(e^{S}fe^{-S}) = 0 \,. \ee
We have a few equations involving $B$, which acts as a derivation. 
In particular,  recalling that $[\Delta, B ] = 0$ we have 
\be
[\Delta_\Phi,  B]  =  \operatorname{ad}_{B\Phi} \,.
\ee
A state $S_B$, built from the gauge parameter $S$ and $B$, plays a useful role:
\be
\label{sbdefined}
S_B \equiv  e^{-S} Be^{S}
\, \ \ \ \to \ \    Be^S =  e^S  S_B\,, \ \ \  B e^{-S} = - S_B e^{-S} \,. 
\ee
It is simple to show that
\be
\label{bbbskk}
BS_B + S_B S_B = 0 \,. 
\ee
With a bit of work one can show that
\begin{equation}
\label{badsb} 
		e^{-S}B(e^{S}fe^{-S})e^{S} = (B+ \operatorname{ad}_{S_{B}}) f\,.
\end{equation}
On account of~\eqref{gtofx2}, we have
\be
\label{gtofx2e} 
\begin{split}
B\Phi  = B\Phi_1 -  B \bigl[  e^{S}  \Delta_{\Phi} e^{-S} \bigr]  = &\   B\Phi_1- e^S S_B  \Delta_\Phi e^{-S}  - e^S [ B, \Delta_\Phi] e^{-S} + e^S  \Delta_\Phi  B e^{-S} \\[0.5ex]
=& \  B\Phi_1- e^S S_B  \Delta_\Phi e^{-S}  - e^S \operatorname{ad}_{B\Phi}  e^{-S} - e^S  \Delta_\Phi  (S_B e^{-S})\\[0.5ex]
=& \  B\Phi_1 - e^S (\Delta_\Phi  S_B) e^{-S}  - e^S (B\Phi) e^{-S} + B\Phi\,.
\end{split}
\ee
The $B\Phi$ terms on the left and right-hand sides cancel and we are left with the identity
\be
\label{aux1}
B\Phi + \Delta_\Phi S_B  =   e^{-S}B\Phi_1 e^S \,.
\ee

The equation we wish to solve is the Maurer-Cartan equation~\eqref{3ouhrfdk}.
 For computational efficiency, it is better to define $\chi$ as 
\begin{equation}
\label{thanszrwl}
    \Psi = \Phi + S_B + e^{-S}\chi e^S\, ,
\end{equation}
where $\Phi$ solves the Maurer-Cartan equation $\Delta \Phi + \Phi\Phi = 0$, and $S_B$ is defined as in~\eqref{sbdefined}. 
Substituting~\eqref{thanszrwl} into~\eqref{3ouhrfdk}, and collecting terms
using the definition of adjoint `ad' we get
\be
\begin{split}
0 =  &\  \Delta \Phi + \Phi \Phi  \ \ + B\Phi + \Delta_\Phi S_B \  +   BS_B + S_B S_B \\[0.5ex]  
& \ + (B + \operatorname{ad}_{S_B} )  ( e^{-S} \chi e^S )   + \Delta_\Phi ( e^{-S} \chi e^S )  + e^{-S} \chi \chi e^S \,.
\end{split}
\ee 
We use the Maurer-Cartan equation for $\Phi$, and equations \eqref{aux1} 
and~\eqref{bbbskk} to deal with the terms on the first line above. 
We then use~\eqref{badsb} and~\eqref{pairofdeltas} to deal with the second line.
All in all we find the equation 
\begin{equation}\label{eq:forchi}
    \Delta_{\Phi_1}\chi + B\chi + B\Phi_1+ \chi\chi =0\,.
\end{equation}
If one tried to solve~\eqref{3ouhrfdk} or~\eqref{eq:forchi} order by order in factor number, knowledge of the
homology of $B$ would be needed.  Since this homology depends on the particular theory,
we will not try this route. The equation will be solved in perturbation in $B$.  For this
we write
\be\label{reconstructY}
\Psi=  \sum_{n=0}^\infty  \Psi^{(n)} = \Phi + S_B + \sum_{n=1}^\infty e^{-S} \chi^{(n)}e^{S}\,,
\ee
with $n$ counting the (formal) order in $B$.  All $\Psi^{(n)}$ and $\chi^{(n)}$ include all values
of the factor number. Here we identify
\be
\begin{split} 
\Psi^{(0)} = & \ \Phi\,,\\
\Psi^{(1)} = & \ S_B + e^{-S}\chi^{(1)}e^{S} \,, \\
\Psi^{(n)} = & \  e^{-S} \chi^{(n)}e^{S}, \ \ n\geq 2\,.
\end{split} \ee
The vector $\Phi$ above is the solution~\eqref{gtofx11} of the Maurer-Cartan equation without $B$. 
We can set $\Psi^{(0)} = \Phi$ because the order zero part of equation~\eqref{3ouhrfdk} is
$\Delta \Psi^{(0)} + \Psi^{(0)}  \Psi^{(0)} = 0$.
In equation~\eqref{3ouhrfdk} only the term $B\Psi$ contributes at factor number one. The vanishing of this factor one term is an independent equation
\begin{equation}
    (B\Psi)_1 = B\Psi_1 = B\Phi_1 + B\chi_1 = 0\, ,
\end{equation}
and we call it the subsidiary condition. The subsidiary condition must also hold perturbatively in $B$, which in particular requires
\begin{equation}\label{susbsusd}
    B\Phi_1 = 0\, ,
\end{equation}
because $\chi$ is at least linear in $B$.

Back in~\eqref{3ouhrfdk} to zeroth order we have satisfied the equation,  and to higher order we get 
\be
\label{mstbtfljnpss} 
\Delta_\Phi \Psi^{(n)} +  B \Psi^{(n-1)}  + \sum_{{i+j = n}\atop {i,j \geq 1}}  \Psi^{(i)}  \Psi^{(j)}  = \ 0 \,, \ \  n \geq 1\, , 
\ee
or equivalently, with $\chi^{(0)} = 0$,
\be
\label{mstbtfljnpsschi} 
\Delta_{\Phi_1} \chi^{(n)} +  B \chi^{(n-1)}  + \sum_{{i+j = n}\atop {i,j \geq 1}}  \chi^{(i)} \chi^{(j)}  = \ 0 \,, \ \  n \geq 1\, .
\ee
The subsidiary conditions are, in addition to~\eqref{susbsusd},
\be 
\label{bcondvnvg}
B \Psi^{(n)}_1  = B \chi^{(n)}_1= 0  \,,  \ \ n =1,2, \ldots \,. 
\ee
We are going to use the triviality of the homology of $\Delta_{\Phi_1}$ at factor number greater than one to solve for $\chi^{(n)}$ and thus $\Psi^{(n)}$.

\subsection{Solving for
\texorpdfstring{$\Psi$}{Psi}}
\textbf{ Solving for $\chi^{(1)}$}. The $n=1$ equation in~\eqref{mstbtfljnpsschi} is
\be
\Delta_{\Phi_1} \chi^{(1)} = 0  \,. 
\ee
This means $\chi^{(1)}$ equals an exact term plus a nontrivial class $W^{(1)}_1$: 
\be\label{y1finform}
\chi^{(1)}  =  \Delta_{\Phi_1}M^{(1)}  +  W_1^{(1)}\, .
\ee
Here $W^{(1)}_1$ is of the form indicated in~\eqref{detx1coh}. We then have
\be 
\Psi^{(1)}  =  S_B +  e^{-S} \chi^{(1)}  e^{S}\,. 
\ee 
We now check the subsidiary condition $B \Psi^{(1)}_1 = 0$.  For this we must find the factor-number-one part of $\Psi^{(1)}$. Since multiplication by 
exponentials cannot decrease the factor number, we have
\be
\Psi^{(1)}_1  = (S_B)_1   + (\chi^{(1)})_1 =  (B e^S )_1  +  W^{(1)}_1 =   B S_1 + W^{(1)}_1 \,.   
\ee
Since $B^2=0$, the condition $B\Psi^{(1)}_1=0$ simply implies
\be
\label{secondsubs}
 B W^{(1)}_1 = 0  \,.
\ee
This completes the analysis of the solution to first order in $B$.

\bigskip
\noindent\textbf{ Solving for $\chi ^{(2)}$}.  Equation~\eqref{mstbtfljnpsschi} for $n=2$ gives
\be
\Delta_{\Phi_1} \chi^{(2)}  + B \chi^{(1)} + \chi^{(1)}  \chi^{(1)} = 0 \,.
\ee
On account of the earlier result for $\chi^{(1)}$ we have
\be
\label{a2newterms}
\Delta_{\Phi_1} \chi^{(2)}  =- B (\Delta_{\Phi_1} M^{(1)} +  W^{(1)}_1 )  - (\Delta_{\Phi_1} M^{(1)} +  W^{(1)}_1 ) (\Delta_{\Phi_1} M^{(1)} +  W^{(1)}_1 ) \,.
\ee
Note that in this equation the only term with factor number one 
is $B W^{(1)}_1$, confirming why the subsidiary condition~\eqref{secondsubs} must hold. It is easy to see that all the terms on the right-hand side are $\Delta_{\Phi_1}$-closed using the fact that $\Delta_{\Phi_1}$ is
a derivation of the product.
Since $\Delta_{\Phi_1}$ has no homology at factor number two or higher, we must have that
\be
 W^{(1)}_1    W^{(1)}_1  = \Delta_{\Phi_1} \overline{W}^{(2)}_1 \,,
\ee
for some factor-number-one state $\overline{W}^{(2)}_1$ at second order in $B$.  We can now write the right-hand side
of~\eqref{a2newterms} as follows
\be
\label{aXnewterms}
\Delta_{\Phi_1} \chi^{(2)}  =   \Delta_{\Phi_1} \Bigl(  B M^{(1)} - M^{(1)} \Delta_{\Phi_1} M^{(1)}   + W^{(1)}_1 M^{(1)} 
- M^{(1)} W^{(1)}_1 -  \overline{W}^{(2)}_1 \Bigr) \,.
\ee
Therefore, $\chi^{(2)}$ equals the expression in parentheses up to a trivial term $\Delta_{\Phi_1} M^{(2)}$ 
and a homology class $W^{(2)}_1$:
\be
\chi^{(2)} =  B M^{(1)} - M^{(1)} \Delta_{\Phi_1} M^{(1)}   + W^{(1)}_1 M^{(1)} 
- M^{(1)} W^{(1)}_1 -  \overline{W}^{(2)}_1  + \Delta_{\Phi_1} M^{(2)}  + W^{(2)}_1 \,.
\ee
We rewrite this as 
\be
\label{a2form3oi}
\chi^{(2)} =  B M^{(1)} - M^{(1)} \Delta_{\Phi_1} M^{(1)}   +  W^{(1)}_1 M^{(1)} 
- M^{(1)} W^{(1)}_1 \  + \Delta_{\Phi_1} M^{(2)}  + \widetilde W^{(2)}_1 \,, 
\ee
with
\be
\widetilde W^{(2)}_1  \equiv   W^{(2)}_1 - \overline{W}^{(2)}_1 \,. 
\ee
Note that 
\be
\Delta_{\Phi_1}\widetilde W^{(2)}_1 + W^{(1)}_1 W^{(1)}_1 = 0\,. 
\ee 
This time $\chi^{(2)}_1 = B M^{(1)}_1 + \widetilde W^{(2)}_1$
and thus the subsidiary condition  $B \chi_1^{(2)} = 0$ gives
\be
B \widetilde W^{(2)}_1 = 0 \,. 
\ee
Next, we solve for $\chi^{(3)}$.  The result is long and will be given below.   
In fact, $\chi^{(1)}$,  $\chi^{(2)}$, and $\chi^{(3)}$ are generated
by the following closed-form expression
\be
\label{guessforA}
\chi =   {1\over 1+M}  ( \Delta_{\Phi_1} + B) (1 + M)  +  {1\over 1+M}  \widetilde W_1  (1 + M) 
\,.
\ee
Here $\widetilde W_1$ is a  factor-number-one state satisfying two conditions:
\be  
\label{conditionsonW11}
\begin{split}
B \widetilde W_1 = & \ 0 \,, \\
\Delta_{\Phi_1}\widetilde W_1 + \widetilde W_1\widetilde W_1 = & \ 0\,.
\end{split}
\ee  
Moreover, $\widetilde W_1$ has a $B$-expansion 
\be
\widetilde W_1  = 
\sum_{n=1}^\infty  \widetilde W^{(n)}_1 \,, \ \ \ \ \hbox{with} \ \ \ 
\widetilde W^{(1)}_1 := W^{(1)}_1 \,. 
\ee
One can verify that with this expansion, equations~\eqref{conditionsonW11} demand, to first and
second order in $B$, that
\be
B \widetilde W_1^{(1)} = B \widetilde W_1^{(2)} = 0, \ \   \Delta_{\Phi_1}\widetilde W_1^{(1)} = 0 
\,, \ \ \ \Delta_{\Phi_1}\widetilde W_1^{(2)} + \widetilde W_1^{(1)}\widetilde W_1^{(1)} = 0 \,,
\ee
all of which have already been verified above.  

The state $M$ is also to be
expanded in $B$ as follows
\be 
M =  \sum_{n=1}^\infty  M^{(n)}   \,. \ee 
To reproduce our results for $\chi^{(1)}$,  $\chi^{(2)}$, and $\chi^{(3)}$ 
from~\eqref{guessforA}, we expand first the prefactor: 
\be
{1\over 1 + M} = {1\over 1 + M^{(1)} + M^{(2)} +\cdots} =  1 - M^{(1)} + M^{(1)} M^{(1)}  - M^{(2)} + {\cal O} (B^3) \,.  
\ee
As a result, the expansion of $\chi$ truncated through $\order{B^3}$ is 
\be
\chi =  \bigl( 1 - M^{(1)} + M^{(1)} M^{(1)}  - M^{(2)} \bigr)  \Bigl(  \Delta_{\Phi_1} + B 
+ \widetilde  W^{(1)}_1   + \widetilde W_1^{(2)} + \widetilde W_1^{(3)}   \hskip-1pt\Bigr) 
\Bigl( 1 + M^{(1)} + M^{(2)} + M^{(3)}    \Bigr) + {\cal O}(B^4)  
\ee
Expanding this out, we find
\be
\begin{split}
\chi =   & \ \  \  \Delta_{\Phi_1} M^{(1)}  +\widetilde  W^{(1)}_1  \\[1.5ex]
& \  +  \Delta_{\Phi_1} M^{(2)} - M^{(1)} \Delta_{\Phi_1} M^{(1)} + B M^{(1)} + \widetilde  W^{(1)}_1   M^{(1)}     - M^{(1)} \widetilde  W^{(1)}_1  + \widetilde W^{(2)}_1 \\[2.0ex]
& \  +  \Delta_{\Phi_1} M^{(3)} - M^{(2)} \Delta_{\Phi_1} M^{(1)}- M^{(1)} \Delta_{\Phi_1} M^{(2)}  + M^{(1)} M^{(1)}\Delta_{\Phi_1} M^{(1)}\\[1.0ex]
& \, + B M^{(2)}  -  M^{(1)} B M^{(1)}  
+ \widetilde  W^{(1)}_1   M^{(2)}     - M^{(2)} \widetilde  W^{(1)}_1   
+ \widetilde W_1^{(2)} M^{(1)}     - M^{(1)} \widetilde W_1^{(2)} \\[1.0ex]
& \, + M^{(1)} ( M^{(1)} \widetilde W_{1}^{(1)} -\widetilde  W^{(1)}_1   M^{(1)}) 
+ \widetilde W^{(3)}_1 +   {\cal O} (B^4) \,. 
\end{split}
\ee
The first line on the right-hand side is indeed $\chi^{(1)}$ as given in~\eqref{y1finform}.
The next line on the right-hand side is $\chi^{(2)}$, as given in~\eqref{a2form3oi}.
Finally, the last three lines comprise $\chi^{(3)}$, which we computed independently.
Having checked the validity of the ansatz~\eqref{guessforA} for the first few terms in $\chi$, a
general proof that the ansatz follows from the equations to be solved is obtained by
induction, along the lines of the proof in Section~\ref{30dlkr0sd}.  This proof is sketched in
Appendix~\ref{inductproofxx9}.

\medskip
Redefining $M$ in~\eqref{guessforA} as $ M = e^{N}-1$,  we have
\be
\chi =    e^{-N}  ( \Delta_{\Phi_1} + B  +   \widetilde W_1) e^N   \,.
\ee
We reconstruct $\Psi$ starting from~\eqref{reconstructY}, and using the above expression for $\chi$: 
\be
\label{reconstructYx}
\Psi= \Phi + S_B +  e^{-S} e^{-N} \bigl[  ( \Delta_{\Phi_1} + B  +   \widetilde W_1) e^N \bigr] \, e^{S}\,.
\ee
Since $\Delta_{\Phi_1} + B $ is a derivation,
\be
[(\Delta_{\Phi_1} + B ) e^N ]  e^S =  (\Delta_{\Phi_1} + B ) e^N e^S  - e^N (\Delta_{\Phi_1} + B ) e^S  \,.
\ee
As a result, 
\be
\label{reconstructYxz}
\Psi= \Phi + S_B +  e^{-S} e^{-N}  ( \Delta_{\Phi_1} + B + \widetilde W_1) e^N e^S  - e^{-S} (\Delta_{\Phi_1} + B ) e^S   \,.
\ee
From~\eqref{gtofx11} we have $\Phi-\Phi_1 =e^{-S}  (\Delta_{\Phi_1} e^S ) $, and we recall the definition~\eqref{sbdefined} of $S_B$ to find that $\Psi$ becomes
\be
\label{reconstructYxzz}
\Psi= \Phi_1 +  e^{-S} e^{-N}  ( \Delta_{\Phi_1} + B + \widetilde W_1) e^N e^S  \,.
\ee
Writing out $\Delta_{\Phi_1} = \Delta + \operatorname{ad}_{\Phi_1}$, we get 
\be
\label{finealelrk}
\Psi= e^{-S} e^{-N} (\Delta + B ) e^{N} e^{S} + e^{-S} e^{-N} (\Phi_1 + \widetilde W_1 ) e^{N} e^{S} \,. 
\ee
Writing $e^{\Lambda} := e^{N}e^S$, and comparing with~\eqref{gaugetansnffull}, we find that the general solution $\Psi$ is a gauge transformation of the {\em canonical factor-number-one} solution $\Psi'$ given by 
\be
\Psi' =  \Psi'_1 =  \Phi_1 + \widetilde W_1\,. 
\ee
For a factor-one state $\Psi'$ to satisfy $(\Delta + B) \Psi' + \Psi' \Psi' = 0$, given that $B\Psi'$ carries factor number one, whereas $\Delta \Psi'$ and $\Psi' \Psi'$ carry factor number two, we must have 
\begin{equation}
B \Psi'= 0 \,, \qquad \text{and} \qquad \Delta \Psi' + \Psi' \Psi' = 0\,.
\end{equation}
The first condition holds because $B$ annihilates both $\Phi_1$ and $\widetilde W_1$ 
-- see~\eqref{susbsusd} and~\eqref{conditionsonW11}.  The second condition, expanded gives
\be
\Delta \Phi_1 + \Phi_1 \Phi_1 + \Delta_{\Phi_1} \widetilde W_1 + \widetilde W_1 \widetilde W_1 = 0\,,
\ee 
which holds on account of $\Delta \Phi_1 + \Phi_1 \Phi_1 =0$ and equation~\eqref{conditionsonW11}. 
This concludes our proof that the general solution is a gauge transformation of a canonical solution.

\subsection{Gauge transformations of canonical solutions}

As we have defined above, a canonical solution of 
\be (\Delta + B) \Psi  + \Psi  \Psi = 0\,,
\ee
is a factor one solution  $\Psi= \Psi_1$ that satisfies the above equation and, as a result,
satisfies the two following conditions: 
\be   B\Psi_1 = 0 \,, \ \ \ \Delta \Psi_1  + \Psi_1 \Psi_1 = 0 \,.
\ee
Consider now a gauge transformation of this $\Psi_1$ solution, necessarily mapping it to another
solution that we call $\Psi'$. 
Here, for simplicity, we work with infinitesimal transformations with gauge
parameter $\Lambda$ in which case we get, from~\eqref{infgaugbbb},
\be
\Psi' =  \ \Psi_1 + (\Delta + 
B) \Lambda  + \Psi_1 \Lambda - \Lambda \Psi_1  
=  \ \Psi_1 +   \Delta_{\Psi_1} \Lambda  + B \Lambda \,.
\ee
This $\Psi'$ is by construction a solution of the equations of motion (to order $\Lambda$).
We want to consider those gauge transformations for which $\Psi'$ is still a canonical solution.
This just requires that the factor number two or higher components
be zero.  Filtering the above equation
by factor number we have the following two equations:
\be\begin{split}
 \Psi'_1 = & \  \Psi_1 + B \Lambda_1 \,. \\
 \Psi'_n = &    \   B\Lambda_n + \Delta_{\Psi_1} \Lambda_{n-1} \,, \ \  n \geq 2 \,. 
\end{split}
\ee
We need to have the $\Psi'_{n\geq 2}=0$ for the new solution to be canonical:
\be
 B\Lambda_n + \Delta_{\Psi_1} \Lambda_{n-1} = 0  \,, \ \ \  n\geq 2 \,.
 \ee
We can set $\Lambda_{n\geq 2} = 0$ without affecting the gauge transformed $\Psi'_1$, and we will focus here on this subclass of gauge transformations.
Doing so the only condition we get from the above equation~is
\be
\Delta_{\Psi_1} \Lambda_1 = 0 \,.\ee
Being a factor one gauge parameter, $\Lambda_1$ cannot be $\Delta_{\Psi_1}$-exact, and the only solution is the nontrivial homology class.
To write it recall from~\eqref{phi1solguessed}, 
that with a solution $\Psi_1$ written as 
\be
\Psi_{1,1} = s^{-1}\psi\,, \ \ \ \Psi_1 = s^{-1} E(\psi)\,,  \ \  E(\psi) =  {\psi\over 1-\psi} \,,
\ee
 the homology class, as described in~\eqref{detx1coh} takes the form:
\be\Lambda_1 = {1\over 1 + h \operatorname{ad}_{\Psi_1} }  \Lambda_{1,1}  =  s^{-1}  \Bigl( {1\over 1-\psi}\, \lambda \, {1\over 1-\psi} \Bigr) \,, \ \ \hbox{with} \ \ \Lambda_{1,1} = s^{-1} \lambda \,. 
\ee
We see that 
\be
\begin{split} 
\Lambda_{1,1}  = & \ \phantom{-} \ s^{-1} ( \lambda)
\,,\\
\Lambda_{1,2} = & \ \phantom{-}\ s^{-1} 
(  \psi\lambda + \lambda \psi   )\,, \\
\Lambda_{1,3}=  & \ \phantom{-}\  s^{-1} 
(  \psi^2 \lambda  + \psi\lambda \psi  +   \lambda \psi^2 )\,, \\
\vdots \ \   = & \ \ \qquad  \vdots \\
\Lambda_{1,n+1} = & \ \phantom{-} \,   s^{-1} ( \psi^n \lambda  + \psi^{n-1} \lambda \psi   + \cdots +  \lambda  \psi^n )\,.
\end{split}
\ee
The gauge transformations are now 
\be
\delta \Psi_1 =  B \Lambda_1  = 
(B_1 + B_2 + \cdots ) 
( \Lambda_{1,1} + \Lambda_{1,2} + \Lambda_{1,3} + \cdots) \,.
\ee
Since the information of $\Psi_1$ lies at length equal one, we have 
\be
\delta \Psi_{1,1}  = B_1 \Lambda_{1,1}+ B_2 \Lambda_{1,2}+ B_3 \Lambda_{1,3} + \cdots  \,. 
\ee
Recalling that $B_k s^{-1} = s^{-1}\bar B_k$, and that $\bar B_k$ acting on $V^{\otimes k}$ is $b_k$, we conclude that
\be
s^{-1}\, \delta \psi  = s^{-1} \bigl( 
b_1( \lambda) + b_2 (  \psi\lambda + \lambda \psi   )+ b_3 ( \psi^2 \lambda  + \psi\lambda \psi  +   \lambda \psi^2 )+ \cdots \bigr)\,,
\ee
and we finally read
\be
\delta \psi  = b_1( \lambda) + b_2 (  \psi\lambda + \lambda \psi   )+ b_3 ( \psi^2 \lambda  + \psi\lambda \psi  +   \lambda \psi^2 )+ \cdots\,. 
\ee
These are the $A_\infty$ gauge transformations, derived here from a natural subclass of the gauge transformations of the bar-cobar theory. 
Indeed, more compactly, the above is written in an abbreviated as 
\be\delta \psi
=  \sum_{n=0}^\infty   \sum_{i  =0 }^n  b_{n+1} ( \psi^{i}, \lambda, \psi^{n-i} ) \, ,
\ee
which coincides with~\eqref{ainftygauge}.

\section{Bar-cobar formulation of scalar field theory} \label{sec:ScalarField}
In this section we illustrate our analysis by applying it to a $D$-dimensional scalar field theory with cubic and quartic interactions. Although this theory lacks gauge symmetry, and is therefore particularly simple,
its bar-cobar construction 
still yields nontrivial structural information and clearly exhibits  multilocality. We will solve the bar-cobar quadratic equations of motion explicitly, establishing that these solutions coincide precisely with those of the original cubic equation of motion.

	We discuss a $D$-dimensional field theory
	\be
	S =  \int \df^Dx\ \bigl( \, 
    \tfrac{1}{2} \phi(\partial^2 - m^2) \phi  +\tfrac{1}{3}  g \phi^3  +  \tfrac{1}{4}  \, \lambda \phi^4\bigr)  \,.
	\ee
    The equation of motion is cubic: 
    \begin{equation}\label{seom}
        \begin{split}
            (\partial^2 - m^2) \phi  +g \phi^2  +  \, \lambda \phi^3\, = 0.
        \end{split}
    \end{equation}
	 For the $A_\infty$ formulation of the theory, we follow~\cite{Okawa:2022sjf}
    and work in the $b$-picture, 
    with vector space $V$:
	\be
	V = V_0 \oplus V_{-1} \,, \qquad V_0  = \operatorname{span}_{\int} \{ c(x)\} \,, \quad \quad V_{-1} = \operatorname{span}_{\int} \{ d(x)\} \,.
	\ee
	The subscript in $V$ indicates degree, so deg$(c(x))$= 0 and deg$(d(x)) = -1$. By the notation $\operatorname{span}_{\int}\{c(x)\}$, we mean that elements are given by integrals
	\be
	    \int \df^D x \, f(x) \, c(x)\, , 
	\ee
    with $f(x)$ the expansion coefficient. 
The $A_\infty$ field $\varphi$  at degree zero is given by an integral
involving the component field $\phi$:
    \be
	\varphi=  \int \df^D x\, \phi(x) \, c(x)\, .
	\ee
	The list of {\em nonvanishing} products, all of degree minus one, is the following: 
	\be
	\begin{split}
	    b_1 (c(x)) & =\,  (\partial^2-m^2) \, d(x) \,, \ \ \ \\[0.5ex]
        b_2 (c(x_1)\otimes c(x_2)) & =\,    g\, \delta^{D}(x_1-x_2) \, d(x_1) \,,\\[0.5ex]
        \ \ \  b_3 (c(x_1)\otimes c(x_2)\otimes c(x_3)) & 
        =\,  \lambda \, \delta^{D}(x_1-x_2)\delta^{D}(x_1-x_3)\, d(x_1)\,. 
	\end{split}
	\ee
    The $A_\infty$ relations are satisfied trivially. 
    With these, the equation of motion~\eqref{seom} from the $A_\infty$ perspective is
    \be
    b_1 (\varphi)  + b_2 (\varphi, \varphi) + b_3 (\varphi, \varphi, \varphi) = 0\, .
    \ee
	This $A_\infty$ theory has no gauge symmetry because there is no degree one space $V_1$ such that 
	the $b$ action yields an element in $V_0$ where the fields are. 
	The multilinear maps $b_k$ lift to coderivations $\bar B_k$ on the coalgebra $C= T^c(V)$, and their sum is the
    coderivation $\bar B$:
	\be
    \bar B = \bar B_1 +\bar B_2  + \bar B_3 \,.
    \ee
We write $B$ for the derivation on $T (s^{-1} C)$ built from $\bar B$.  
In the bar-cobar theory, we work with $T(s^{-1} T^c(V))$.  For $s^{-1} V$ we have 
\be
\label{degonedeg2}
\begin{split}
\hbox{degree} = -1 \,, \ \ \   & \  s^{-1} (c(x))\in \, s^{-1} V_0\,, \\
\hbox{degree} = -2 \,, \ \ \   &  \  s^{-1} (d(x) ) \in \, s^{-1} V_{-1}\, .
\end{split}
\ee
The field $\Phi$ should be of degree minus one, thus $\Phi \in s^{-1} T^c(V)$ and it is expanded in the states
\be
\Phi =  \sum_{i=1}^\infty\int \prod_{n=1}^i\df^D x_n  \,   \phi_i(x_1,\ldots,x_i) \, s^{-1} 
(c(x_1)\cdots c(x_i)) \,,
\ee
where $ c(x_1)\cdots c(x_i)\equiv c(x_1)\otimes\cdots\otimes c(x_i)$. We cannot have elements with factor numbers greater than $1$, as each factor has degree 
$\leq -1$, so two or more factors give degree $\leq -2$. 
The fields do not have gauge transformations as there are no degree zero vectors in $T(s^{-1} T^c(V))$.

The bar-cobar
    equation of motion reads:
	\be
	\label{twoeqns0dim} 
	\begin{split}
	(\Delta + B) \Phi  + \Phi \Phi = 0\,.
	\end{split}
	\ee
Since $\Phi = \Phi_1$, the equation of motion gives just two independent equations 
\be 
B\Phi = 0\,, \ \ \  \hbox{and} \ \ \  \Delta \Phi  + \Phi \Phi = 0\, .
\ee
For the first equation, recalling that $B s^{-1} = s^{-1} \bar B$, 
we have
		\begin{equation}
		\begin{split}
			 & B \Phi
 = 	s^{-1} \sum_{i=1}^\infty\int \prod_{n=1}^i\df^D x_n  \,
 \phi_i(x_1,\ldots,x_i) \, (\bar B_1 + \bar B_2 +\bar B_3)c(x_1)\cdots c(x_i)\,, \\
& = s^{-1} \sum_{i=1}^\infty\int \prod_{n=1}^i\df^D x_n  \,
\phi_i(x_1,\ldots,x_i) \, \Big(\sum_{k=1}^i(\partial_k^2 -m^2)c(x_1)\cdots c(x_{k-1}) d(x_k) 
 c(x_{k+1}) \cdots c(x_i) \\
&\hskip40pt +  g \ \sum_{k=1}^{i-1}\    \delta^D(x_k-x_{k+1})c(x_1)\cdots c(x_{k-1}) d(x_k) c(x_{k+2}) \cdots 
 c(x_i) \\ 
&\hskip40pt +\lambda\ \sum_{k=1}^{i-2}\delta^D(x_k-x_{k+1})\delta^D(x_k-x_{k+2}) 
c(x_1)\cdots c(x_{k-1}) d(x_k)  c(x_{k+3})\cdots 
 c(x_i) \Big) \,, \\
 & = s^{-1} \sum_{i=1}^\infty\int \prod_{n=1}^i\df^D x_n \sum_{k=1}^{i}
 \Big((\partial_k^2 -m^2)\phi_i(x_1,\ldots,x_i)  
 + g\ \phi_{i+1}(x_1,\ldots,x_k,x_k,\ldots,x_i) \\   
&\hskip60pt  + \lambda\ \phi_{i+2}(x_1,\ldots,x_k,x_k,x_k,\ldots,x_i)\Big) 
c(x_1)\cdots c(x_{k-1}) d(x_k)c(x_{k+1}) \cdots c(x_i)\, .
		\end{split}
	\end{equation}
In passing to the last expression, we integrated by parts the spatial derivatives, and  we shifted the $i$ index noting that the terms proportional to $g$ start with $i=2$ and the
terms proportional to $\lambda$ start with $i=3$. 
In the last expression, 
the factor multiplying the basis vectors $c(x_1)\cdots c(x_{k-1}) d(x_k)c(x_{k+1}) \cdots c(x_i)$  must separately vanish for all values of $i$, and for any fixed $i$, must vanish separately for all allowed values of $k$.   
It follows that the equations for the component fields are
\begin{equation}\label{eqnhej}
    \begin{split}
       (\partial_k^2 -m^2)&\phi_i(x_1,\ldots,x_i) + g\ \phi_{i+1}(x_1,\ldots,x_k,x_k,\ldots,x_i)\\[1.0ex]
       & \qquad \qquad  + \lambda\ \phi_{i+2}(x_1,\ldots,x_k,x_k,x_k,\ldots,x_i) = 0\,, \ \ \  \forall  \ k \leq i = 1,2,3, \ldots \,. 
    \end{split}
\end{equation}
For example, the lowest equation here reads
\be
(\partial_1^2 - m^2) \phi_1 (x_1) + g \, \phi_2 (x_1, x_1) + \lambda \, \phi_3 (x_1, x_1, x_1) = 0 \,. 
\ee
For the second equation, $\Delta \Phi  + \Phi \Phi = 0$, we start writing out the term $\Delta \Phi$.  Noting that
$\bar\Delta c(x_1) = 0$ we get,
\begin{equation}
\begin{split}
    \Delta \Phi & = 	\Delta\sum_{i=2}^\infty\int \prod_{n=1}^i\df^D x_n   \,   \phi_i(x_1,\ldots,x_i) \, s^{-1} 
(c(x_1)\cdots c(x_i))\\
& =  -s^{-1} \otimes'  s^{-1} \sum_{i=2}^\infty\int \prod_{n=1}^i\df^D x_n  \,    \phi_i(x_1,\ldots,x_i) \, 
\bar \Delta (c(x_1)\cdots c(x_i))\\
& = -\sum_{i=2}^\infty\int \prod_{n=1}^i\df^D x_n   \sum_{j=1}^{i-1} \phi_i(x_1,\ldots,x_j,\ldots,x_i)C_{i,j}(x_1,\ldots,x_{i+j})\, ,
\end{split}
\end{equation}
where $ C_{i,j}(x_1,\ldots,x_{i+j}) = s^{-1} (c(x_1)\cdots c(x_j)) \otimes's^{-1} (c(x_{j+1})\cdots c(x_i))$.
With the change of indices $i\rightarrow i+ j, j \rightarrow i$, we get
	\begin{equation}
			\Delta \Phi=-\sum_{i,j=1}^\infty\int \prod_{n=1}^{i+j}\df^D x_n  
            \, \phi_{i+j}(x_1,\ldots,x_i,x_{i+1},\ldots,x_{i+j}) C_{i,j}(x_1,\ldots,x_{i+j})\,. 
	\end{equation}
From the quadratic term, we have
\begin{equation}
        \Phi\Phi 
 = \sum_{i,j=1}^\infty\int \prod_{n=1}^{i+j}\df^D x_n    \phi_i(x_1,\ldots,x_i)\phi_j(x_{i+1},\ldots,x_{i+j})C_{i,j}(x_1,\ldots,x_{i+j})\,.
\end{equation}
The second equation then becomes 
	\begin{equation} \begin{split}
	  0 = \Delta \Phi  + \Phi \Phi  & = \sum_{i,j=1}^\infty\int \prod_{n=1}^{i+j}
     \df^D x_n   \ C_{i,j}(x_1,\ldots,x_{i+j})\\
             &\Big(\hskip-5pt -\phi_{i+j}(x_1,\ldots,x_i,x_{i+1},\ldots, x_{i+j}) + \phi_i(x_1,\ldots,x_i)\phi_j(x_{i+1},\ldots, x_{i+j})\Big) \,,
		\end{split}
	\end{equation} 
and the equations for the component fields are 
\begin{equation}
 \phi_{i+j}(x_1,\ldots,x_i,x_{i+1},\ldots, x_{i+j}) - \phi_i(x_1,\ldots,x_i)\phi_j(x_{i+1},\ldots, x_{i+j}) = 0\,,   \ \ \ \forall \ i, j 
 \geq 1  \,.
\end{equation}
For $i=j=1$ this gives $\phi_2(x_1, x_2) = \phi_1(x_1) \phi_1(x_2)$.  It is straightforward to see that the unique
solution to this set of equations is
	\be
	\phi_i(x_1,\ldots, x_i) =  \phi_1(x_1)\cdots\phi_1(x_i)\, ,  \ \ i \geq 1\,. 
\ee
In particular, the $\phi_i$ are fully symmetric in all their arguments. 

Back into the $B\Phi=0$ equation~\eqref{eqnhej}
we find
	\begin{equation}
		\begin{split}
			(\partial_k^2 -m^2)&\phi_1(x_1)\cdots\phi_1(x_i) + g\, \phi_1(x_1)\cdots\phi_1(x_k)\phi_1(x_k)\cdots\phi_1(x_i)\\[1.0ex]
       & + \lambda\, \phi_1(x_1)\cdots\phi_1(x_k)\phi_1(x_k)\phi_1(x_k)\cdots\phi_1(x_i)\\[1.0ex]
       &\hskip-60pt = \phi_1(x_1)\cdots\phi_1(x_{k-1})\Big((\partial_k^2 -m^2)\phi_1(x_k) + g\,\phi_1^2(x_k)+ \lambda \phi_1^3(x_k)\Big)\phi_1(x_{k+1})\cdots\phi_1(x_i) = 0 \, . 
		\end{split}
	\end{equation}
All equations for all allowed values of $i,k$ now reduce to the same equation,
\be
(\partial_k^2 -m^2)\phi_1(x_k) + g\,\phi_1^2(x_k)+ \lambda\, \phi_1^3(x_k) = 0\,. 
\ee
This is precisely the equation of motion of the original scalar field theory, confirming the equivalence
of this theory to its bar-cobar formulation.

\section{Discussion and open questions} \label{sec:Discussion}
 In this paper, we gave a universal quadratic field equation for arbitrary field theories based on the bar-cobar construction. 
 We discussed  
 the space in which the extended set of fields lives and the two types of multilocality. We first studied the universal structures, which are the coproduct $\Delta$ and the product $\otimes'$ in the extended space $T(s^{-1}T^c(sA))$. We then used these universal structures to analyze the solutions to the universal quadratic equation and found the relation between the Maurer-Cartan space of the bar-cobar construction and the original Maurer-Cartan space. We showed this by studying the universal quadratic field equation perturbatively in $B$, the original $A_\infty$ products. Furthermore, we showed how the $A_\infty$ gauge transformations
 arise from a subclass of
 gauge transformations of the universal quadratic field equation. We illustrated the construction with the simple example of quartic scalar field theory.

 The auxiliary fields in the bar-cobar construction are multilocal. The bar construction produces the type-I multilocal fields and the cobar construction produces type-II multilocal fields. When working with canonical solutions, the equations of motion split into
\be\label{eq:CanSol}
B \Phi = 0 \, , \qquad \Delta \Phi + \Phi \otimes' \Phi = 0 \, ,
\ee
where $\Phi$ is a type-I multilocal field, thus of factor number one.
The first equation $B(\Phi_{1,1} + \Phi_{1,2} + \Phi_{1,3} + \ldots) = 0$ is linear and is related to the original nonlinear Maurer-Cartan equation $m_1(\varphi) + m_2(\varphi,\varphi) + \ldots = 0$. 
The second equation, 
which takes 
values at factor number two,
relates the type-I fields in a set of
universal quadratic relation
\be
\Phi_{1,n}(x_1,\ldots,x_n) = \Phi_{1,k}(x_1,\ldots,x_k) 
\Phi_{1,n-k}(x_{k+1},\ldots,x_n)\,,  \quad 1\leq k \leq n-1\, .
\ee
The solution of  these recursive relations is
\be
\Phi_{1,n}(x_1,\ldots,x_n) = \varphi(x_1) \cdots \varphi(x_n)\,, 
\ee
and when plugged into $B\Phi = 0$ reproduces the original Maurer-Cartan equation.

\medskip
One may wonder if there are other constructions that produce quadratic equations of motion for a given theory, in particular simpler ones that introduce fewer auxiliary fields. The power of the bar-cobar construction is that it is universal: it can be applied to any $A_\infty$ or $L_\infty$ algebra.
The large number of auxiliary fields seems to be necessary for this level of generality in the construction. Simpler constructions most likely require more knowledge about the theory at hand, for example, whether the theory is polynomial to a fixed order or whether an underlying principle determines the theory, like the decomposition of moduli spaces in string field theory. In many practical field theory examples, a finite number of extra fields is sufficient. The Hubbard-Stratonovich transform of a quartic theory only requires an extra bilocal field, as discussed in the introduction. 
 Yang-Mills theory, for example,  can be made cubic with a single extra two-form $B$ and this replacement in fact induces  a quasi-isomorphism both in the $L_\infty$ and the $A_\infty$ case (cf.~chapter 6 of \cite{Costello:2011reft} for a discussion of this from the BV point of view).

We now discuss a few open questions that may be suitable for further research on this subject.   
\begin{enumerate}
\item 
    {\bf Bilinear form and action.}
    In terms of field theories, the bar-cobar construction gives a universal quadratic field equation. It does not provide us with  an associated action generating that universal field equation. In the context of homotopy algebras, the additional data required for an action is a cyclic structure, which is a nondegenerate, degree three bilinear form $\omega: A \otimes A \rightarrow \mathbb{R}$. Given a cyclic structure, the action takes the form $S(\varphi) = \sum_{n \ge 1}{1\over n+1} \omega(m_n(\varphi,\cdots,\varphi),\varphi)$. This is a Batalin-Vilkovisky (BV) action if the degree of $\varphi$ is unconstrained, so it encodes also the gauge structure of the theory. The bar-cobar construction has a state space that does not 
    allow for a nondegenerate, degree three bilinear form. This is the case, for example, in the scalar field theory example, where the 
    bar-cobar vector space $T(s^{-1} T^c (V))$ has vectors of all negative degrees,
    given that $s^{-1} V$ has degree minus one and minus two elements~\eqref{degonedeg2}.
    Since there are no vectors of zero and higher degrees, a nondegenerate
    bilinear form cannot exist in that complex. 
    
    Still, one could have (even in the scalar model) an action principle 
    following from an bilinear form coupling 
    degree minus two and degree minus one states as in writing the term $\langle \Phi, (\Delta + B) \Phi \rangle$.  But this runs 
    into complications, too. If the bilinear form was diagonal in factor number and length,
    cyclicity seems unlikely to hold: $\langle \Phi_2 , \Phi_1 \otimes' \Phi_1\rangle $ could be nonzero, but the cyclically permuted
    $\langle \Phi_1 , \Phi_2 \otimes' \Phi_1 \rangle$ would be zero. Nor would it seem possible to have a self-adjoint $\Delta$ operator as $\langle \Phi_f , \Delta \Phi_{f-1}\rangle$ could be nonzero, but then moving $\Delta$ we would get $\langle \Delta \Phi_f , \Phi_{f-1} \rangle $ which is zero. 

    An important clue is the presence of redundancies in the equations of motion of the bar-cobar theory. For example, defining the degree minus two equation of motion ${\cal E} \equiv D \Phi + \Phi \otimes'\Phi$, with $D = \Delta + B$,
    we quickly find the redundancy 
    \be
    \label{bianchiiddlddl}
    D {\cal E} + \Phi \otimes' {\cal E} - {\cal E} \otimes' \Phi = 0\, .
    \ee
    If the equation of motion comes from an action $S$, the redundancy arises due to a gauge symmetry such that the variation of the action under the gauge transformation takes the form
    \be
    \delta_\Lambda  S = \langle \delta_\Lambda \Phi, {\cal E}  \rangle = 0 \,.
    \ee 
    With $\delta_\Lambda \Phi = D \Lambda    + \Phi \otimes' \Lambda - \Lambda\otimes' \Phi$, and $\Lambda$ of degree zero, a suitable bilinear form would imply that the Bianchi identity~\eqref{bianchiiddlddl} holds. 
    In the scalar field theory example, there is no gauge parameter and a modification of the theory would be needed to explain the redundancies.

  \item  {\bf Action within BV.}
  It could be fruitful to explore the existence of cubic actions within the BV formalism. At the perturbative level, cyclic $L_\infty$ algebras correspond to classical BV theories \cite{Zwiebach:1992ie,Grigoriev:2023lcc}.
    Similarly, cyclic $A_\infty$ algebras can be described as BV theories in the language of formal noncommutative geometry; see, for example~\cite{Gaberdiel:1997ia,Kajiura:2003ax}. 
    
    Classical BV theories come with an odd symplectic structure $\omega$ and a degree one vector field $Q$ satisfying $Q^2 = 0$, such that the action $S$ is the Hamiltonian of the vector field, i.e.~ 
    \be
    \text d S = i_Q \omega \, .
    \ee
    By nondegeneracy of $\omega$, the equations of motion $\text d S = 0$ are the points where $Q = 0$.

    $A_\infty$ and $L_\infty$ algebras without cyclic structure are encoded in the cohomological vector fields $Q$. The algebra relations are given by $Q^2 = 0$. In the BV language, the bar-cobar construction would produce a vector field $Q$ with coefficients that are polynomials up to quadratic order. To our knowledge, there is no result on whether there is also a cubic BV action $S$.

  \item 
    {\bf A picture for bar-cobar string fields.} Since the $A_\infty$ and $L_\infty$ formulations of string field theory have a natural geometric basis in terms of moduli spaces of punctured Riemann surfaces and a Batalin-Vilkovisky algebra of such spaces~\cite{Sen:1994kx}, it is reasonable to wonder if the bar-cobar construction, with its associated quadratic equations of motion, has a compelling geometrical picture in string field theory. 

    We claim that the type-I multilocal fields
    correspond to entangled states
    of tensor products of the associated CFT state space. Such entangled states are inserted across multiple punctures on a single Riemann surface. To see this consider the factor-number-one equations we have
    \be \label{dflkepi}  B \Phi_1   = 0 \,, \ \ \   \Delta \Phi_1 + \Phi_1 \otimes' \Phi_1 =0\,.
    \ee
        The first equation gives constraints at various lengths, as it demands that 
    \be
   B ( \Phi_{1,1} + \Phi_{1,2} + \Phi_{1,3} + \cdots ) = 0\,.
    \ee
    Since  $\Phi_{1,\ell} = s^{-1} \phi_{1,\ell} $ and $B s^{-1} = s^{-1} \bar B$
    we have
    \be
    \bar B ( \phi_{1,1} + \phi_{1,2} + \phi_{1,3} + \cdots ) = 0\,.
    \ee 
    The terms at length equal one must vanish, thus with $\bar B = \sum_n \bar B_n$,
    and the coderivations $\bar B_n$ defined by the multilinear products $b_n$, we have 
    \be
    \label{sftrealeqn}
    b_1 \phi_{1,1} + b_2 \phi_{1,2}  + b_3 \phi_{1,3} + \cdots = 0 \,.
    \ee 
    Let ${\cal H}$ denote the conformal field theory (CFT) state space spanned by vectors $V^i$ corresponding to local operators. 
    The first few string fields above take the form
    \be
    \phi_{1,1} = \sum_i \psi_i V^i \,, \ \ \phi_{1,2}= \sum_{i,j} \psi_{ij} V^i \otimes V^j    \,, \ \  \phi_{1,3} = \sum_{i,j,k} \psi_{ijk} V^i \otimes V^j \otimes V^k \,.
    \ee
    For $\ell \geq 2$ the $\phi_{1,\ell}$ are, in general, \emph{entangled} states in ${\cal H}^{\otimes \ell}$.  To see the role of Riemann surfaces consider, for example, the second term in equation~\refb{sftrealeqn}
    \be
    b_2 \phi_{1,2}=  
    b_2 \Bigl( \sum_{ij} \psi_{ij} V^i \otimes V^j\Bigr) 
     =  \sum_{ij} \psi_{ij} b_2 (V^i \otimes V^j) \,.\ee
    The picture of $b_2(V^i\otimes V^j)$ has a disk with three boundary punctures with $V^i$ inserted at one puncture, $V^j$ at another one, and the output product comes out as the state on the remaining puncture.  Similarly, in  $b_\ell \phi_{1,\ell}$ we have a chain in the  moduli space of disks with $(\ell+1)$--boundary punctures (or $(\ell+1)$--punctured Riemann spheres in the $L_\infty$ case) with the entangled state $\phi_{1,\ell}$ inserted across $\ell$ of the punctures.  Thus the type-I multilocal fields are indeed inserted across multiple punctures on a given Riemann surface.

    Entangled type-I states are off-shell states of the theory.  In canonical solutions, where type-II fields must vanish, we have seen that $\phi_{1, \ell} = (\phi_{1,1})^{\otimes \ell}$,
    entanglement disappears and the states become product states.  The $\phi_{1,\ell}$ become rank one tensors in ${\cal H}^{\otimes \ell}$. With $\phi_1\in T^c({\cal H})$
    a formal sum of elements $\phi_{1,\ell}$ of all lengths, $\phi_1$ is a direct sum of rank one elements, sometimes called a pure simple tensor.  The general solution of the bar-cobar equation of motion is thus gauge equivalent to a simple pure tensor. 

    The entanglement picture of type I states holds for arbitrary field theories: such states 
    are entangled states of tensor products of $V$, the vector space of the original field theory. It is in string theory, however, that the underlying vector space is itself the state space of a quantum field theory, 
    the two-dimensional CFT that describes the string background.  Thus, string theory type-I states
    are generally genuine quantum mechanical entangled states of the~CFT.

     This leaves the question of the interpretation of multilocal 
      type-II states---those associated with factor number greater than one.  Since these are (primed) tensor products
     of type I multilocal fields, general type-II states are entangled states of type I states.  If more than one $B$ operator acts
     on such states as in
     \be
     \sum \psi_{i_1 \cdots i_k; j_1\cdots j_p; 
     \cdots } \,  s^{-1} \bar B (V^{i_1} \otimes \cdots \otimes V^{i_k}) \ \otimes'  s^{-1} \bar B  (V^{j_1} \otimes \cdots \otimes V^{j_p}) 
     \otimes'  \cdots\,  , \ee 
     the natural interpretation of the above term would feature disconnected Riemann surfaces, with as many connected components as there are $B$ factors, and each one having  multi-puncture insertions.
    It would seem natural that a type-II state of factor number $f$ would be inserted on a disconnected surface comprising $f$ connected Riemann surfaces. 
     
     While more work would be needed to refine this picture and explain its relevance, we note that in the BV algebra of surfaces, disconnected surfaces are relevant and arise via the BV
     algebra dot product of connected surfaces, having a natural map into the space of string functionals~\cite{Sen:1994kx}.

\item
{\bf  A local bar-cobar construction?} An obvious question is whether it is possible to generalize the bar-cobar construction in a way so that the resulting differential graded associative algebra only contains local fields. For this, 
one must begin with a \emph{local} $A_\infty$ or $L_\infty$ 
algebra, the latter defined  
in~\cite{Costello:2011reft} and \cite{Butson:2016jys}. We will briefly try to convey the idea for $A_\infty$ algebras and outline how a \emph{local} bar-cobar construction could be performed within this setup, which could be used for local theories if one wants to preserve the locality property. 

To speak of locality 
we assume that fields $\Phi(x)$ are sections of some vector bundle on some spacetime manifold $M$. In that case, the fields can be pointwise multiplied by smooth functions $f \in C^\infty(M)$
\be
(f \Phi)(x) = f(x) \Phi(x) \, .
\ee
A field space $V$ with this structure is called a $C^\infty(M)$-module. Modules generalize vector spaces, where rescaling with scalars is replaced by rescaling elements of some algebra -- in our case, the algebra of functions. The tensor product of $C^\infty(M)$ modules, which is denoted by $\otimes_{loc}$, is the pointwise tensor product, i.e.
\be
(\Phi_1 \otimes_{loc} \Phi_2)(x) := \Phi_1(x) \otimes \Phi_2(x) \, ,
\ee
where the right-hand side uses the tensor product of vector spaces. The local tensor product has the property that
\be\label{eq:flineart}
f(\Phi_1 \otimes_{loc} \Phi_2) = (f \Phi_1) \otimes_{loc} \Phi_2 =  \Phi_1 \otimes_{loc} (f\Phi_2) \, .
\ee 
The normal tensor product has this property only for constant functions, i.e.~scalars.

Using the local tensor product in the bar and cobar construction instead of the normal tensor product, the fields in the bar-cobar theory will again be local. Further, the universal product $\otimes'_{loc}$ will also be local. However, in order for the differential to have a well-defined lift, first to $\bar B$ and then to $B$, the products $b_n$ are required to be multilinear not only with respect to scalars, but also with respect to functions. This means that we should require that
\be\label{eq:flinearity}
b_n(\Phi_1,\ldots,f\Phi_l,\ldots,\Phi_n) = f b_n(\Phi_1,\ldots,\Phi_l,\ldots,\Phi_n)\, ,
\ee
for all $l = 1,\ldots,n$ and $f \in C^\infty(M)$. 
One can readily confirm, for example, that one would need $b_1 (f \Phi) = f b_1(\Phi)$ in order to have
$\bar B_1(f\Phi_1 \otimes_{loc} \Phi_2) =\bar B_1(\Phi_1 \otimes_{loc} f\Phi_2)$.
Even for local field theories, however, the $b_n$ do not satisfy \eqref{eq:flinearity} because they usually contain derivatives. In particular, the differential $b_1$ is essentially never linear in functions, since it is given by the kinetic operator. In local field theories the $b_n$ are in general polydifferential operators of some finite order, meaning, they act as differential operators on each entry.

If the $b_n$ are polydifferential operators, they become linear in functions after passing to the so-called jet space, a space that includes derivatives of fields as new fields.  An example is useful.
Consider a $b_2$ defined as
\be
b_2 (\phi_1, \phi_2 ) (x) :=  \partial_x \phi_1 (x)  \partial_x \phi_2 (x) \,.
\ee  
This clearly fails to be linear in functions multiplying the fields. 
Here one defines the first jet space $j_x^1 \phi$ as a tuple containing the field and its derivatives as the two entries:
\be
j_x^1 \phi (x) = ( \Phi_{(0)}, \Phi_{(1)} )(x) :=  ( \phi, \partial_x \phi) \,. \ee
The new product $\tilde b_2$ is then defined as follows:
\be
\tilde b_2 ( j^1_x \phi_1, j^1_x \phi_2)  =  \tilde b_2 \Bigl(
( \Phi_{1(0)}, \Phi_{1(1)} )\,, ( \Phi_{2(0)}, \Phi_{2(1)} ) \Bigr)
=  \Phi_{1(1)} \Phi_{2(1)} \,.
\ee
The new product naturally replaces the old one since  $\tilde b_2 ( j_x^1 \phi_1 , j_x^1 \phi_2) = b_2 (\phi_1, \phi_2) $.
Moreover, given that $f j_x^1 \phi = ( f \Phi_{(0)} , f \Phi_{(1)})$  we see that $\tilde b_2$ is linear in $f$:  
\be \tilde b_2 ( f j_x^1 \phi_1 , j_x^1 \phi_2) = 
\tilde b_2 ( j_x^1 \phi_1 , f j_x^1 \phi_2) = f \, \Phi_{1(1)} \Phi_{2(1)} =   f \, \tilde b_2  ( j_x^1 \phi_1 , j_x^1 \phi_2)\,. 
\ee

More generally, the jet space will include fields $\Phi_M$, replacing
field derivatives encoded by a multiindex $M = (\mu_1\cdots\mu_n)$
\be
\Phi_M (x) := \ \partial_M \Phi(x) := \partial_{\mu_1} \cdots \partial_{\mu_n} \Phi(x)\, ,
\ee 
and the products $b_n$ contain no derivatives once expressed in terms of the $\Phi_M$. The bar-cobar construction can then be performed using the local tensor product and so this will result in a 
differential graded associative 
algebra that is also local. Since in the jet space formulation, derivatives of fields are essentially treated as new fields, the bar-cobar construction will generate type-I and type-II fields also for derivatives. We therefore expect that the local bar-cobar construction trades multilocal fields
\begin{equation}
    \Phi_{1,\ell}(x_1\ldots,x_\ell) = \phi_1(x_1) \otimes \cdots \otimes \phi_\ell(x_\ell)\, ,
\end{equation}
for local fields
\begin{equation}
    \Phi_{1,M_1,\ldots,M_\ell}(x) = \partial_{M_1} \phi_1(x) \otimes \cdots \otimes \partial_{M_\ell} \phi_\ell(x)=  \Phi_{1,M_1} \otimes \cdots \otimes \Phi_{\ell, M_\ell} \, , 
\end{equation}
that ultimately depend on the jets of the original algebra,
as indicated above for fields in factor number one.
    
\end{enumerate}
  
The bar-cobar construction applied to arbitrary field theories is remarkable in that it provides quadratic equations in which the interactions are universal and, in some sense, theory independent. The linear term is manifestly theory dependent, and the interactions of the original field theory are encoded there.  This is not all that different from the way string field theory is structured: the equation of motion, of type $QA + A\star A=0$ of a cubic string field theory has the background information encoded in the linear term through $Q$, and the interaction is formally background independent, if chosen to be of contact type.   The bar-cobar construction provides a conceptually clear universal construction that may give new insights into field theory.  It would be very intriguing if, in particular, the multilocal type-II fields turn out to have an important role to play.

\section*{Acknowledgments}
We would like to thank Atakan Fırat, Olaf Hohm, Branislav  Jur\v{c}o, J\'an Pulmann, Ivo Sachs, and Ashoke Sen for discussions and comments. We acknowledge the use of ChatGPT for proofreading this paper, and for improving the clarity of our presentation in Appendices A and C. 
 The research of C.C. is funded by the Deutsche Forschungsgemeinschaft (DFG, German Research Foundation), ``Homological Quantum Field Theory'', Projektnummer 9710005691. The
work of R.M. is supported by the MIT Dean of Science Fellowship and MIT Department of Physics.
The work of B.Z. was supported by the U.S. Department of Energy, Office of Science, Office of High
Energy Physics of U.S. Department of Energy under grant Contract Number DE-SC0012567.(High Energy Theory research). 

\appendix
\section{Properties of the homotopy contraction 
$\tilde{\mathbf{\emph{h}}}$}\label{appA}   

In this appendix, we prove two identities used in  
Section~\ref{thehomofDanddef}: the nilpotency
$\tilde h^2=0$, stated below~\eqref{hhiszero}, and the formula
\eqref{4tidn} for the graded commutator $[\Delta,\tilde h]$.
The proof becomes clear if we regard a state as a tensor word
with a specified set of cuts. We will keep track of both
degree and length of words.

Let $W$ be a state written as 
\begin{equation}
 W=s^{-1}A_1\mid\cdots\mid s^{-1}A_f,
 \qquad A_i\in C=T^c(V),
\end{equation}
where all $A_i$ are homogeneous tensor word of fixed length. We write
\begin{equation}
 \epsilon_i:=\deg A_i,
 \qquad
 \ell_i:=\operatorname{length}(A_i),
 \qquad
 \ell:=\sum_{i=1}^f\ell_i.
\end{equation}
As usual, all exponents of $(-1)$ are understood modulo two.  Now flatten the words $A_i$ in $W$ into a single word,
\begin{equation}
 A_1\cdots A_f=v_1\cdots v_\ell.
\end{equation}
Define a set $S$ of positions at which the vertical bars occur.
For this consider a set of integers $N_i$ taking 
values in the set $\{ 1, \ldots, \ell-1\}$, with $N_i$ representing
the number of vectors to the left of $A_{i+1}$.  Indeed, defining $ N_0:=0$,
\begin{equation}
 N_i:=\sum_{k=1}^i\ell_k,
 \qquad
 S:=\{N_1,\ldots,N_{f-1}\}\subseteq\{1,\ldots,\ell-1\}.
\end{equation}
We denote the same underlying word, with bars at the positions in an
arbitrary subset $S'\subseteq\{1,\ldots,\ell-1\}$, by $W_{S'}$.
In particular, $W=W_S$. Ordinary tensor products between letters
inside each factor are left implicit.

For every possible cut position $p=1,\ldots,\ell-1$, we define
$\Gamma_p$ as the sum of degrees of the vectors up to and including $v_p$,
we define $\nu_S(p)$ as  the number of cuts to the left of the cut $p$, 
and define $\alpha_p(S)$ as the total degree of the entries in the state
$W$ (that includes factors of $s^{-1}$) to the left of the cut $p$:
\begin{equation}
 \Gamma_p:=\sum_{r=1}^p\deg v_r,
 \qquad
 \nu_S(p):=\#\{q\in S\mid q<p\},
 \qquad
 \alpha_p(S):=\Gamma_p+\nu_S(p)+1.
\end{equation}
We introduce two local operators on the $W$ states. The operator $d_p$ inserts a bar at
$p$ or annihilates the state if there is already a bar at $p$, whereas $r_p$ removes a bar at $p$ or annihilates the state if there is no bar at $p$.  Including signs, they are defined as follows:
\begin{align}
 d_pW_S&:=
 \begin{cases}
 (-1)^{\alpha_p(S)}W_{S\cup\{p\}},&p\notin S,\\
 0,&p\in S,
 \end{cases}
 \label{eq:AppendixCutInsertion}\\[1ex]
 r_pW_S&:=
 \begin{cases}
 (-1)^{\alpha_p(S)}W_{S\setminus\{p\}},&p\in S,\\
 0,&p\notin S.
 \end{cases}
 \label{eq:AppendixCutRemoval}
\end{align}
With the sign conventions of the main text, these operators give
\begin{equation}
 \Delta W_S=\sum_{p=1}^{\ell-1}d_pW_S,
 \qquad
 \tilde hW_S=\sum_{p=1}^{\ell-1}r_pW_S.
 \label{eq:AppendixDeltaHByCuts}
\end{equation}
Let us verify the signs in~\eqref{eq:AppendixDeltaHByCuts}. 
The sign in $d_p$ is precisely the sign 
obtained by extending $\Delta$ as a
derivation and using~\eqref{deltaonCs}; this sign is $(-1)$ raised to the total degree of the part of $W_S$ left of the new bar. 
Moreover, the sign in $r_p$, when there is a bar to be deleted at $p$,
is precisely the sign given for $\tilde h$ in~\eqref{hhiszero}; 
$(-1)$ raised to the total degree of the part of $W_S$ left 
of the bar to be removed. 
Thus inserting a bar reproduces
$\Delta$, while removing a bar reproduces~$\tilde h$.

\medskip
\noindent
\textbf{Anticommutation at distinct cut positions.}
Suppose $p<q$. Toggling  the bar at $p$  (that is, either removing the bar if present, or adding the bar if not present) changes the number of bars to
the left of $q$ by one, modulo two, whereas toggling the bar at $q$
does not change the number of bars to the left of $p$. Hence
\begin{equation}
 \alpha_p(S)+\alpha_q(S\mathbin\triangle\{p\})
 =\alpha_q(S)+\alpha_p(S\mathbin\triangle\{q\})+1
 \qquad (\mathrm{mod}\ 2),
 \label{eq:AppendixToggleSign}
\end{equation}
where $\triangle$ denotes symmetric difference (union minus intersection of sets). In the above equation the left-hand side represents the total sign factor (the exponent of $(-1)$) when toggling 
at $p$ before toggling at $q$, and the right-hand side represents the total sign factor when toggling at $q$ before toggling at $p$. Therefore, whenever
two operations at distinct positions are allowed, performing them in
the opposite order gives the same bar pattern with the opposite sign.
In particular,
\begin{equation}
 r_pr_q+r_qr_p=0,
 \qquad
 d_pr_q+r_qd_p=0
 \qquad (p\neq q),
 \label{eq:AppendixLocalAnticommutation}
\end{equation}
when acting on any $W_S$. Equation~\eqref{eq:AppendixToggleSign} is the
explicit sign relation that pairs the two orders of operation; no
endpoint conventions are needed.

\medskip
\noindent
\textbf{Nilpotency of $\tilde h$.}
A given bar cannot be removed twice, so $r_p^2=0$. Using
\eqref{eq:AppendixDeltaHByCuts} and pairing the two possible orders of
removing any two distinct bars, we find
\begin{equation}
 \tilde h^2W_S
 =\sum_{\substack{p,q\in S\\p<q}}
 \bigl(r_qr_p+r_pr_q\bigr)W_S
 =0.
\end{equation}
This proves $\tilde h^2=0$ for every factor number, including the
endpoint cases $f=1$ and $f=2$.

\medskip
\noindent
\textbf{The commutator $[\Delta,\tilde h]$.}
Both $\Delta$ and $\tilde h$ have odd degree, and therefore
\begin{equation}
 [\Delta,\tilde h]=\Delta\tilde h+\tilde h\Delta.
\end{equation}
Contributions involving two distinct cut positions cancel by the second relation 
in~\eqref{eq:AppendixLocalAnticommutation}. At a fixed position $p$,
exactly one of the two compositions $d_pr_p$ and $r_pd_p$ is nonzero:
if $p\in S$, the first removes and then restores the bar; if
$p\notin S$, the second inserts and then removes it. Moreover,
$\nu_S(p)$ counts only bars strictly to the left of $p$, so it is
unchanged when the bar at $p$ itself is toggled. The two sign factors
are therefore equal, and their product is one. Consequently,
\begin{equation}
 \bigl(d_pr_p+r_pd_p\bigr)W_S=W_S
 \qquad
 (p=1,\ldots,\ell-1).
 \label{eq:AppendixSameCut}
\end{equation}
Summing over all possible cut positions gives
\begin{equation}
 [\Delta,\tilde h]W_S
 =\sum_{p=1}^{\ell-1}W_S
 =(\ell-1)W_S.
\end{equation}
Returning to the notation $\abs{A_i}=\ell_i$ and
$\abs{A}=\sum_i\abs{A_i}=\ell$ used in the main text, we obtain
\begin{equation}\label{4tidnx}
 [\Delta,\tilde h]
 \bigl(s^{-1}A_1\mid\cdots\mid s^{-1}A_f\bigr)
 =\bigl(\abs{A}-1\bigr)
 \bigl(s^{-1}A_1\mid\cdots\mid s^{-1}A_f\bigr).
\end{equation}
This is the identity~\eqref{4tidn}.

\section{Contracting homotopy for \texorpdfstring{$\Delta_\Phi$}{Delta-Phi}}\label{appB}

The proof that $h_\Phi$ is a contracting homotopy for $\Delta_\Phi$, 
as stated in~\eqref{conjconhom},
is given below. It follows as a special case from the homological perturbation lemma, for example, see \cite{crainic}. One version is the following. Suppose we are given a differential graded vector space $(V,\Delta)$ together with a homotopy satisfying
\begin{equation}\label{eq:SDRData}
    [\Delta,h] = 1-\boldsymbol{\pi} \, , \quad \boldsymbol{\pi}^2 = \boldsymbol{\pi} \, , \quad h\boldsymbol{\pi} = \boldsymbol{\pi}h = h^2 = 0 \, .
\end{equation}
Suppose further that we are given a deformation $\Delta \rightarrow \Delta_M = \Delta + M$, such that $\Delta_M^2 = 0$ and $1+h M$ is invertible. Then there is a homotopy $h_M$ and a projection $\boldsymbol{\pi}_M$ defined by
\begin{equation}
    h_M = 
    \frac{1}{1 + hM}h
    \, , \quad \boldsymbol{\pi}_M = \frac{1}{1 + hM}\boldsymbol{\pi}\frac{1}{1 + Mh} \, ,
\end{equation}
such that \eqref{eq:SDRData} is satisfied by replacing $(\Delta,h,\boldsymbol{\pi})$ with $(\Delta_M,h_M,\boldsymbol{\pi}_M)$.

In our case, the homotopy satisfies
\begin{equation}
    [\Delta,h] = 1 - \boldsymbol{\pi}_{1,1}\,,
\end{equation}
and the perturbation is given by $\Delta \rightarrow \Delta_\Phi := \Delta + \operatorname{ad}_\Phi$. We have $\Delta_\Phi^2 = 0$ because $\Phi$ is assumed to satisfy the Maurer-Cartan equation.
 From the definition of $\Delta_\Phi$, and equation~\eqref{chainmxx}, we have
 \begin{equation}
     \Delta_\Phi = \Delta + \operatorname{ad}_\Phi = \frac{1}{1+h\operatorname{ad}_\Phi}\Delta(1+h\operatorname{ad}_\Phi)\,.
 \end{equation}
 We then have, with $h_\Phi$ given in~\eqref{conjconhom},  
 \begin{equation}
     \Delta_\Phi h_\Phi = \frac{1}{1+h\operatorname{ad}_\Phi}\Delta(1+h\operatorname{ad}_\Phi) \frac{1}{1+h\operatorname{ad}_\Phi} h = \frac{1}{1+h\operatorname{ad}_\Phi}\Delta h\, .
 \end{equation}
We also have
\begin{equation}
    \begin{split}
        h_\Phi\Delta_\Phi = & \frac{1}{1+h\operatorname{ad}_\Phi} h(\Delta + \operatorname{ad}_\Phi) = \frac{1}{1+h\operatorname{ad}_\Phi} h\Delta + \frac{1}{1+h\operatorname{ad}_\Phi}h\operatorname{ad}_\Phi\\
        & = \frac{1}{1+h\operatorname{ad}_\Phi} h\Delta + 1 -\frac{1}{1+h\operatorname{ad}_\Phi}\, .
    \end{split}
\end{equation}
Now we can write the bracket, which happens to be an anticommutator 
\begin{equation}
    \begin{split}
        [\Delta_\Phi, h_\Phi] & = \frac{1}{1+h\operatorname{ad}_\Phi} [\Delta, h] + 1 -\frac{1}{1+h\operatorname{ad}_\Phi}\\
        & = 1 -\frac{1}{1+h\operatorname{ad}_\Phi}\,
  \boldsymbol{\pi}_{1,1}\,.
    \end{split}
\end{equation}
This indeed shows that $h_\Phi$ is the contracting homotopy for the differential $\Delta_\Phi$, as written in~\eqref{conjconhom}.

\section{Inductive proof of the general solution}\label{inductproofxx9}

We now prove that~\eqref{guessforA},
\begin{equation}
\label{guessforAapp}
    \chi
    =
    \frac{1}{1+M}
    \bigl(
        \Delta_{\Phi_1}+B+\widetilde W_1
    \bigr)
    (1+M)\,,
\end{equation}
is the most general formal solution of the equation of
motion~\eqref{eq:forchi},
\begin{equation}
\label{eqtosolveapp}
    \Delta_{\Phi_1}\chi+B\chi+\chi\chi=0\,,
\end{equation}
when \(B\Phi_1=0\). In~\eqref{guessforAapp} the factor-number-one state  \(\widetilde W_1\) acts by left
multiplication. 
Recall that \(M\) and \(\widetilde W_1\) are expanded in orders in \(B\): $
    M=\sum_{i=1}^{\infty}M^{(i)}\,,
    \widetilde W_1
    =
    \sum_{i=1}^{\infty}\widetilde W_1^{(i)}\,,
    \widetilde W_1^{(1)}=W_1^{(1)}\,.
$
For any positive integer \(r\), we introduce the notation
\begin{equation}
\label{truncateddataapp}
    M_{\leq r}
    :=
    \sum_{i=1}^{r}M^{(i)}\,,
    \qquad
    \widetilde W_{1,\leq r}
    :=
    \sum_{i=1}^{r}\widetilde W_1^{(i)}\,.
\end{equation}
The superscript on a parenthesized expression will denote its
component of the indicated order in \(B\). The operator
\(\Delta_{\Phi_1}\) preserves this order, whereas \(B\) raises it by
one.

We proceed by induction. Our induction hypothesis is that, through
order \(n\), the components of \(\chi\) are given by
\begin{equation}
\label{solnnnn}
    \chi^{(k)}
    =
    \left[
    \frac{1}{1+M_{\leq n}}
    \left\{
        \bigl(\Delta_{\Phi_1}+B\bigr)(1+M_{\leq n})
        +
        \widetilde W_{1,\leq n}(1+M_{\leq n})
    \right\}
    \right]^{(k)},
    \qquad
    k\leq n\,.
\end{equation}
The data entering~\eqref{solnnnn} are required to satisfy the two
conditions~\eqref{conditionsonW11} through order \(n\):
\begin{equation}
\label{condhypxx}
    B\widetilde W_1^{(k)}=0\,,
    \qquad
    \left[
        \Delta_{\Phi_1}\widetilde W_{1,\leq n}
        +
        \widetilde W_{1,\leq n}
        \widetilde W_{1,\leq n}
    \right]^{(k)}
    =
    0\,,
    \qquad
    k\leq n\,.
\end{equation}

We shall prove that this hypothesis determines the component
\(\chi^{(n+1)}\), together with new data \(M^{(n+1)}\) and
\(\widetilde W_1^{(n+1)}\) that satisfy~\eqref{condhypxx}, so that
\eqref{solnnnn} continue to hold at order
\(n+1\). The base cases for \(\chi^{(1)}\) and \(\chi^{(2)}\) were
established in \eqref{y1finform} and \eqref{a2form3oi},
respectively.

The equation of motion~\eqref{eqtosolveapp} at order \(n+1\) reads
\begin{equation}
\label{nplus1}
    \Delta_{\Phi_1}\chi^{(n+1)}
    =
    -B\chi^{(n)}
    -
    (\chi\chi)^{(n+1)}\,.
\end{equation}
We note that $(\chi\chi)^{(n+1)}$ involves only components $\chi^{(k)}$ with $k \leq n$, so
the full right-hand side above is completely determined by the induction hypothesis.

We denote by \(\Xi^{(n+1)}\) the order-\((n+1)\) term obtained by
inserting only the already determined data into the proposed
solution:
\begin{equation}
\label{Xidefapp}
    \Xi^{(n+1)}
    :=
    \left[
    \frac{1}{1+M_{\leq n}}
    \left\{
        \bigl(\Delta_{\Phi_1}+B\bigr)(1+M_{\leq n})
        +
        \widetilde W_{1,\leq n}(1+M_{\leq n})
    \right\}
    \right]^{(n+1)}.
\end{equation}
In particular, \(\Xi^{(n+1)}\) contains neither \(M^{(n+1)}\) nor
\(\widetilde W_1^{(n+1)}\).

To get going we need to show that for arbitrary \(M\) and \(\widetilde W_1\), we have
\begin{equation}
\label{identityapp}
\begin{split}
&
\bigl(\Delta_{\Phi_1}+B\bigr)
\left[
\frac{1}{1+M}
    \bigl(\Delta_{\Phi_1}+B +\widetilde W_1\bigr)(1+M)
\right]
+
\left[
\frac{1}{1+M}
    \bigl(\Delta_{\Phi_1}+B+ \widetilde W_1\bigr)(1+M)
\right]^2
\\
&\qquad \qquad \qquad \qquad
=
\frac{1}{1+M}
\left\{
    \Delta_{\Phi_1}\widetilde W_1
    +
    B\widetilde W_1
    +
    \widetilde W_1\widetilde W_1
\right\}
(1+M)\,.
\end{split}
\end{equation}
This equation makes sense: the right-hand side of~\eqref{guessforAapp} is a gauge transformation of $\widetilde W_1$
and should satisfy the Maurer-Cartan equation if $\widetilde W_1$ does.   
To verify this identity, one uses the derivation property of $\Delta_{\Phi_1} + B$ and
\([B,\Delta_{\Phi_1}]=0\), which holds because
\(B\Phi_1=0\).

We now apply~\eqref{identityapp} with
\begin{equation}
\label{auxeihjh}
    M=M_{\leq n}\,,
    \qquad
    \widetilde W_1=\widetilde W_{1,\leq n}\,.
\end{equation}
Since the induction hypothesis requires
\begin{equation}\label{inductivehypothesis}
    B\widetilde W_{1,\leq n}=0\,,
\qquad
    \left[
        \Delta_{\Phi_1}\widetilde W_{1,\leq n}
        +
        \widetilde W_{1,\leq n}
        \widetilde W_{1,\leq n}
    \right]^{(k)}
    =
    0\,,
    \qquad
    k\leq n\,,
\end{equation}
the first possibly nonzero component of the expression
in braces on the right-hand side of~\eqref{identityapp} occurs at
order \(n+1\).
Because \(\widetilde W_{1,\leq n}\) contains no
order-\((n+1)\) component,
\begin{equation}
\label{quadraticremainderapp}
\begin{split}
&
\left[
    \Delta_{\Phi_1}\widetilde W_{1,\leq n}
    +
    \widetilde W_{1,\leq n}
    \widetilde W_{1,\leq n}
\right]^{(n+1)}
=
\left[
    \widetilde W_{1,\leq n}
    \widetilde W_{1,\leq n}
\right]^{(n+1)}
=
\sum_{\substack{i+j=n+1\\ i,j\geq1}}
\widetilde W_1^{(i)}
\widetilde W_1^{(j)}\,.
\end{split}
\end{equation}
This expression involves only previously determined data. Moreover,
for the full right-hand side of~\eqref{identityapp}
\begin{equation}
 \Bigl(  \frac{1}{1+M_{\leq n}}
\left\{
   \Delta_{\Phi_1}\widetilde W_1
    +
    B\widetilde W_1
    +
    \widetilde W_1\widetilde W_1
\right\}
(1+M_{\leq n})  \Bigr)^{(n+1)} =  \sum_{\substack{i+j=n+1\\ i,j\geq1}}
\widetilde W_1^{(i)}
\widetilde W_1^{(j)}\,,
\end{equation}
since multiplication on either side by a positive-order term from
\((1+M_{\leq n})^{-1}\) or \(1+M_{\leq n}\) would raise the order
to at least \(n+2\). 

The components through order \(n\) of the expression in
\eqref{solnnnn} agree with the corresponding components of
\(\chi\). It follows from~\eqref{identityapp}, with~\eqref{auxeihjh}, that
\begin{equation}
\label{approxeomapp}
    \Delta_{\Phi_1}\Xi^{(n+1)}
    +
    B\chi^{(n)}
    +
    (\chi\chi)^{(n+1)}
    =
    \left[
        \widetilde W_{1,\leq n}
        \widetilde W_{1,\leq n}
    \right]^{(n+1)}.
\end{equation}
Subtracting~\eqref{approxeomapp} from the exact order-\((n+1)\)
equation~\eqref{nplus1}, we obtain
\begin{equation}
\label{leftoverapp}
    \Delta_{\Phi_1}\chi^{(n+1)}
    =
    \Delta_{\Phi_1}\Xi^{(n+1)}
    -
    \left[
        \widetilde W_{1,\leq n}
        \widetilde W_{1,\leq n}
    \right]^{(n+1)}.
\end{equation}

The quadratic remainder in~\eqref{leftoverapp} is
\(\Delta_{\Phi_1}\)-closed. To see this explicitly, recall that
every \(\widetilde W_1^{(i)}\) has odd degree. The graded Leibniz
rule gives
\begin{align}
&
\Delta_{\Phi_1}
\left[
    \widetilde W_{1,\leq n}
    \widetilde W_{1,\leq n}
\right]^{(n+1)}
=
\left[
    \bigl(
        \Delta_{\Phi_1}\widetilde W_{1,\leq n}
    \bigr)
    \widetilde W_{1,\leq n}
    -
   \widetilde W_{1,\leq n}
    \bigl(
        \Delta_{\Phi_1}\widetilde W_{1,\leq n}
    \bigr)
\right]^{(n+1)} = 0,
\label{closedremainderapp}
\end{align}
where the second equality uses the second condition of~\eqref{inductivehypothesis}.
Since the homology of
\(\Delta_{\Phi_1}\) is trivial above factor number one, there exists
a factor-number-one state \(\overline W^{(n+1)}_1\) such that
\begin{equation}
\label{Wbardefapp}
    \left[
        \widetilde W_{1,\leq n}
        \widetilde W_{1,\leq n}
    \right]^{(n+1)}
    =
    \Delta_{\Phi_1}\overline W^{(n+1)}_1\,. 
\end{equation}
Substituting~\eqref{Wbardefapp} into~\eqref{leftoverapp} gives
\begin{equation}
    \Delta_{\Phi_1}
    \left(
        \chi^{(n+1)}
        -
        \Xi^{(n+1)}
        +
        \overline W^{(n+1)}_1
    \right)
    =
    0\,.
\end{equation}
The expression in parentheses is therefore the sum of a
factor-number-one homology representative \(W_1^{(n+1)}\) and an
exact term \(\Delta_{\Phi_1}M^{(n+1)}\). Hence
\begin{equation}
\label{chinp1app}
    \chi^{(n+1)}
    =
    \Xi^{(n+1)}
    +
    \Delta_{\Phi_1}M^{(n+1)}
    -
    \overline W^{(n+1)}_1
    +
    W_1^{(n+1)}\,.
\end{equation}
We now define
\begin{equation}
\label{newdataapp}
    \widetilde W_1^{(n+1)}
    :=
    W_1^{(n+1)}
    -
    \overline W^{(n+1)}_1\,,  
\end{equation}
so that
\begin{equation}
\label{chinp1appXX}
    \chi^{(n+1)}
    =
    \Xi^{(n+1)}
    +
    \Delta_{\Phi_1}M^{(n+1)}
    +
    \widetilde W_1^{(n+1)}\,, 
\end{equation}
and enlarge the truncated data according to
\begin{equation}
\label{enlargeddataapp}
    M_{\leq n+1}
    =
    M_{\leq n}
    +
    M^{(n+1)}\,,
    \qquad
    \widetilde W_{1,\leq n+1}
    =
    \widetilde W_{1,\leq n}
    +
    \widetilde W_1^{(n+1)}\,.
\end{equation}
With this enlarged data the proposed form of the solution 
at order $n+1$ from the induction hypothesis would be 
\begin{align}
&
\left[
\frac{1}{1+M_{\leq n+1}}
\left\{
    \bigl(\Delta_{\Phi_1}+B\bigr)(1+M_{\leq n+1})
    +
    \widetilde W_{1,\leq n+1}(1+M_{\leq n+1})
\right\}
\right]^{(n+1)}
\nonumber\\[1.0ex]
&\qquad =
\Xi^{(n+1)}
+
\Delta_{\Phi_1}M^{(n+1)}
+
\widetilde W_1^{(n+1)}
=
\chi^{(n+1)}\,,
\end{align}
In passing to the second line we used the definition of $\Xi^{(n+1)}$ in~\eqref{Xidefapp} and accounted for the new terms that appear at order \(n+1\) on account of the enlargement of the data. The new terms are  
\(\Delta_{\Phi_1}M^{(n+1)}\) and
\(\widetilde W_1^{(n+1)}\), which are manifest. There are
no more terms generated solely between the braces, and
the change in the formal inverse contributes only at
order \(n+2\) or higher.
The last equality in the above equation follows from
\eqref{chinp1appXX}. This proves the first item 
\eqref{solnnnn} in the induction at order \(n+1\).

The quadratic condition in~\eqref{condhypxx} is satisfied by construction as can be seen by acting on~\eqref{newdataapp} with $\Delta_{\Phi_1}$ and using~\eqref{Wbardefapp}.  The linear condition,  which guarantees that the subsidiary condition~\eqref{bcondvnvg} holds, requires us to impose $ B\widetilde W_1^{(n+1)} = 0$.

This completes the inductive step. Together with the base cases, it
shows that every formal solution of~\eqref{eqtosolveapp} is of the
form~\eqref{guessforA}, with \(\widetilde W_1\) satisfying
\eqref{conditionsonW11}.

\bibliographystyle{JHEP}

\bibliography{ref}

\end{document}